\documentclass[aps,pre,nofootinbib,reprint,superscriptaddress]{revtex4-2}
\usepackage[utf8]{inputenc}
\usepackage{graphicx}
\usepackage{subfigure}
\usepackage{xcolor}
\usepackage{subfigure}
\usepackage{placeins}
\usepackage{float}

\usepackage[
colorlinks=true,        
citecolor=blue,         
linkcolor=blue,         
urlcolor=blue           
]{hyperref}             
\usepackage{xcolor}

\usepackage{enumitem}

\newcommand{\be}{\begin{eqnarray}}
\newcommand{\ee}{\end{eqnarray}}

\newcommand{\wbe}{\begin{widetext}}
\newcommand{\wee}{\end{widetext}}

\newcommand{\eq}[1]{(\ref{#1})}

\usepackage{bbm}
\usepackage{amsmath,leftidx}
\usepackage{graphicx}
\usepackage{times}
\usepackage{CJK}
\usepackage{color,colortbl}
\usepackage[normalem]{ulem}

\begin{document}

\title{Subsystem Thermalization Hypothesis in Quantum Spin Chains with Conserved Charges}

\author{Feng-Li Lin}
\email{fengli.lin@gmail.com}
\affiliation{Department of Physics, \\
National Taiwan Normal University, Taipei, 11677, Taiwan}

\author{Jhh-Jing Hong}
\email{hongjhhjing06132061@gmail.com}
\affiliation{Department of Physics, \\
National Taiwan Normal University, Taipei, 11677, Taiwan}

\author{Ching-Yu Huang}
\email{cyhuangphy@thu.edu.tw}
\affiliation{Department of Applied Physics, \\
Tunghai University, Taichung 40704, Taiwan}

\begin{abstract}

We consider the thermalization hypothesis of pure states in {non-integrable quantum spin chains with conserved charges, including the non-intergable variants of the} Ising chain with $Z_2$ symmetry, XXZ chain with $U(1)$ symmetry, and XXX chain with $SU(2)$ symmetries. Two kinds of pure states are considered:  the energy eigenstates and the typical states evolved unitarily from the random product states for a long enough period. We further group the typical states by their expectation values of the conserved charges and consider the fine-grained thermalization hypothesis. We compare the locally (subsystem) reduced states of typical states/eigenstates with the ones of the corresponding thermal ensemble states. Besides the usual thermal ensembles such as the (micro-)canonical ensemble without conserved charges and the generalized Gibbs ensemble (GGE) with all conserved charges included, we also consider the so-called partial-GGEs (p-GGEs), which include only part of the conserved charges in the thermal ensemble. Moreover, in the framework of p-GGE, the Hamiltonian and other conserved charges are on an equal footing. The introduction of p-GGEs extends quantum thermalization to a more general scope. The validity of the subsystem thermalization hypothesis can be quantified by the smallness of the relative entropy of the reduced states obtained from the GGE/p-GGE and the typical states/eigenstates. We examine the validity of the thermalization hypothesis by numerically studying the relative entropy demographics. We show that the thermalization hypothesis holds generically for small enough subsystems for various p-GGEs. Thus, our framework extends the universality of quantum thermalization.

\end{abstract}

\date{\today}

\maketitle


\section{Introduction}

The thermalization of a classical many-body system is governed by the second law of thermodynamics and is also closely linked to chaos and ergodicity. For quantum systems, a long-evolved open subsystem interacting with the environment displays similar thermalization behaviors, which can also be understood as the spreading of entanglement across the boundary between the subsystem and the environment. On the other hand, the isolated quantum system evolves unitarily and appears unable to thermalize. Among them, the energy eigenstate is stationary and should not be able to thermalize. Inspired by von Neuman's idea of trying to generalize the ergodicity theorem to quantum mechanics \cite{von2010proof}, the so-called eigenstate thermalization hypothesis (ETH) was proposed \cite{Deutsch:1991msp, Srednicki:1994mfb,  Srednicki:1995pt, Srednicki_1999}, which states that, even in isolated quantum system, the expectation value of a local observable in a single energy eigenstate is well approxemated by the microcanonical ensemble average with finite energy density, up to corrections that vanish in the thermodynamic limit. Since then, the thermalization of isolated quantum systems has been intensively studied over the past few decades; see \cite{rigol2008thermalization, DAlessio:2015qtq, Gogolin:2015gts, Mori:2017qhg, deutsch2018eigenstate} and the references therein. Surprisingly, ETH seems quite general. The ETH implies that the reduced state in a small region of an energy eigenstate can be approximated by the (reduced) state of some thermal ensemble state, such as the (micro-)canonical ensemble. This can be viewed as the subsystem version of ETH \cite{Dymarsky:2016ntg, Lashkari:2016vgj, He:2017vyf}, i.e., by treating a local region as a subsystem of the entire system. The validity of (subsystem) ETH implies that an energy eigenstate locally looks like thermal.

Energy eigenstates comprise only a tiny portion of the total Hilbert space of an isolated many-body system. It is natural to ask if the (subsystem) thermalization also happens to the typical states, which can be practically obtained by the long-time evolution of some initial non-eigenstate.  This was proposed in \cite{Goldstein:2005aib} as the {\it canonical typicality} to conjecture that a typical state is locally thermal.\footnote{{The typical states adopted in \cite{Goldstein:2005aib, popescu2006entanglement, Mueller:2013bww} for non-integrable systems are the Haar-random states from the system's Hilbert (sub-)space. However, due to the limited computing power and the purpose of this paper, we will prepare the typical states by evolving the product of on-site random states, which sample the system's Hilbert space well, for a sufficiently long time to saturate the local entanglement. This should be sufficient for considering the thermalization of local observables due to the similar mechanism of scrambling by unitary evolution proposed in \cite{Hayden:2007cs, Sekino:2008he}, with supporting examples as demonstrated in \cite{rigol2008thermalization, santos2010onset, kaufman2016quantum}}.} The subsystem thermalization of a typical state of an isolated system may be understood by treating the subsystem as an open system surrounded by the rest. The open system is expected to equilibrate with the surrounding environment due to their mutual interaction, allowing the entanglement to grow and spread across the boundary. Despite that, the equilibrium state may not be some thermal ensemble state, but some pointer state preferred by the environment's einselection \cite{Zurek:1981xq}. This is expected because the environment here is just complementary to the subsystem of a typical state and may not behave as a thermal bath. However, inspired by the subsystem ETH, the reduced state of a subsystem in a typical state can resemble the reduced state of some thermal ensemble state \cite{popescu2006entanglement, Mueller:2013bww}.

It has been known that the above (subsystem) ETH fails\footnote{Or, it holds only weakly (weak ETH) \cite{biroli2010effect,  mori2016weak, iyoda2017fluctuation, He:2017vyf}, i.e., the distance between two reduced states vanishes by the power law of the number of degrees of freedom, not by the exponential decay as for the strong ETH. {Besides, there exist atypical states, which comprise an exponentially small fraction in system size and violate weak ETH \cite{mori2016weak}.}} for the integrable systems \cite{biroli2010effect, mori2016weak, iyoda2017fluctuation, He:2017vyf, He:2017txy, Basu:2017kzo, Guo:2018pvi, Muralidharan:2016acb, Byju:2018eyb} due to the infinite number of conserved charges prohibiting quantum chaos for thermalization. It also fails for many-body localized systems \cite{Abanin_2019, Nandkishore_2015, Sierant_2025, Protopopov_2020} and chaotic systems in the presence of scars \cite{Serbyn_2021, Moudgalya_2018, Moudgalya_2021, Shiraishi_2017, Moudgalya_2022, Chandran_2023}, with different underlying mechanisms, such as strong disorders for the former, or Hilbert space fragmentation due to emerging spectrum generating algebra for the latter. One expects the ETH to hold for the generalized Gibbs ensemble (GGE), which includes all conserved charges \cite{Dymarsky:2019etq, chen2024subsystem}. However, such GGE states are hard to implement due to the infinite number of conserved charges, {and one may want to adopt some truncated GGE states. Indeed, some studies on the approximation of GGE states by the truncated GGEs have been done in \cite{fagotti2013reduced, essler2017truncated, piroli2017integrable} by including some semi-local charges or imposing integrable initial conditions.} Similarly, {we may expect the integrable systems' typical states also satisfy the subsystem thermalization hypothesis with GGE or some truncated GGE.} This then raises the issue of the role of the conserved charges in the thermalization of energy eigenstates or typical states.

With the additional conserved charges other than Hamiltonian, there are more options when considering the thermalization hypothesis in various aspects. Firstly, the energy eigenstates will be classified into the superselection sectors of the conserved charges. When considering ETH, we can choose to restrict to the superselection sectors or not. We can also generalize the concept of superselection sectors to the typical states. Although typical states, in general, are not eigenstates of energy and charges, we can still classify them into almost superselection sectors by their expectation values of the charges. 

Secondly, we can choose different thermal ensemble states to examine the thermalization hypothesis for either eigenstates or typical states. A conserved quantity and its chemical potential form a conjugate pair when constructing a thermal ensemble; i.e., we can fix either charge or chemical potential, but not both. Or, we can ignore both.  For example, when there is no conserved quantity other than the Hamiltonian, we can either fix energy to yield a microcanonical ensemble state or fix (inverse) temperature to yield a canonical ensemble state. We can adopt either to consider the thermalization hypothesis. Thus, with more conserved quantities, we can fix either one for each conjugate pair or ignore both to yield different thermal ensembles. If we fix all the chemical potentials, including the inverse temperature, we will obtain the GGE. Otherwise, we will obtain the so-called partial-GGE (p-GGE).  The freedom to choose GGE or various p-GGE implies that one can choose to ignore specific information when considering the thermalization hypothesis. As most of the studies on the thermalization hypothesis are based on GGE \cite{Dymarsky:2019etq, chen2024subsystem, yunger2016microcanonical,  2020PhRvR...2c3403F, Murthy:2022dao, PhysRevE.101.042117, Majidy:2023xhm, Lasek:2024ess, Upadhyaya:2023kjv}, we propose in this paper to study the thermalization hypothesis of typical states and energy eigenstates by comparing them with various p-GGE ensemble states. Our results indicate that the thermalization hypothesis is effective in most cases. This implies that the subsystem thermalization hypothesis can be extended from GGE to p-GGEs. This then enlarges the scope of quantum thermalizations for the systems with conserved charges.

Finally, as we have the option of choosing (almost) superselection sectors for the quantum states and the p-GGEs, we should not just examine the thermalization of a single state but shall evaluate the thermalization hypothesis by the demographics of the relative entropy measures for a given set of states in some (almost) superselection sector and a chosen p-GGE. Here, we adopt the term "demographic" to mean the population distribution of the chosen ensemble typical states/eigenstates over the values of the relative entropy. If the demographic mostly peaks around the regime of vanishing relative entropy, the thermalization hypothesis works for the chosen ensemble states and p-GGE. 


{
In particular, experiments such as those by Kaufman et al.\cite{Kaufman2016} have demonstrated the applicability of the quantum thermalization framework in cold atom and Rydberg atom platforms, where coherent many-body dynamics and local observables can be directly probed. 
Other platforms, including trapped ion systems\cite{Govinda2016} and superconducting qubit arrays~\cite{Neill2016}, also provide a high degree of control over isolated quantum many-body systems. 
These setups enable the preparation of specific initial states, coherent unitary evolution, and local measurements, providing an environment for investigating thermalization phenomena.

In parallel, recent experimental advances have made it possible to explore quantum many-body systems with multiple conserved quantities or near-integrable dynamics. For example, Kinoshita, Wenger, and Weiss~\cite{Kinoshita:2004fnt} observed the absence of thermalization in a 1D Bose gas, a hallmark of integrability often referred to as a "quantum Newton’s cradle." Gring et al.\cite{Gring2012} studied prethermalization in isolated quantum systems, while Langen et al.\cite{Langen2015} provided experimental evidence for the emergence of a generalized Gibbs ensemble (GGE), which incorporates multiple conserved quantities beyond energy. More recently, Scholl et al.~\cite{Scholl2022} used programmable Rydberg atom arrays to simulate two-dimensional antiferromagnetic spin models, enabling the study of constrained dynamics and engineered local symmetries. These results highlight that nontrivial conserved structures are not only theoretically relevant but also experimentally realizable, thereby motivating further investigations into their role in quantum thermalization.
}

In this paper, we will study quantum thermalization in the above setup by considering the non-integrable spin chain systems with either discrete or continuous Abelian and non-Abelian symmetries. In particular, the thermalization hypothesis for systems with non-Abelian conserved charges has been studied recently \cite{yunger2016microcanonical, 2020PhRvR...2c3403F, Murthy:2022dao, PhysRevE.101.042117, Majidy:2023xhm, Lasek:2024ess, Upadhyaya:2023kjv}, among which a GGE-like state called non-Abelian thermal state (NATS) has been proposed for ETH. In our framework with p-GGE states, we will numerically study the subsystem thermalization hypothesis for the typical states, (almost-)superselection states, and energy eigenstates of non-integrable variants of the quantum Ising and XXZ chains by comparing their locally reduced states with those of various p-GGE thermal states. The demographics of such comparisons will reveal how general the thermalization of isolated quantum systems with conserved charges is.

The rest of the paper is organized as follows. In Section \ref{sec2}, we will outline our framework for examining the subsystem thermalization hypothesis and its relationship with the generalized second law of thermodynamics, as well as our numerical implementation methodology. We then present our numerical results of the thermalization demographics for the (non-integrable variants of) Ising chain with $Z_2$ conserved charges, the XXZ chain with $U(1)$ charges, and the XXX chain with $SU(2)$ charges, in Section \ref{sec3:Ising}, \ref{sec4:XXZ_U1} and \ref{sec5:XXX}, respectively. We will see that the subsystem thermalization hypothesis holds well for the local region of small sizes in most cases.  Finally, we conclude our paper in Section \ref{sec:fin}. In Appendix \ref{App_A}, we conduct a size-scaling study of the thermalization demographics of Ising chains by varying the system size. To make the main text more concise, we put some typical-state thermalization demographics for the cases of XXZ chains in Appendix \ref{App_B} with $U(1)$ charge and in Appendix \ref{App_C} with parity symmetry.

\section{Thermalization of Quantum Systems with Conserved Charges}\label{sec2}

We want to consider the thermalization of typical states in a system with conserved charges. In this section, we first review the basics of thermalization for systems without conserved charges and then generalize to those with conserved charges. 

There are many ways to consider the thermalization of a quantum system, and the key point is to pick up a thermal ensemble state to compare with the given typical state on the agreement of local observables or, more directly, the reduced density matrices of some local regions. Take the eigenstate thermalization hypothesis (ETH) \cite{Deutsch:1991msp, Srednicki:1995pt, Srednicki_1999}  as an example. One compares a given energy eigenstate $|E \rangle$ of energy $ E$ with a thermal ensemble state, which can be either a microcanonical or canonical ensemble. It is natural to choose the energy of the micro-canonical ensemble to center around  $E$ with a small window $\Delta E \le E$. The ETH works if both diagonal and off-diagonal matrix elements of a set of local operators $\mathcal{O}$ in the basis of eigenstates satisfy the following Srednicki's ansatz \cite{Srednicki_1999},
\begin{align}\label{S_ETH}
    \langle E_\alpha| \mathcal{O} |E_\beta\rangle \simeq  \langle \mathcal{O \rangle_{\rm mce}}(\overline{E}) \delta_{\alpha\beta} + e^{-S_{\rm th}(\overline{E})/2} R_{\alpha\beta} f(\omega, \overline{E}),
\end{align}
where ${\rm mce}$ denotes the abbreviation of micro-canonical ensemble, $\overline{E} = (E_\alpha + E_\beta)/2$ denotes the average eigenenergy, $\omega = E_\alpha - E_\beta$ is the energy difference, $S_{\rm th}(\overline{E})$ is the thermodynamics entropy, $R_{\alpha\beta}$ is a normal distributed random number, and $f(\omega, \overline{E})$ is a smooth spectral function.

On the other hand, we can consider the so-called subsystem ETH \cite{Dymarsky:2016ntg, Lashkari:2016vgj,  He:2017vyf} based on the canonical ensemble
\be\label{rho_ce}
\rho_{\rm ce}={e^{-\beta H} \over {\rm Tr} \; e^{-\beta H}} 
\ee
which is specified by the system's Hamiltonian $H$ and the inverse temperature $\beta$. The latter should be fixed by requiring the average energy to be the energy $E$ of the energy eigenstate $|E\rangle$, i.e.,  
\be
E=\langle E| H |E\rangle ={\rm Tr}[\rho_{\rm ce} H]\;.
\ee
Unlike the microcanonical ensemble, the canonical ensemble allows energy fluctuations but fixes the inverse temperature.  The subsystem ETH holds if the reduced density matrices for a small local region $A$ (with its complement denoted by $\overline{A}$) agree, i.e.,
\be
S\Big[ {\rm Tr}_{\overline{A}} |E\rangle \langle E \vert \Big\Vert  {\rm Tr}_{\overline{A}} \rho_{\rm ce} \Big] \simeq 0 
\ee
where $S\big[ \rho_1\Vert \rho_2 \big]$ is the relative entropy between $\rho_1$ and $\rho_2$. A  tighter measure than the relative entropy for the difference between $\rho_1$ and $\rho_2$ is the trace distance $t_{12}:={\rm tr}|\rho_1-\rho_2|$. However, it is more difficult to evaluate. Despite that, by Pinsker's inequality \cite{Cover2006}
\be
t^2_{12} \le 2 S\big[ \rho_1\Vert \rho_2 \big]\;,
\ee
we can evaluate the relative entropy as the upper bound to the trace distance. 

Besides (subsystem) ETH, one can consider the thermalization for a typical pure state  $|\Psi \rangle$, which is not an eigenstate or a product state. Then, the canonical ensemble or its $\beta$ can be fixed by
\be
\langle \Psi| H |\Psi\rangle ={\rm Tr}[\rho_{\rm ce} H]
\ee
and the typical state $|\Psi \rangle$ behaves thermally inside $A$ if 
\be
S\Big[{\rm Tr}_{\overline{A}} |\Psi \rangle \langle \Psi| \Big\Vert {\rm Tr}_{\overline{A}} \rho_{\rm ce}  \Big] \simeq 0 \;.
\ee
In \cite{caceres2024generic}, the above thermalization hypothesis is proposed and termed the generic ETH, which has been verified for long-time evolved product states. In this work, we refer to it as the subsystem thermalization hypothesis, a generalization of the subsystem ETH to include both eigenstates and typical states. For the subsystem thermalization hypothesis to hold, the size of $A$ should be significantly smaller than that of the system. In Appendix A, we conduct a limited size-scaling study to provide evidence that the thermalization hypothesis is more effective for larger systems.


\begin{figure*}
\includegraphics[width=0.75\textwidth]{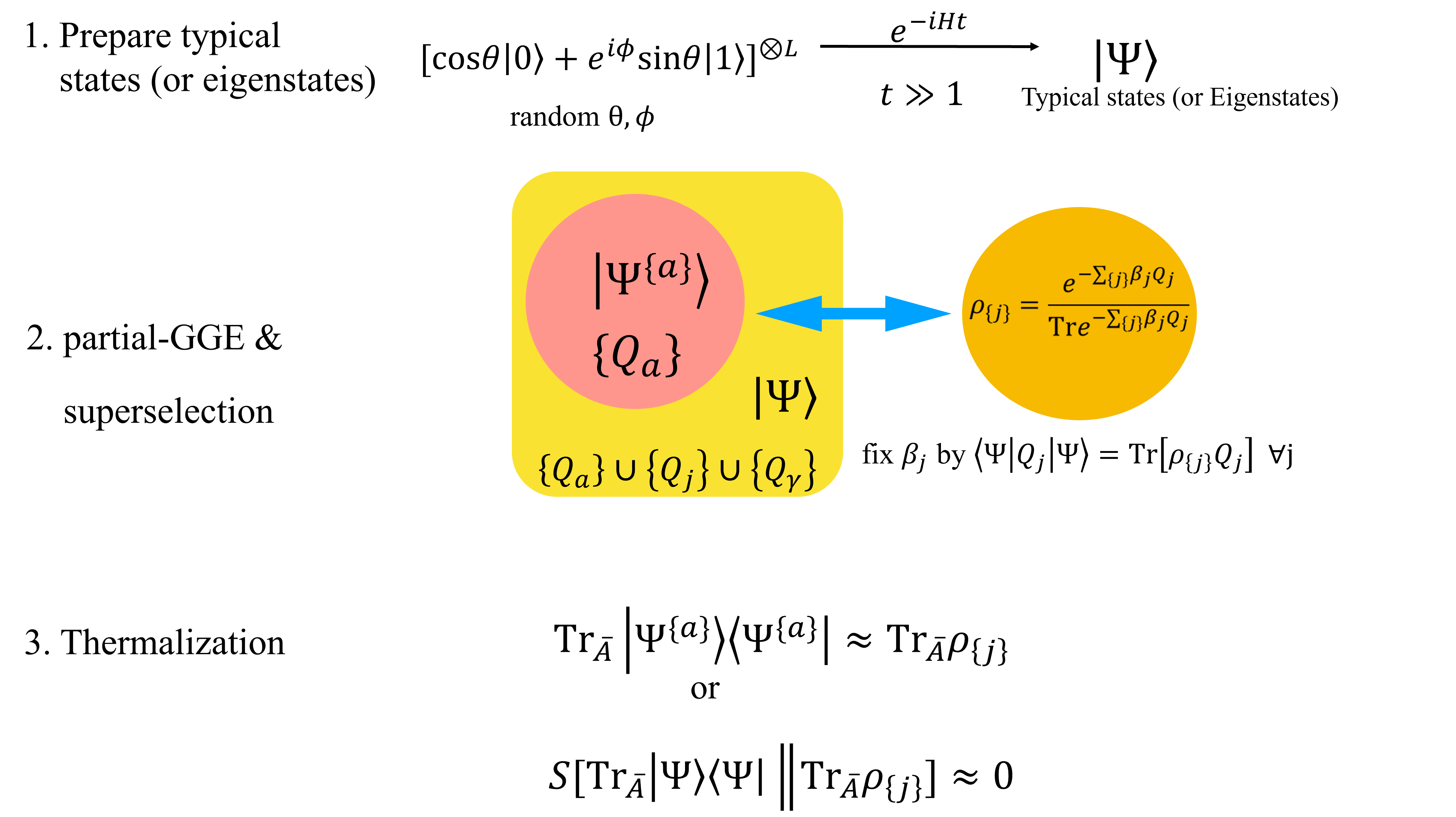}
\caption{\small Scheme of subsystem thermalization hypothesis for the systems with conserved charges by comparing the reduced states on a region $A$ of a typical state (or eigenstate) and a partial-GGE state.}
\label{scheme}
\end{figure*}

We can generalize the above subsystem thermalization hypothesis to systems with conserved charges $\{Q_I \}$ where $I$ labels the charges. For simplicity, we set $Q_0=H$ so that $[Q_0, Q_{I\ne 0}]=0$. In statistical mechanics, the conserved charge $Q_I$ and its corresponding chemical potential $\beta_I$ form the conjugate pair. When fixing a thermal ensemble, we can fix either $Q_I$ or $\beta_I$.  Denote the set of fixed charges as $\{ Q_a \}$ and the set of fixed chemical potentials as $\{\beta_j\}$. These two sets are mutually exclusive, i.e., $\{a \} \cap \{ j \} =\emptyset$. However, there are remaining charges labeled by index $\gamma$, for which we fix neither $Q_{\gamma}$ nor $\beta_{\gamma}$. Note that $\{a \} \cup \{ j \} \cup \{ \gamma \} = \{ I \}$. With the above classification of the conserved charges, we define the so-called partial generalized Gibbs ensemble (p-GGE) thermal state as 
\be\label{p-GGE}
\rho_{\{ j \}} = { e^{-\sum_j \beta_j Q_j} \over {\rm Tr} e^{-\sum_j \beta_j Q_j} }    \qquad \forall{j}\;.
\ee
If $\{j\}=\{I\}$, it is just the GGE state, i.e., $\rho_{\rm gge}=\rho_{\{I\}}$. 

After introducing the p-GGE, the subsystem thermalization hypothesis can be extended to systems with conserved charges. For a given state  $|\Psi\rangle$, a chosen p-GGE ensemble state can be fixed by  requiring
\be\label{beta_j}
\langle \Psi | Q_j | \Psi \rangle = {\rm Tr}\big[ \rho_{\{ j \}} Q_j \big] \qquad \forall{j}
\ee
to solve for $\{ \beta_j \}$ to fix $\rho_{\{ j \}}$. The subsystem thermalization hypothesis for the chosen p-GGE holds if 
\be\label{p-GGE_thermal}
S_A:=S\Big[{\rm Tr}_{\overline{A}} |\Psi \rangle \langle \Psi | \Big\Vert {\rm Tr}_{\overline{A}} \rho_{\{ j \}}  \Big] \simeq 0 \;.
\ee
This scheme of subsystem thermalization hypothesis for the systems with conserved charges is summarized in Fig. \ref{scheme}.

As the construction of p-GGE only involves the set of conserved charges labeled by $\{j\}$, this leads to the following variations of the subsystem thermalization hypothesis by picking up different sets of states $\{ |\Psi\rangle\}$ related or unrelated to the remaining sets of conserved charges labeled by $\{a\}$ or $\{\gamma \}$.  

\begin{enumerate}
    \item Typical-state thermalization. In this case, the states $\{|\Psi \rangle \}$ do not belong to any superselection sector associated with $\{ Q_I\}$. 

    \item Extended ETH. In this case, we choose $\{ Q_a\}$ to be mutually commuting and examine the thermalization hypothesis of their eigenstates $|\{q_a\} \rangle$, defined by
    \be
    Q_a |\{q_b\} \rangle = q_a |\{q_b\} \rangle \qquad \forall{a} \;.
    \ee

    \item Thermalization of the almost $\{Q_a\}$-superselection states $\{ |\Psi^{\{a\}} \rangle \}$, defined by  
    \be\label{almost_super}
    \langle \Psi^{\{a\}} | Q_a | \Psi^{\{a\}} \rangle \simeq \langle Q_a \rangle  \qquad \forall{a}
    \ee
    for a given set $\{\langle Q_a \rangle \}$. In our work, we define the "almost superselection states" by the condition that the difference between the corresponding quantities is approximately $0.1 \%$ of $\langle Q_a \rangle$, which we consider to be a reasonably small and acceptable uncertainty for our analysis. This case is the fine-grained version of the typical-state thermalization hypothesis, which subdivides the typical states into almost superselection sectors specified by $\langle Q_a \rangle$.      
    
    Note that we call $\{ |\Psi^{\{a\}}\rangle \}$ the almost superselection states to distinguish from the genuine superselection states, i.e., the eigenstates of $\{ Q_a\}$. The latter cannot be defined if $\{ Q_a\}$ are not mutually commuting. 
    
\end{enumerate}

Besides the above choices of states to test the thermalization hypothesis, we have some remarks to extend the scope of the subsystem thermalization hypothesis further, which we will consider in this paper. 

\begin{enumerate}[label=(\alph*)]
    \item Even though the Hamiltonian $Q_0$ is a special one among the conserved charges, it is interesting to consider the p-GGE with $Q_0$ excluded in the set $\{j\}$. 

    \item As the p-GGE is constructed by fixing $\{\beta_j\}$ instead of $\{Q_j\}$, we can consider the p-GGE for either  commuting or non-commuting $\{Q_j\}$. The latter can be applied to systems with non-Abelian symmetry, such as the XXX chains, and can be compared with the GGE one, which is the NATS proposed and studied in \cite{Murthy:2022dao, PhysRevE.101.042117, Majidy:2023xhm, Lasek:2024ess}. 

    \item When considering the subsystem thermalization hypothesis for the non-Abelian symmetry systems, we can also choose the almost superselection states $\{|\Psi^{\{a\}}\rangle \}$ for the non-commuting $\{Q_a\}$, or the eigenstates $\{|\{q_a \} \rangle \}$ for commuting $\{Q_a\}$. These provide new ways to examine the thermalization hypothesis of typical states and ETH with non-Abelian symmetries.

    \item The framework discussed so far should be applied to systems with continuous or discrete symmetries. The eigenvalues of charges of discrete symmetries take only discrete values. However, the values of $\{\langle Q_a \rangle\}$ of \eq{almost_super} for the {almost} superselection states can also take continuous values for discrete symmetries.   

    \item When considering the integrable models, it is impractical to include all the conserved charges in the hypothetical thermal ensemble states. In this case, we can form a p-GGE of a small set $\{a\}$ to test the thermalization hypothesis and ignore most integrable charges, i.e., a large set of $\gamma$.
    
    \item With the numerical implementation to check the thermalization hypothesis by \eq{p-GGE_thermal}, we can study the dependence of $S\Big[{\rm Tr}_{\overline{A}} |\Psi \rangle \langle \Psi |  \Big\Vert {\rm Tr}_{\overline{A}} \rho_{\{ j \}}  \Big]$ on the size of $A$.
    
\end{enumerate}

The primary objective of this paper is to investigate the thermalization demographics for each possible scenario discussed above and determine if the thermalization is sufficiently generic. As mentioned, we will adopt the ensemble states' demographics over the relative entropy to characterize the validity of the thermalization hypothesis for the corresponding p-GGE. The associated notations used later for various scenarios are summarized in Table \ref{table_S_A}.

In summary, we will consider the subsystem thermalization hypothesis for two types of spin-$1/2$ chains with conserved charges: the quantum Ising chains and the quantum XXZ chains. The nonintegrable Ising chains have a conserved $Z_2$ charge. Interestingly, as we will see, the subsystem thermalization hypothesis also works for the p-GGE of discrete charge. The symmetries of XXZ chains are richer than Ising's, but we will mainly focus on $U(1)$ and $SU(2)$. The latter is non-Abelian, and it is interesting to examine the subsystem thermalization hypothesis for typical states or eigenstates of systems with such non-Abelian symmetries by comparing them with p-GGE.

That is, we construct an extensive set of eigenstates or typical states $\{\Phi \}$ labeled by $\{ \langle Q_a \rangle \}$, and choose some chosen p-GGE labeled by $\{\beta_j\}$. We then present the demographic distribution of the relative entropy of $S\Big[{\rm Tr}_{\overline{A}} |\Psi \rangle \langle \Psi | \Big\Vert {\rm Tr}_{\overline{A}} \rho_{\{ j \}}  \Big]$ as functions of $\{ \langle Q_a \rangle \}$ and $\{\beta_j\}$. By examining these thermalization demographics, we can examine the validity of the subsystem thermalization hypothesis. 

\begin{table}
\begin{center}
\begin{tabular}{||c c c c||} 
 \hline
   & GGE & p-GGE & p-GGE (almost) \\ [0.5ex] 
 \hline\hline 
 typical state  & $S_A^t$ & $S_A^t$ & $S_A^a$ \\ [1ex]
 \hline 
 eigenstate & $S_A^e$ & $S_A^e$ &  \\ [1ex] 
 \hline
\end{tabular}
\caption{\small Notations of realtive entropy $S_A$ defined in  \eq{p-GGE_thermal} and used in different scenarios of the thermalization hypothesis. The superscript $t$ denotes the typical state, $e$ is the eigenstate, and $a$ is the almost superselection or just superselection state. The size of $A$ ranges from one site to half of the spin chain.}
\label{table_S_A}
\end{center} 
\end{table}

Before presenting the thermalization demographics, we have two remarks about the generality of quantum thermalization and how to prepare the typical states numerically. 
Firstly, we consider the generality of quantum thermalization 
by extending an argument for GGE in \cite{guryanova2016thermodynamics} to the cases for p-GGE. This can provide some insight into quantum thermalization through a second-law-like argument. The key idea of \cite{guryanova2016thermodynamics} is to define the so-called free entropy for a generic state $\rho$ for a given set of chemical potentials $\{\beta_j \}$ as follows \cite{guryanova2016thermodynamics},
\be
\tilde{F}[\rho] =\sum_j \beta_j {\rm Tr}(\rho Q_j) - S[\rho]
\ee
where $S[\rho]:=-{\rm Tr}\big[\rho \ln \rho \big]$ is the von Neumann entropy of $\rho$. Note that
\be\label{F_e_p-gge}
\tilde{F}[\rho_{\{ j \}}]=-\ln {\rm Tr} \; e^{-\sum_j \beta_j Q_j} \;.
\ee
It can be shown that \cite{guryanova2016thermodynamics}
\be\label{free_e_d}
\tilde{F}[\rho] - \tilde{F}[\rho_{\{ j \}}] = S[\rho \Vert \rho_{\{ j \}}]\;.
\ee
Thus, p-GGE is the state of minimal free entropy because $S[\rho \Vert \rho_{\{ j \}}] \ge 0$, which can be shown to be equivalent to maximalizing von Neumann entropy with fixed ${\rm Tr}(\rho Q_j)$ for all $j$ \cite{guryanova2016thermodynamics}. By \eq{free_e_d}, the thermalization condition \eq{p-GGE_thermal} can also be understood as the free entropy difference for the reduced states of the region $A$ inherited from $|\Psi \rangle \langle \Psi|$ and $\rho_{\rm gge}$, respectively.

One can put the above discussion in the context of the resource theory of quantum thermodynamics \cite{ng2019resource}. A resource theory is determined by two key ingredients: the free states and the free operations. Given that, one also needs a monotone quantity to define the condition for state transitions. To be specific in quantum thermodynamics, the p-GGE states described above are the free states, and the free entropy defined above is the monotone so that the state transition: $\rho \rightarrow \rho'$ is possible only if 
\be
\Delta \tilde{F}: = \tilde{F}[\rho'] - \tilde{F}[\rho] \le 0\;.
\ee
This is the statement of the second law of thermodynamics. Given free entropy as the monotone for state transitions, the free operations are just the evolution conserving $Q_j$'s, e.g., $U(t)=e^{- i t \sum_j \alpha_j Q_j }$, which will not change the free entropy. Otherwise, the operations are referred to as thermal operations (TOs) \cite{ng2019resource}.

Secondly, we will mention how we numerically prepare and define the typical states. The subsystem thermalization hypothesis can be applied to systems even without taking the thermodynamic limit, as long as the subsystem size is significantly smaller than that of the system. We will primarily employ the exact diagonalization method (ED) to determine the eigenstates of spin chain models with approximately ten sites. We will also use it to evolve the initial product states, preparing the typical states for checking the subsystem thermalization hypothesis. Explicitly, we first prepare the initial entangled state by $|\psi\rangle = \otimes_{i=1}^L \Big( \cos( \theta_i/2) | 0 \rangle +e^{i \phi_i}   \sin( \theta_i/2) | 1 \rangle  \Big) $  with $\theta_i \in [ 0, \pi ]$,  $\phi_i \in [ 0, 2\pi ]$ 
generated uniformly and randomly. Then, we evolve these states with the Hamiltonian of the spin chain model by ED for chains of 10 sites or so. If the evolution is long enough, we will expect the final states to be the typical states, {leading to the thermalization of local observables due to the scrambling by the long-time unitary evolution \cite{Hayden:2007cs, Sekino:2008he}.} This can be checked by examining the saturation of the entanglement entropy.  
Although 10-site chains seem small, as we will see, they are sufficient to achieve subsystem thermalization, at least for $A=1$. One would expect $A_c$ to increase as the system size increases. This could be a future study using matrix product states (MPS) to obtain the typical states for longer chains. In our previous work \cite{lin2024work}, we demonstrated that it is possible to accurately evolve product states or prepare thermal states using matrix product states (MPS) for spin chains with a few hundred sites.

\begin{figure}[t!]
\includegraphics[width=0.4\textwidth]{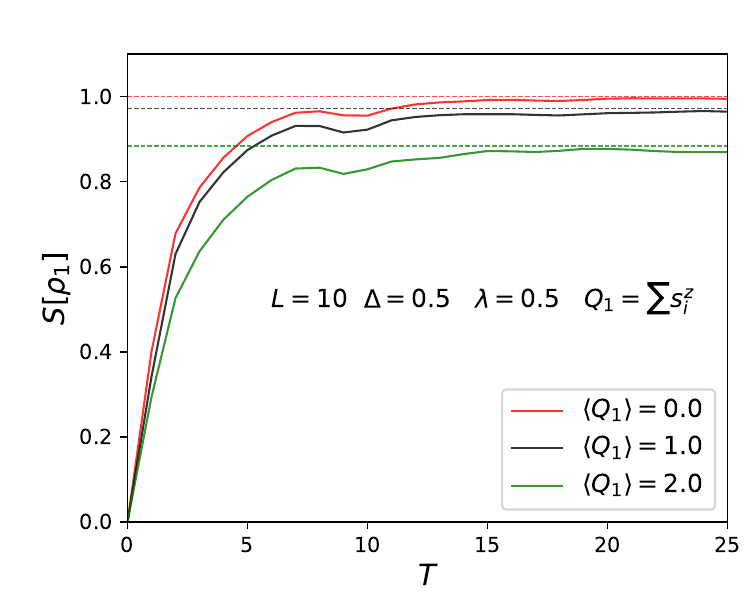}
\caption{\small  
The evolving pattern of one-site entanglement entropy $S[\rho_1]$ of a $10$-site  nonintegrable XXZ chain specified by \eq{H_XXZ} and \eq{XXZ_cc}. This model has a $U(1)$ conserved charge denoted by  $Q_1$, see \eq{XXZ_ch}. Each solid line represents the average of $S[\rho_1]$ over 500 states at each time step, which are evolved from the initial product states with fixed $\langle Q_1 \rangle$. In this figure, we consider  $\langle Q_1 \rangle= 0.0$ (red), $1.0$ (blue) and $2.0$ (green). The dashed lines are the values of $S[\rho_1]$ of the corresponding GGE states.  The result shows that the values of $S[\rho_1]$ saturate to a non-thermal value around $T=25$.  
}
\label{entropy_1DXXZ}
\end{figure} 

There is no unique definition of a typical state. In this work, we adopt the following definition used in \cite{caceres2024generic}: its entanglement entropy of a single site with the rest of the spin chain, denoted by $S[\rho_1]$, is about the maximal value given by the time evolution of an initial product state. This will exclude some transient states, which may not be typical regarding entanglement spreading. We give an example for this construction by considering a ten-site nonintegrable XXZ spin chain with a $U(1)$ conserved charge $Q_1$. We then use ED to evolve a randomly generated product state with zero $S[\rho_1]$ but fixed $\langle Q_1\rangle$. In Fig. \ref{entropy_1DXXZ}, we show the evolving pattern of the averaged $S[\rho_1]$ over $500$ initial states. The result shows that the (averaged) $S[\rho_1]$ values saturate to a maximum after $T \simeq 25$. However, these saturation values may differ from those given by the corresponding GGE states. 
{Besides, the dependence on $\langle Q_1 \rangle$ of these saturation values implies that the typical states can be classified by $\langle Q_1 \rangle$. Each class corresponds to the almost $\{ Q_1 \}$-superselection sector discussed earlier.}
In this work, the typical states we collect for the demographics of the subsystem thermalization hypothesis are obtained by evolving randomly generated product states over a time duration of $T=1000$. This should be long enough to ensure the typicality. Based on this, we will prepare the typical states by choosing $T=1000$.

\section{Thermalization demographics for quantum Ising spin chains with $Z_2$ symmetry}\label{sec3:Ising}


As a starter, we consider the quantum Ising spin chain in the transverse field $h_x$ and longitudinal field $h_z$  {with open boundary condition, given by the Hamiltonian  
\be\label{H_Ising}
 H   = \sum_{i=1}^L (  h_x \sigma_i^x  +   h_z \sigma_i^z  )+ \sum_{i=1}^{L-1}  \sigma_i^z   \sigma_{i+1}^z, 
\ee
}
where $\sigma_i^x$,  $\sigma_i^z$ are Pauli matrix operator on the site $i$, and $L$ is the number of spins. This model possesses a $Z_2$ reflection symmetry $\Pi$, which swaps the order of spins by sending site $i$ to $L+1-i$, i.e.,
\be 
\Pi:=
  \begin{cases}
    P_{1,L} P_{2,L-1}\cdots P_{\frac{L}{2}, \frac{L+2}{2}}       & \quad \text{if } L= \text{ is even}\\
    P_{1,L} P_{2,L-1}\cdots P_{\frac{L-1}{2}, \frac{L+3}{2}}  & \quad \text{if } L= \text{ is odd}
  \end{cases}  \qquad \label{Z2_Pi}
\ee
where $P_{k,l} = ( \sigma_k^x  \sigma_l^x  + \sigma_k^y  \sigma_l^y  + \sigma_k^z \sigma_l^z  + I  )/2$ is the permutation operator and $I$ is the identity operator. $P_{k,l}$ can swap the state of $k^{th}$ and $l^{th}$ sites. To avoid introducing additional conserved charge due to translational invariance, in this work, we will consider open spin chains without imposing periodic boundary conditions.

\begin{figure}[t!]
\includegraphics[width=0.48\textwidth]{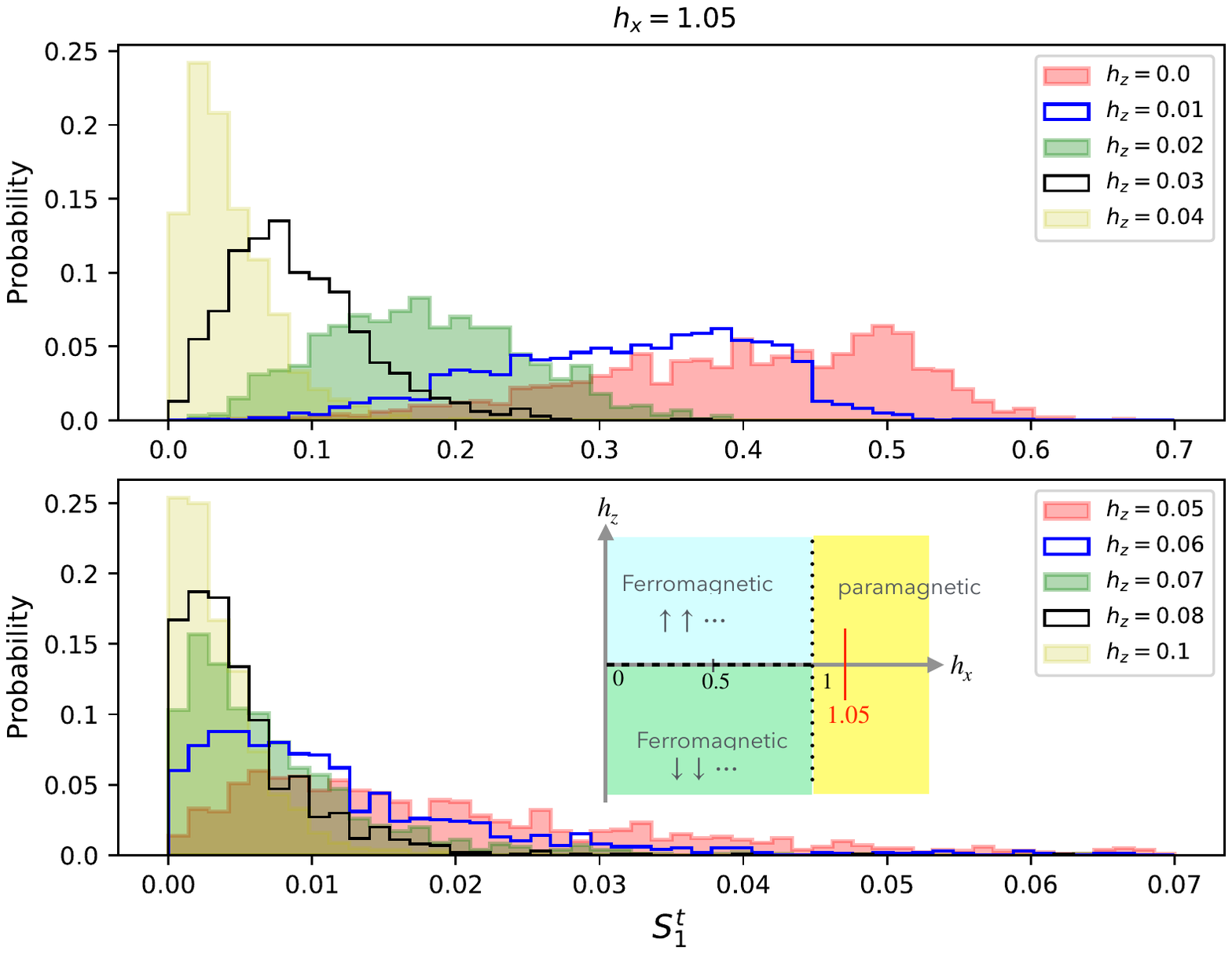}
\caption{\small  Patterns of typical-state thermalization demographics of $S^t_{A=1}$ of a $10$-site Ising chain of \eq{H_Ising} near an integrable point when comparing with the corresponding GGE state specified by the chemical potentials $(\beta_0,\beta_1)$ of the conserved charges \eq{Ising_ch}. 
The phase diagram of the Ising chain is shown in the inset of the bottom subfigure, where a red line indicates the region we plot for thermalization demographics. The integrable point on the red line is $h_x = 1.05$ and $h_z = 0$. We plot $10$ demographics from $h_z=0$ to $h_z=0.1$. At the integrable point, the GGE state is simply a p-GGE state, as all conserved charges other than $Q_1$ are neglected. The results show that the thermalization hypothesis breaks down significantly near the integrable point, but gradually holds better as one moves away from it.
}
\label{PD_hx1.05}
\end{figure} 

This model is integrable if $h_x=0$ or $h_z=0$. Otherwise, it is nonintegrable. Its phase diagram is shown in the inset of the bottom subfigure of Fig.~\ref{PD_hx1.05}, e.g., see~\cite{Yuste_2018}. It was shown that the level-spacing statistics of the superselection sectors with $\Pi=\pm 1$ match with the Wigner-Dyson distribution for the nonintegrable Ising chain \cite{caceres2024generic}. Despite that, we can examine how the pattern of the demographics of $S^t_A$ changes by tuning the coupling constant to make the model more nonintegrable. 
The thermalization hypothesis holds well if most of the population lives near $S_A^t=0$. Otherwise, we can quantify the violation of the thermalization hypothesis by the spreading of the population away from $S_A^t=0$. A typical result for such pattern changes is shown in Fig. \ref{PD_hx1.05}. We observe that the demographics of $S_1^t$ for the integrable case significantly violate the subsystem thermalization hypothesis. However, as the model moves away from the integrable point, the values of $S_1^t$ gradually converge to zero.

From now on, we will only consider the thermalization for the non-integrable variants of the quantum spin chains. Thus, for the Ising chain, we will set the parameters to the following: 
\be\label{Ising_cc}
L=10\;, \;\; h_x=1.05\;, \;\; h_z=0.5\;.
\ee
This model has two conserved charges denoted by
\be\label{Ising_ch}
Q_0=H\;, \qquad Q_1 =\Pi\;.
\ee
We will present the demographics of $S_A^{t, a, e}$ for various types of p-GGE states.

\subsection{Demographics of typical-state thermalization}

\begin{figure}[t!]
\includegraphics[width=0.40\textwidth]{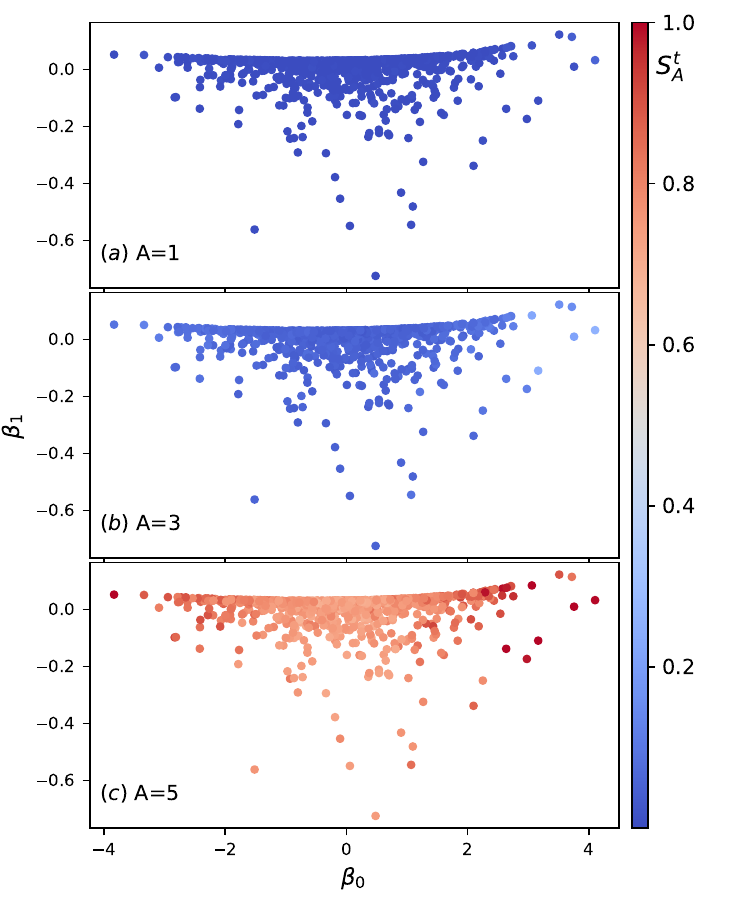}
\caption{\small Density plots of typical-state thermalization demographics: $S^t_{A=1,3,5}$ v.s. ($\beta_0$, $\beta_1$) of a GGE state for the Ising chain of \eq{Ising_cc} with conserved charges $Q_{0,1}$ given by \eq{Ising_ch}. The thermalization hypothesis works well for $A=1$ in subfigure (a) and worsens as $A$ increases, i.e., for $A=3$ in subfigure (b) and $A=5$ in subfigure (c). }
\label{SA_gge_nint_Ising}
\end{figure} 

\begin{figure}[t!]
\includegraphics[width=0.40\textwidth]{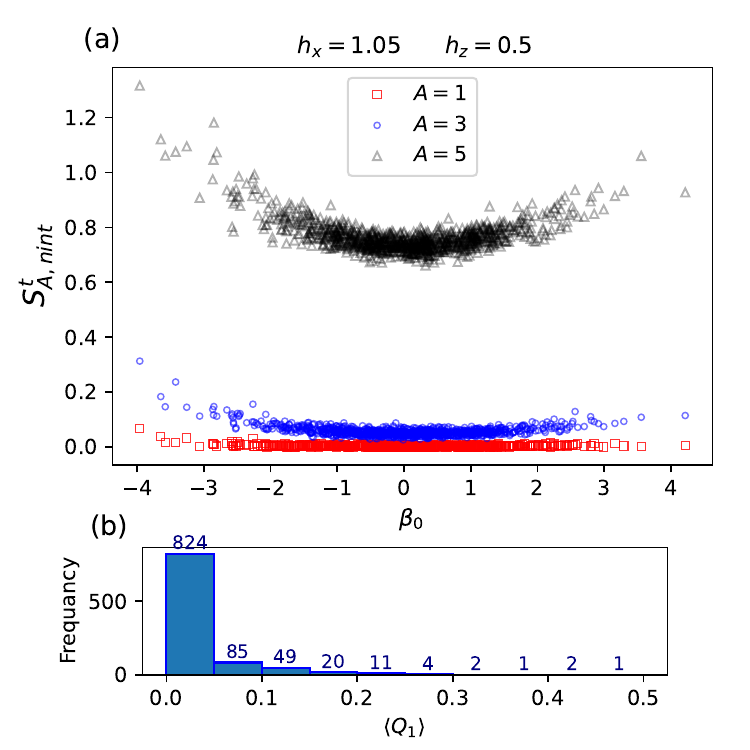}
\caption{\small 
(a) Typical-state thermalization demographics: $S^t_{A=1,3,5}$ v.s. $\beta_0$ for the Ising chain of \eq{Ising_cc} by comparing with a p-GGE specified only by $\beta_0$. The thermalization hypothesis also works well for $A=1$ (red squares), but it worsens as $A$ increases, as seen for $A=3$ (blue circles) and $A=5$ (black triangles). (b) Demographics of the typical states classified by the $\langle Q_1 \rangle$.  }
\label{SA_B0_pgge_nint_Ising}
\end{figure} 

\begin{figure}[b!]
\includegraphics[width=0.40\textwidth]{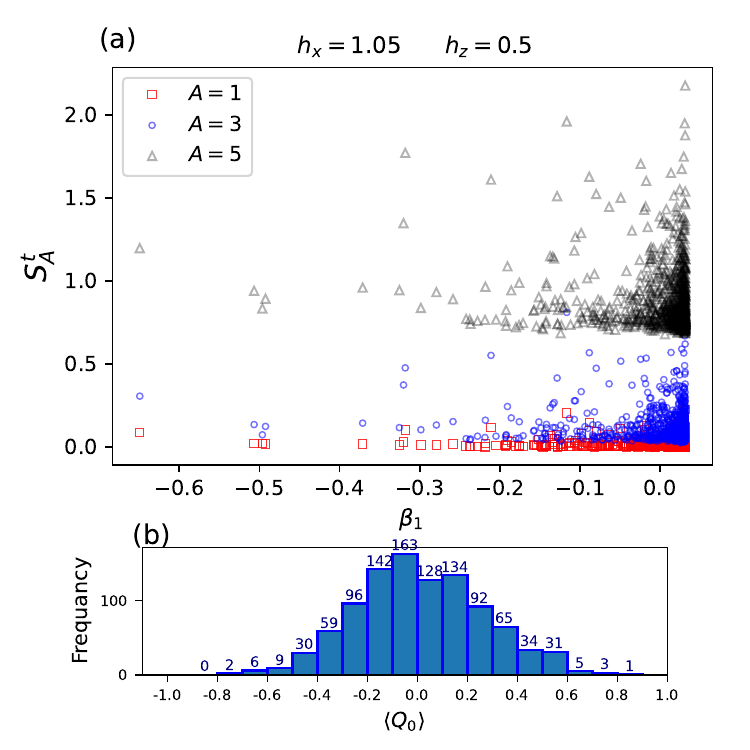}
\caption{\small  (a) Typical-state thermalization demographics: $S^t_{A=1,3,5}$ v.s. $\beta_1$ for the Ising chain of \eq{Ising_cc} by comparing with a p-GGE specified only by $\beta_1$. Again, the thermalization hypothesis works well for $A=1$ and worsens as $A$ increases.  (b) Demographics of the typical states classified by the $\langle Q_0 \rangle$.}
\label{SA_B1_pgge_nint_Ising}
\end{figure} 

We first consider the subsystem thermalization hypothesis for the typical states. In Fig. \ref{SA_gge_nint_Ising}, we show the thermalization demographics of the relative entropy $S^t_A$ of the region $A$ with $1$, $3$, $5$ sites (denoted by $A=1,3,5$) between a typical state and the corresponding GGE specified by $(\beta_0,\beta_1)$. Since the GGE contains two chemical potential parameters $\beta_0$ and $\beta_1$, we present the demographics in the form of density plots. For $A=1$, $S^t_A$'s of all the typical states considered are almost vanishing. This implies that the thermalization hypothesis for typical states is effective for $A=1$ when compared with the corresponding GGE thermal states. On the other hand, for $A=3$, the thermalization hypothesis still looks good, with some exceptions for large $|\beta_0|$ or $|\beta_1|$. However, when $A=5$, the thermalization hypothesis breaks down. The above patterns of dependence of validity of the thermalization hypothesis on the size $A$ are expected and also give an explicitly quantitative picture.  {Moreover, in Appendix \ref{App_B}, we also performed the size-scaling study by varying $L$ to show that the thermalization hypothesis works better for larger systems. This study is not quite extensive due to the limitation of computing power, but the tendency, however, is clear in meeting the expectation.}

Besides checking the typical-state thermalization hypothesis, Fig. \ref{SA_gge_nint_Ising} also reveals the demographics of the typical states on the $(\beta_0,\beta_1)$ of the corresponding GGE ensemble states. Interestingly, the inverse temperature $\beta_0$ can be either positive or negative, and it is almost symmetric about $\beta_0=0$. The GGE states with negative (inverse) temperatures are hypothetical and cannot be realized naturally. However, we see that they can be realized by typical states. Moreover, the thermalization hypothesis works even for such hypothetical GGE thermal ensemble states. On the other hand, the chemical potential $\beta_1$ for reflection symmetry $\Pi$ is almost negative for the typical states we construct. It is unclear why there is such a bias.

After considering the comparison with GGE states, we now compare with the corresponding p-GGE thermal states. In Fig.\ref{SA_B0_pgge_nint_Ising}, we present the typical-state demographics of $S^t_A$ when compared with the corresponding p-GGE states specified only by $\beta_0$ (inverse temperature). On the other hand, in Fig. \ref{SA_B1_pgge_nint_Ising}, we compare the typical states with the corresponding p-GGE states specified only by $\beta_1$. We see that the features of the demographics of p-GGE on either $\beta_0$, i.e., symmetric about $\beta_0$, and $\beta_1$, i.e., favors negative values, are the same as the ones of GGE. Regarding the validity of the typical-state thermalization hypothesis, we see in both Fig. \ref{SA_B0_pgge_nint_Ising} and \ref{SA_B1_pgge_nint_Ising} the similar patterns of dependence on the size $A$ as in Fig. \ref{SA_gge_nint_Ising} for GGE states' comparison. In all three cases, the thermalization hypothesis for the typical states works quite well for $A=1$ and gradually worsens with increasing $A$. 

Although it is generally believed that the typical-state thermalization hypothesis will hold for small subsystems, we emphasize that it is unclear whether it will hold for various thermal ensemble states and how local the subsystem must be for it to be applicable. The results presented here clarify these issues.

\subsection{Demographics of almost superselection-state thermalization}

\begin{figure}
\includegraphics[width=0.40\textwidth] {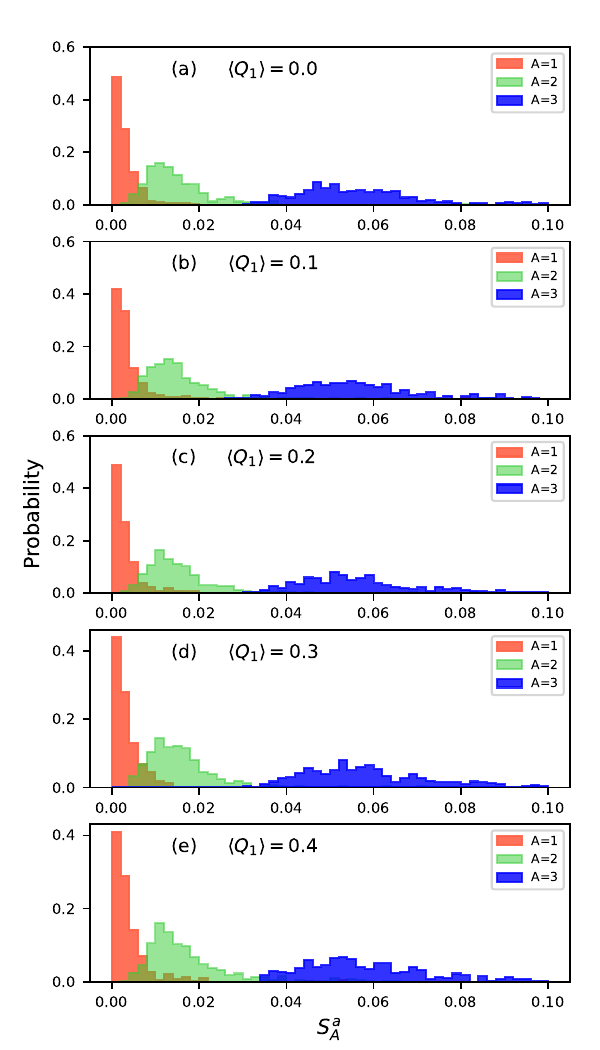}
\caption{\small Fine-grained version of Fig. \ref{SA_B0_pgge_nint_Ising} as the corresponding almost-superselection-state thermalization demographics by showing the population ratios for the collection of typical states with fixed $\langle Q_1 \rangle=$ (a) $0.0$, (b) $0.1$, (c) $0.2$, (d) $0.3$, and (e) $0.4$, respectively. The thermalization hypothesis works well for $A=1$ (blue) but not so for $A=2$ (red) and $A=3$ (green). }
\label{Ising_ass_0}
\end{figure} 

\begin{figure}
\includegraphics[width=0.40\textwidth]{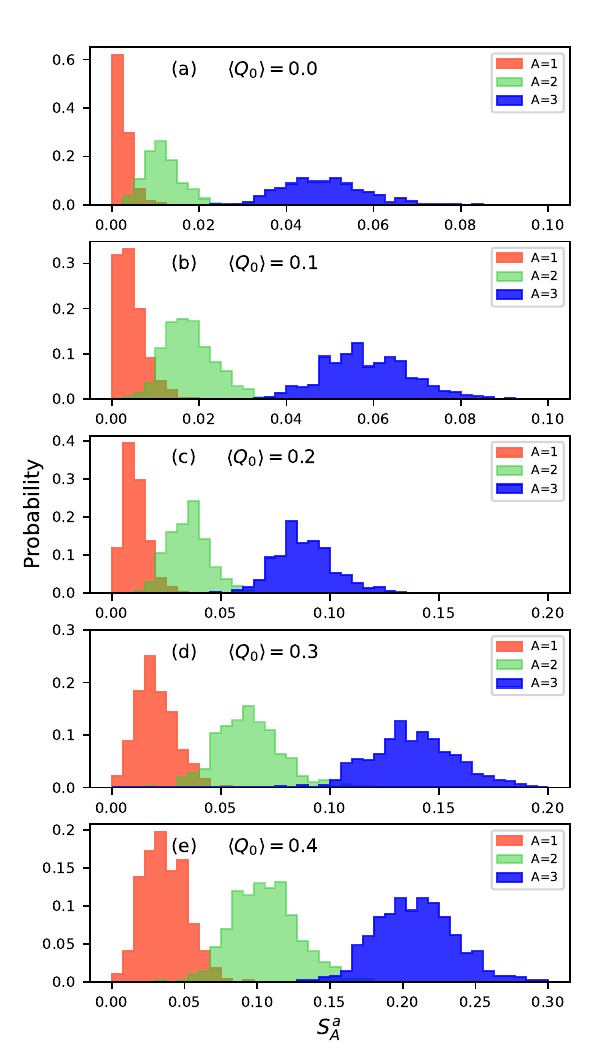}
\caption{\small 
Fine-grained version of Fig. \ref{SA_B1_pgge_nint_Ising} as the corresponding almost-superselection-state thermalization demographics by showing the population ratios for the collection of typical states with fixed $\langle Q_0 \rangle=$ (a) $0.0$, (b) $0.1$, (c) $0.2$, (d) $0.3$, and (e) $0.4$, respectively. Unlike Fig. \ref{Ising_ass_0}, the thermalization hypothesis does not work well even for $A=1$ when $\langle Q_0 \rangle$ increases away from zero.  However, the sector of $\langle Q_0 \rangle=0$ has the largest population as shown in Fig. \ref{SA_B1_pgge_nint_Ising}(b), so the corresponding coarse-grained thermalization in Fig. \ref{SA_B1_pgge_nint_Ising} looks fine.  
}
\label{Ising_ass_1}
\end{figure} 

We now consider the thermalization hypothesis for the almost superselection states. This is to subdivide the typical states into the almost superselection sectors specified by $\langle Q_a \rangle$. This is a fine-grained version of the typical-state thermalization hypothesis considered above. We take snapshots of the population demographics of $S_A^a$ with different values of $\langle Q_a \rangle$ to see how the thermalization hypothesis works at the fine-grained level. Recall the validity of the typical-state thermalization hypothesis shown in Fig. \ref{SA_B0_pgge_nint_Ising} and \ref{SA_B1_pgge_nint_Ising} when compared with p-GGE specified by $\beta_0$ and $\beta_1$, we now examine their fine-grained counterparts to see if the thermalization hypothesis also holds well for each almost superselection sector.   

In Fig. \ref{Ising_ass_0}, we consider the almost superselection typical states with fixed $\langle Q_1 \rangle$ and compare them with the corresponding p-GGE states specified by $\beta_0$. In Fig. \ref{Ising_ass_1}, we swap the role of the $Q_0$ and $Q_1$, that is, compare the almost superselection states with fixed $\langle Q_0 \rangle$ to the corresponding p-GGE states specified by $\beta_1$. In both cases, we show the population ratio demographics against $S_A^a$. The population ratio for a given $\langle Q_a \rangle$ is defined with respect to the total number of all typical states considered. We can quantify the violation of the thermalization hypothesis by the spreading of the population away from $S_A^a=0$. In the first case, as shown in Fig. \ref{Ising_ass_0}, the thermalization hypothesis holds well for each almost superselection sector if $A=1$. Moreover, the patterns of the populations for different $A$'s do not change much as $\langle Q_1 \rangle$ varies. In the second case, as shown in Fig. \ref{Ising_ass_1}, the thermalization holds well only for $\langle Q_0 \rangle=0$ and worsens more and more as $\langle Q_0 \rangle=0$ increases, even for $A=1$. In this case, even the typical-state thermalization holds, but its fine-grained version does not. Unlike in Fig. \ref{Ising_ass_0}, the patterns of populations for different $A$'s change more significantly as $\langle Q_0 \rangle$ varies.

It may look strange why the coarse-grained version of the thermalization hypothesis holds, but most of its fine-grained versions fail. However, if we look into the scales of the population ratio in each subfigure, it can be understood that the population respecting the thermalization hypothesis dominates; see also Fig. \ref{SA_B1_pgge_nint_Ising}(b), where the sector of $\langle Q_0 \rangle=0$, which obeys the thermalization hypothesis as shown in the first subfigure of Fig. \ref{Ising_ass_1}, has the largest population. 

In summary, when comparing the typical states with some chosen thermal ensemble states,  the fine-grained version of the thermalization hypothesis may fail even when the coarse-grained one holds.

\subsection{Demographics of eigenstate thermalization}

\begin{figure}
\includegraphics[width=0.40\textwidth]{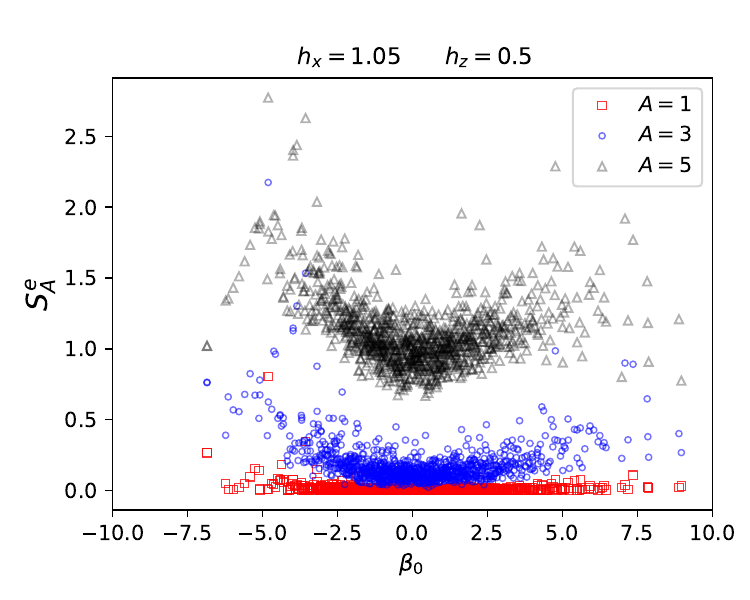}
\caption{\small 
Subsystem eigenstate thermalization demographics: $S^e_{A=1,3,5}$ vs. $\beta_0$ for the Ising chain of \eq{Ising_cc} by comparing with a p-GGE specified only by $\beta_0$. The thermalization hypothesis also works well for $A=1$ (red squares), but it worsens as $A$ increases, i.e., $A=3$ (blue circles) and $A=5$ (black triangles).}
\label{SE_B0_pgge_nint_Ising}
\end{figure} 

\begin{figure}
\includegraphics[width=0.40\textwidth]{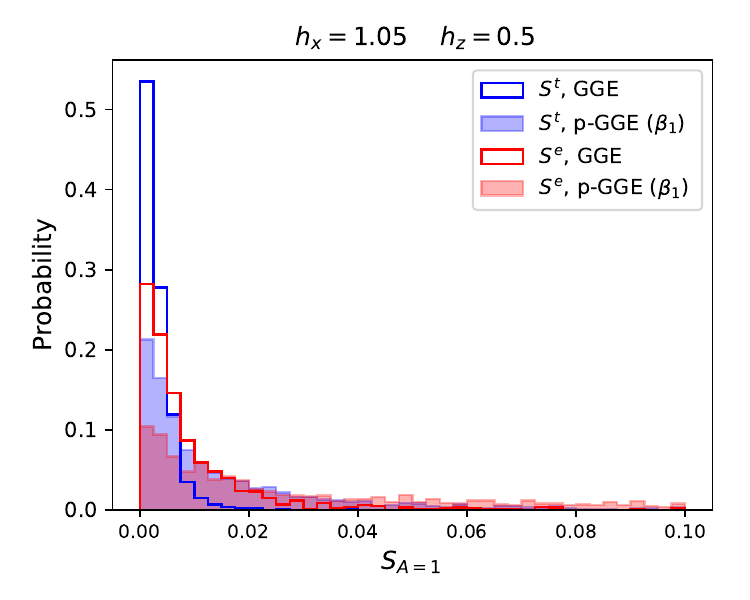}
\caption{\small Subsystem eigenstate thermalization demographics of $S^e_{A=1}$ for the Ising chain of \eq{Ising_cc} by comparing with (i) GGE states (empty red) and (ii) p-GGE states specified by $\beta_1$ (solid pink). We also present the typical-state counterparts (empty blue and solid gray) for comparison. The subsystem thermalization hypothesis is more effective for GGE than p-GGE, and more applicable to typical states than to eigenstates.}
\label{SE_SA_nint_Ising}
\end{figure} 

We finally consider the subsystem ETH \cite{Dymarsky:2016ntg, Lashkari:2016vgj,  He:2017vyf}, which has been extensively studied for integrable $(1+1)$-dimensional conformal field theories \cite{biroli2010effect, mori2016weak, iyoda2017fluctuation, He:2017vyf, He:2017txy, Basu:2017kzo, Guo:2018pvi, Dymarsky:2019etq,  chen2024subsystem}. Here, we will consider the subsystem ETH for the non-integrable Ising chains. Since the conserved charge $\Pi$ commutes with the Hamiltonian $H$, the eigenstates are also the superselection states. As in the usual case for subsystem ETH, we first consider the demographics $S_A^e$ when comparing with the canonical ensemble thermal state, i.e., p-GGE specified by $\beta_0$, and the result is shown in Fig.\ref{SE_B0_pgge_nint_Ising}. We observe that the subsystem ETH performs well for $A=1$ and deteriorates as $A$ increases. We expect this $A$-dependence will be similar when considering the GGE or the other p-GGE. Thus, we now only consider the demographics of the subsystem ETH for $A=1$ only. The demographics for GGE specified by $(\beta_0, \beta_1)$ and p-GGE by $\beta_1$ are shown in Fig.\ref{SE_SA_nint_Ising}, in which we also compare with their counterparts of the typical-state thermalization hypothesis.
Interestingly, for both eigenstates and typical states, the subsystem thermalization hypothesis works better for the GGE than the p-GGE specified by $\beta_1$. On the other hand, the subsystem thermalization hypothesis works better for typical states than for eigenstates of a chosen type of thermal ensemble state. Our results verify the subsystem ETH in a broader context than the canonical ensemble and also show an interesting comparison with the typical-state thermalization hypothesis.

\section {Thermalization demographics of quantum spin-$1/2$ XXZ chains with $U(1)$ symmetry} \label{sec4:XXZ_U1}

To consider spin chains with richer symmetries, such as non-Abelian ones, we consider the following $L$-site spin-$1/2$ XXZ spin chain model with the following Hamiltonian \cite{santos2018nonequilibrium} 
{
\be\label{H_XXZ} 
& H =  J \sum_{i=1}^{L-1} ( S_i^x S_{i+1}^x  + S_i^y S_{i+1}^y  +\Delta S_i^z S_{i+1}^z )  \notag \\  
    & +\lambda J 
    \sum_{i=1}^{L-2} ( S_i^x S_{i+2}^x  + S_i^y S_{i+2}^y  +\Delta S_i^z S_{i+2}^z )  \notag \\ 
    &+d JS^z_{L/2} + \sum_{i=1}^L h_i J S_i^z,
\ee
}
where $\vec{S}_i := \vec{\sigma}_i/2$ is the spin vector operator at the $i$-th site. 
Besides, the parameter $dJ$ denotes the Zeeman splitting at the defect site. The Zeeman splitting  $h_iJ$ reflects onsite disorder due to random static magnetic fields, with $h_i$ being random values drawn from a uniform distribution within $[-h, h]$, where $h$ indicates the disorder strength. 
The anisotropy parameter $\Delta$ determines the interaction type: the system is isotropic when $\Delta=1$, meaning the Ising interaction and the flip-flop term have equal strength. The parameter $\lambda$ characterizes the relative size of the nearest-neighbor (NN) coupling to the next-nearest-neighbor (NNN) coupling.

The symmetry properties of this model can be summarized as follows \cite{santos2018nonequilibrium}:
\begin{enumerate}[label=(\roman*)]

   \item $H$ commutes with $S^z_{\rm tot}:=\sum_{k=1}^L S_k^z$. 

    \item This model is integrable if $d=h=\lambda=0$. If $\Delta=0$, it reduces to a non-interacting XX model.   
   
    \item $H$ is invariant under  reflection $\Pi$ as in quantum Ising chain, if $d=h=0$.

    \item $H$ commutes with $\vec{S}_{\rm tot}:=\sum_{k=1}^L \vec{S}_k$ if $d=h=0$ and $\Delta=1$. This special model, characterized by a non-Abelian $SU(2)$ symmetry, is known as the XXX model. 

    \item $H$ is invariant under a global $\pi$-rotation around the $x$-axis, i.e., $R_{\pi}^x=\sigma_1^x\sigma_2^x\cdots \sigma_L^x$, if $d=h=0$, $L$ is even and the number of up spins in the $z$-direction $N_{\rm up}=L/2$.

    \item This model is chaotic if the NNN coupling $0 < \lambda <1$ and $h=0$ (no random disorder) for both XXZ ($\Delta \neq 0$) and XX ($\Delta = 0$) models. 
    This can be verified by fitting the level spacing statistics to the Wigner-Dyson distribution, as shown in Fig.~\ref{fig:level_spacing} of Appendix \ref{App_A}.
    
\end{enumerate}

\begin{figure}
\includegraphics[width=0.35\textwidth]{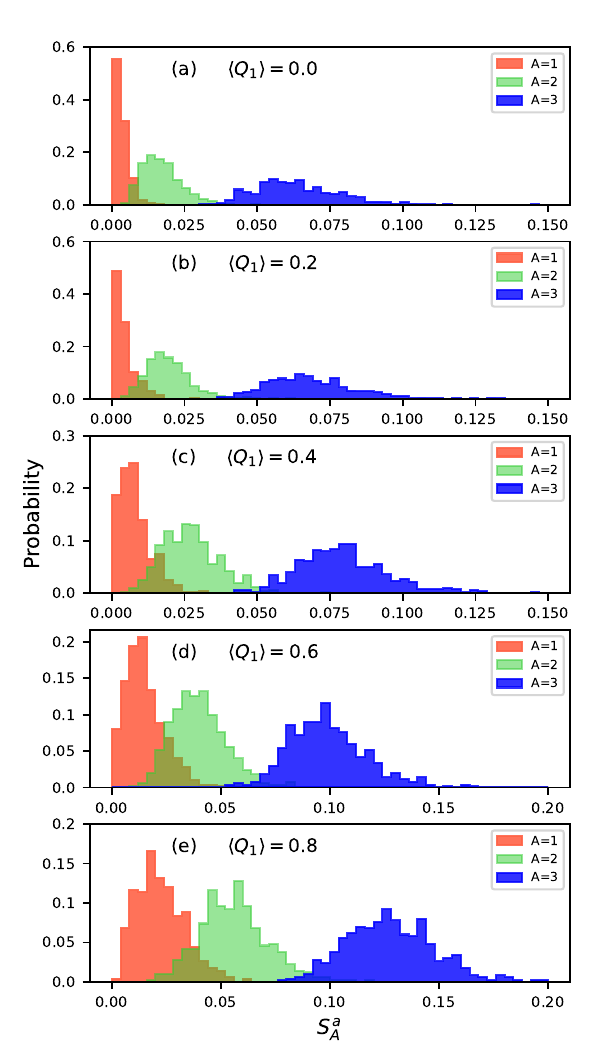}
\caption{\small 
Thermalization demographics of population ratios of $S^a_A$ for the XXZ chain of \eq{XXZ_cc} with $U(1)$ conserved charge for p-GGE specified only by $\beta_0$ for various almost-superselection states with with fixed $\langle Q_1 \rangle=$ (a) $0.0$, (b) $0.2$, (c) $0.4$, (d) $0.6$, and (e) $0.8$, respectively. This is the fine-grained version of Fig. \ref{SA_B1_pgge_nint_XXZ} of Appendix \ref{App_A}. The thermalization hypothesis does not work well even for $A=1$ when $\langle Q_1 \rangle$ increases away from zero. }
\label{ass_1_XXZ}
\end{figure} 

\begin{figure}
\includegraphics[width=0.35\textwidth]{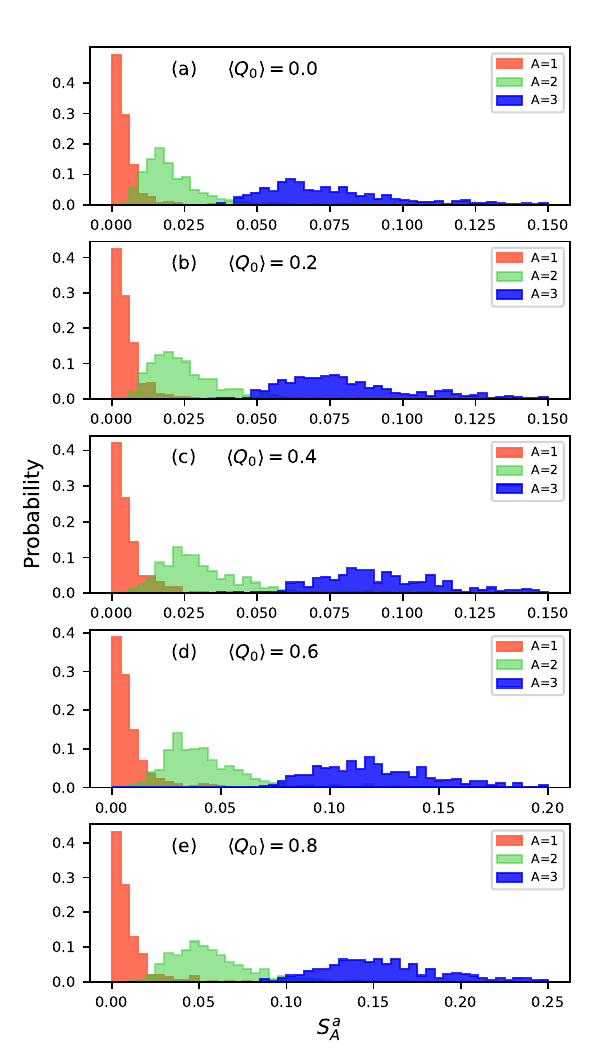}
\caption{\small 
Thermalization demographics of the population ratios of $S^a_A$ for the XXZ chain of \eq{XXZ_cc} with $U(1)$ conserved charge for p-GGE specified only by $\beta_1$ for various almost-superselection states with fixed $\langle Q_0 \rangle=$ (a) $0.0$, (b) $0.2$, (c) $0.4$, (d) $0.6$, and (e) $0.8$, respectively. This is the fine-grained version of Fig. \ref{SA_B0_pgge_nint_XXZ} of Appendix \ref{App_A}. The thermalization hypothesis works well for $A=1$ (blue) but not so for $A=2$ (red) and $A=3$ (green). }
\label{ass_0_XXZ}
\end{figure} 

We will consider only the non-integrable cases. This section will focus on the XXZ chain with only $S_a$ as the $U(1)$ conserved charge. Specifically, we will consider the Hamiltonian \eq{H_XXZ} with 
\be \label{XXZ_cc}
L=10\;, \;\; \Delta = \lambda = 0.5\;, \;\; h=d=0\;, \;\; J=1\;.
\ee
We consider the two conserved charges of this model, denoted by
\be\label{XXZ_ch}
Q_0=H\;, \qquad Q_1=S^z_{\rm tot}\;.
\ee
For completeness, in Fig. \ref{XXZ_with_parity} of Appendix \ref{App_C}, we also present the results with p-GGE, including $Q_2=\Pi$, which demonstrates that the thermalization hypothesis remains effective, as in the Ising chains. In the next section, we will consider the XXX model with $\vec{S}_{\rm tot}$ as the $SU(2)$ conserved charges. Since we have studied the subsystem thermalization hypothesis for the Ising chain with $\Pi$ as the $Z_2$ conserved charge, we will ignore this discrete charge for the following considerations.

As in the cases of quantum Ising chains, we consider the demographics of relative entropies $S_A^{t, a, e}$ for both GGE and p-GGEs. The demographics of $S_A^t$ exhibit similar patterns to those in the Ising cases; that is, the subsystem thermalization hypothesis holds well for $A=1$ and worsens as $A$ increases.  Thus, we will not present the results in the main text, but rather in Appendix \ref{App_A}.

Regarding the thermalization demographics of $S_A^a$, we observe an interesting difference when compared with its Ising counterparts. In the Ising case, the subsystem thermalization hypothesis works well for $A=1$ for all fixed values of $\langle Q_1 \rangle$, but only for small values of $\langle Q_0 \rangle$. However, as shown in Fig. \ref{ass_1_XXZ} and \ref{ass_0_XXZ}, we see the role of $Q_0$ and $Q_1$ swap, that is, the subsystem thermalization hypothesis works well for $A=1$ for all fixed values of $\langle Q_0 \rangle$, but only for small values of $\langle Q_1 \rangle$.

\begin{figure}
\includegraphics[width=0.35\textwidth]{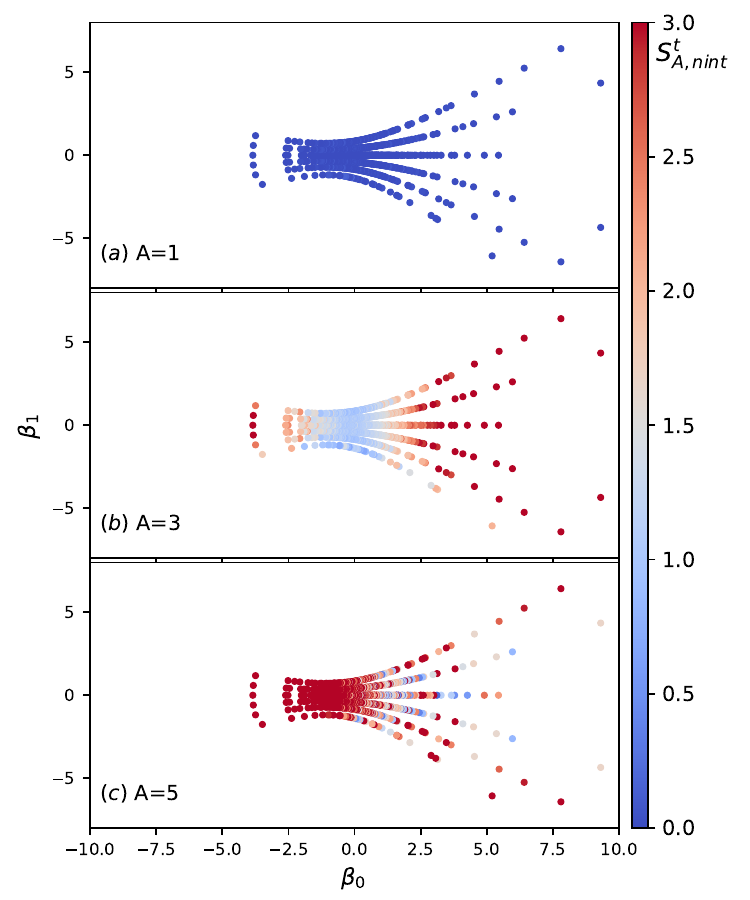}
\caption{\small
Density plot for the subsystem ETH: demographics of $S^e_A$ of the XXZ chain of \eq{XXZ_cc} with $U(1)$ conserved charge for GGE specified by $(\beta_0, \beta_1)$. The subsystem ETH works well for $A=1$ but worsens as $A$ increases. 
 }
\label{sETH_XXZ}
\end{figure} 

Finally, we consider the subsystem ETH for the XXZ chains. With some loss of generality, we only compare with the GGE states specified by $(\beta_0, \beta_1)$. The result is shown in Fig. \ref{sETH_XXZ}. We observe that the subsystem ETH performs well for $A=1$ but deteriorates as $A$ increases. Additionally, the plot patterns for the XXZ chain of $U(1)$ charges differ from those for the Ising chain with $Z_2$ charges.

\section {Thermalization demographics of quantum spin-$1/2$ XXZ chains with $SU(2)$ symmetry}\label{sec5:XXX}

We now will consider the thermalization demographics for the nonintegrable XXX chains with $SU(2)$ conserved charges. Specifically, we will consider the Hamiltonian of \eq{H_XXZ} with 
\be\label{XXX_cc}
L=10\;, \;\; \Delta=J=1\;, \;\; \lambda=0.5\;, \;\; d=h=0\;.
\ee
We consider the four conserved charges denoted as follows:
\be\label{XXX_ch}
Q_0 =H\;, \;\;  Q_1=S^x_{\rm tot}\;, \;\; Q_2=S^y_{\rm tot}\;, \;\;  Q_3=S^z_{\rm tot}\;. \quad
\ee
The charges $Q_{1,2,3}$ do not commute among themselves. Thus, we can only specify the eigenstates by $Q_0$ and one of $Q_i$ with $i=1,2,3$. As usual, we will specify the eigenstates by $(Q_0, Q_3)$. On the other hand, we can further classify the typical states into the almost superselection sectors with fixed values of $\langle Q_{\mu=0,1,2,3} \rangle$. We will consider the subsystem thermalization hypothesis for typical states and energy eigenstates by comparing them locally with GGE states and various types of p-GGE states. Note that the GGE states 
\be
\rho_{\rm GGE}^{\rm SU(2)} = {e^{-\sum_{\mu=0}^3 \beta_{\mu} Q_{\mu}} \over Z}\;, \quad Z:={\rm Tr} e^{-\sum_{\mu=0}^3 \beta_{\mu} Q_{\mu}}\; \quad
\ee
are proposed and studied in \cite{Murthy:2022dao, PhysRevE.101.042117, Majidy:2023xhm, Lasek:2024ess} for the ETH of the system with non-commuting conserved charges, and called non-Abelian thermal states (NATS).

\subsection{Demographics of typical-state thermalization}

\begin{figure}[t!] 
\centering
\subfigure[]{
        \includegraphics[width=0.48\textwidth]{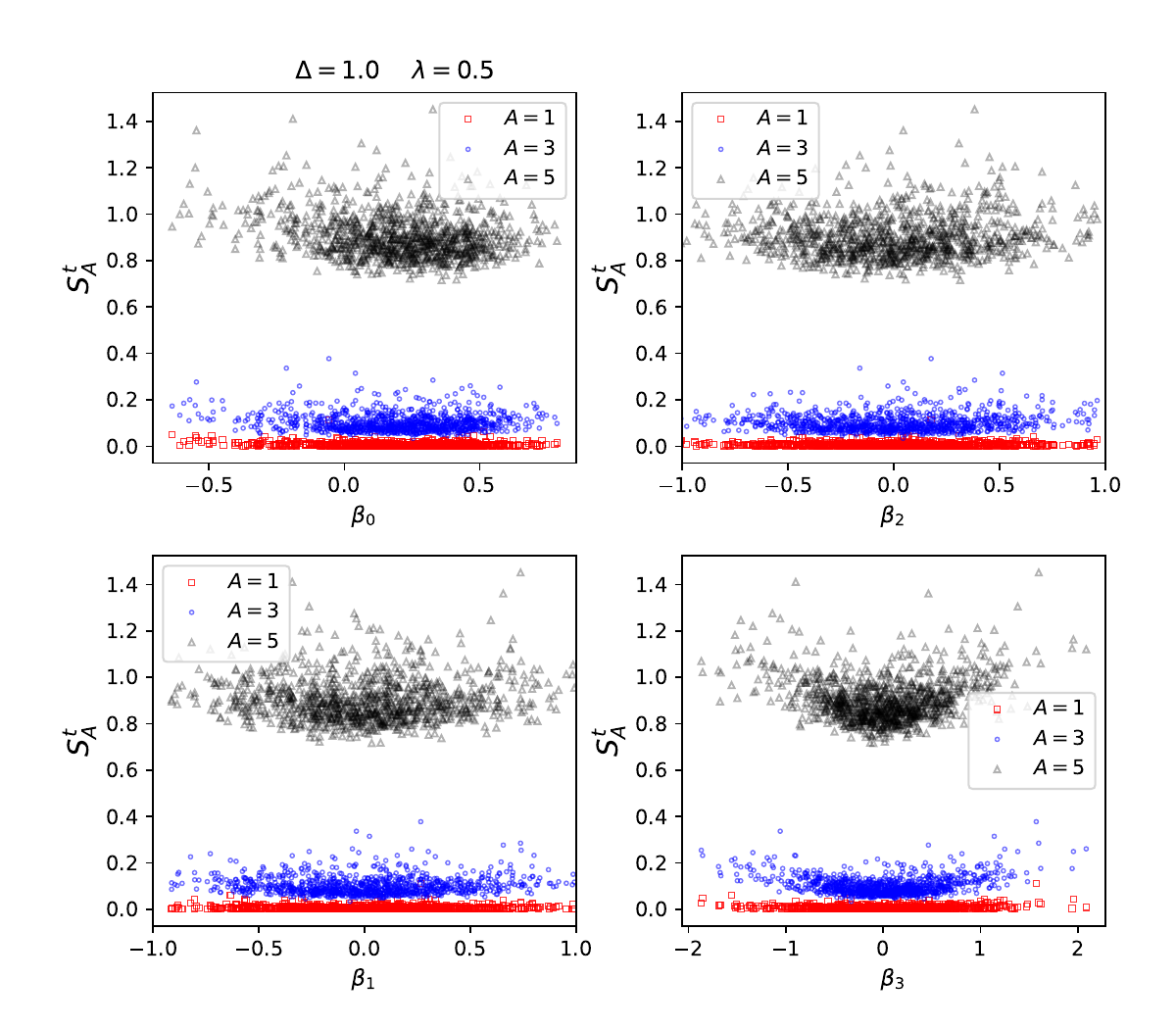}}
\centering
\subfigure[]{
        \includegraphics[width=0.25\textwidth]{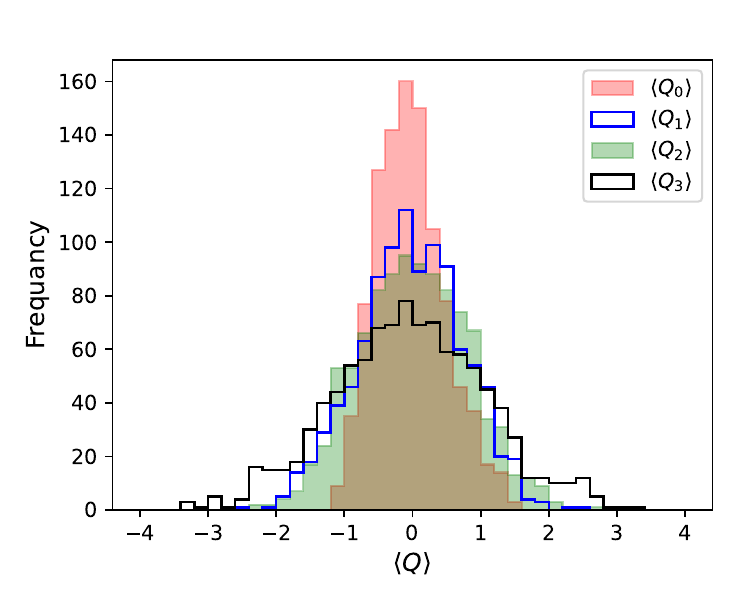}}
\caption{\small Demographics of typical-state thermalization hypothesis for the XXX chain of \eq{XXX_cc} by comparing with the GGE states specified by $\beta_{\mu=0,1,2,3}$, which are the chemical potentials of the conserved charged defined in \eq{XXX_ch}. (a) Density plots of $S^t_{A=1,3,5}$ v.s. $\beta_0$, $\beta_1$, $\beta_2$, and $\beta_3$. The subsystem thermalization hypothesis holds well for $A=1$ and worsens as $A$ increases. (b)
(b)  Demographics of the typical states classified by the $\langle Q_0 \rangle$ (solid pink), $\langle Q_1 \rangle$ (empty blue), $\langle Q_2 \rangle$ (solid green), and $\langle Q_3 \rangle$ (empty black).}
\label{SA_gge_nint_XXX}
\end{figure}

We first consider the subsystem thermalization hypothesis for the typical states compared to GGE states (or NATS), specified by the chemical potentials $\beta_{\mu=0,1,2,3}$. It is impossible to simultaneously express the demographics of $S^t_A$ of four chemical potentials. Instead, we look into the dependence of each chemical potential on the full demographics. The results are shown in Fig. \ref{SA_gge_nint_XXX}, in which we also show the demographics of $\langle Q_{\mu=0,1,2,3} \rangle$ of the typical states. Again, the subsystem thermalization holds well for $A=1$ and worsens as $A$ increases. Our result confirms that NATS/GGE states can be adopted for the thermalization hypothesis of the typical states, even though the conserved charges are noncommuting \footnote{Our result, in fact, confirms the non-Abelian ETH proposed in \cite{Murthy:2022dao} based on the subsystem version of the thermalization hypothesis for typical states. This is because the ETH considered in \cite{Murthy:2022dao} is examined by comparing the time average of the expectation values of the local observables on the evolved pure states with the corresponding thermal average over the NATS/GGE ensemble states. The former ones are equivalent to the average of the expectation values of the local observables on the typical states.}.

\begin{figure}
\includegraphics[width=0.35\textwidth]
{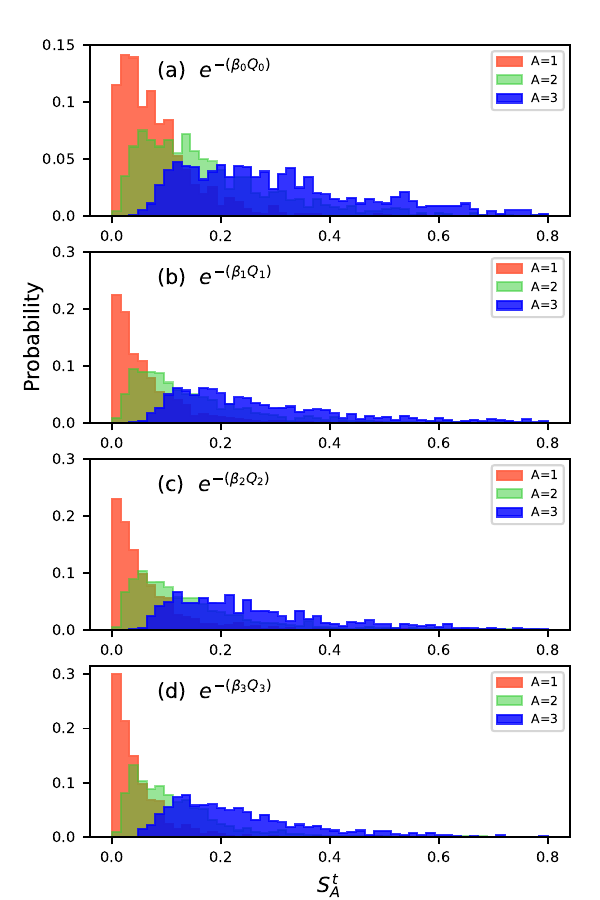}
\caption{\small 
Typical-state thermalization demographics of the population ratios of $S^t_{A=1,2,3}$ for the XXX chain of \eq{XXX_cc} for the p-GGE of a single charge:
(a) $S^t_{A=1,2,3}$  v.s. $\beta_0$, 
(b) $S^t_{A=1,2,3}$  v.s. $\beta_1$,
(c) $S^t_{A=1,2,3}$  v.s. $\beta_2$, and
(d) $S^t_{A=1,2,3}$  v.s. $\beta_3$. }
\label{SU2_typical_one_charge}
\end{figure} 

\begin{figure}
\includegraphics[width=0.35\textwidth]
{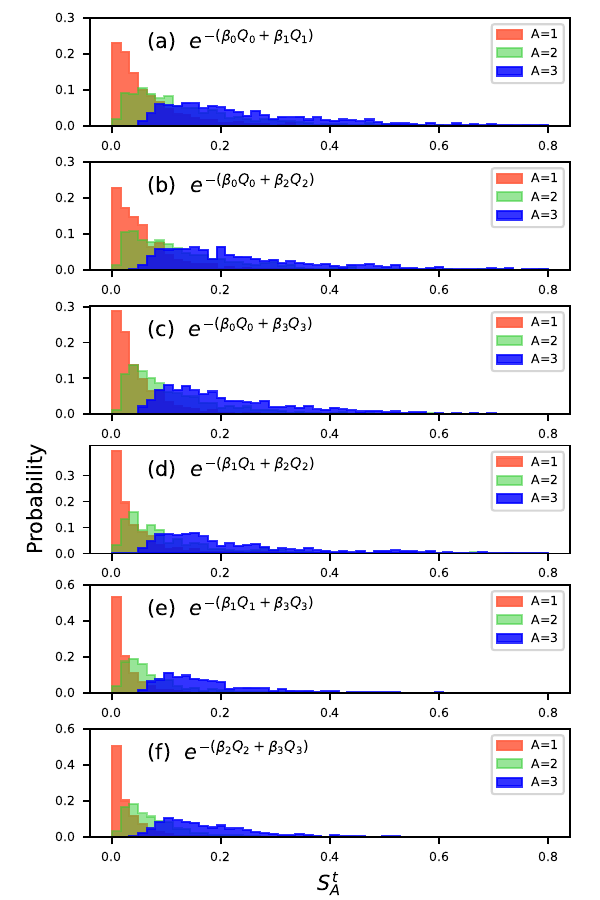}
\caption{\small 
Typical-state thermalization demographics of the population ratios of $S^t_{A=1,2,3}$ for the XXX chain of \eq{XXX_cc} for the p-GGE of two charges: 
(a) $S^t_{A=1,2,3}$  v.s. $(\beta_0, \beta_1)$, 
(b) $S^t_{A=1,2,3}$  v.s. $(\beta_0, \beta_2)$,
(c) $S^t_{A=1,2,3}$  v.s. $(\beta_0, \beta_3)$,
(d) $S^t_{A=1,2,3}$  v.s. $(\beta_1, \beta_2)$, 
(e) $S^t_{A=1,2,3}$  v.s. $(\beta_1, \beta_3)$, and 
(f) $S^t_{A=1,2,3}$  v.s. $(\beta_2, \beta_3)$.   }
\label{SU2_typical_two_charges}
\end{figure} 

\begin{figure}
\includegraphics[width=0.35\textwidth]
{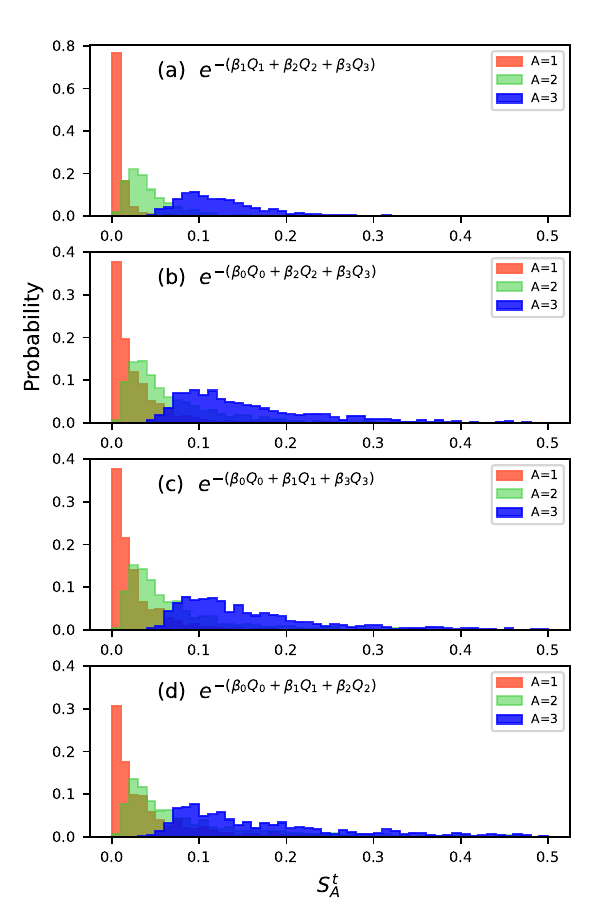}
\caption{\small 
Typical-state thermalization demographics of the population ratios of $S^t_{A=1,2,3}$ for the XXX chain of \eq{XXX_cc} for the p-GGE of three charges:
(a) $S^t_{A=1,2,3}$  v.s. $(\beta_1, \beta_2 , \beta_3)$, 
(b) $S^t_{A=1,2,3}$  v.s. $(\beta_0, \beta_2, \beta_3)$,
(c) $S^t_{A=1,2,3}$  v.s. $(\beta_0, \beta_1, \beta_3)$, and
(d) $S^t_{A=1,2,3}$  v.s. $(\beta_0, \beta_1 , \beta_2)$.   }
\label{SU2_typical_three_charges}
\end{figure} 

We now move to the p-GGEs. To demonstrate the richness of the p-GGEs, we first consider the subsystem thermalization hypothesis by comparing the typical states with the p-GGE states with (i) single conserved charges, (ii) two conserved charges, and (iii) three conserved charges. The results are presented in Fig. \ref{SU2_typical_one_charge}, \ref{SU2_typical_two_charges} and \ref{SU2_typical_three_charges}, respectively. The overall features of these results are as follows. First, the subsystem thermalization hypothesis is more effective for p-GGEs with more conserved charges. Second, for the p-GGEs of the same number of charges, the subsystem thermalization hypothesis works better for the ones involving no $Q_0$. Thus, if we restrict to $A=1$, our results suggest that the typical-state thermalization hypothesis holds quite generically for the p-GGEs with at least two conserved charges without involving $Q_0$. This implies that NATSs are not the only thermal ensemble states adopted for the subsystem thermalization hypothesis for the typical states. Indeed, our framework with p-GGEs extends the thermalization hypothesis to a more general scope. In particular, it is quite interesting to observe that the thermalization hypothesis also applies to those p-GGEs with $Q_0=H$ excluded. This implies that the Hamiltonian may not be essential for considering quantum thermalization. This is also implied by the fact that the (inverse) temperatures for some of the corresponding  GGE or p-GGE are negative, which bears no physical meaning in realistic thermodynamics. This could be related to the thermalization hypothesis working better for the p-GGEs without $Q_0$ than those with $Q_0$.

\subsection{Demographics of eigenstate  thermalization}

\begin{figure}
\includegraphics[width=0.40\textwidth]
{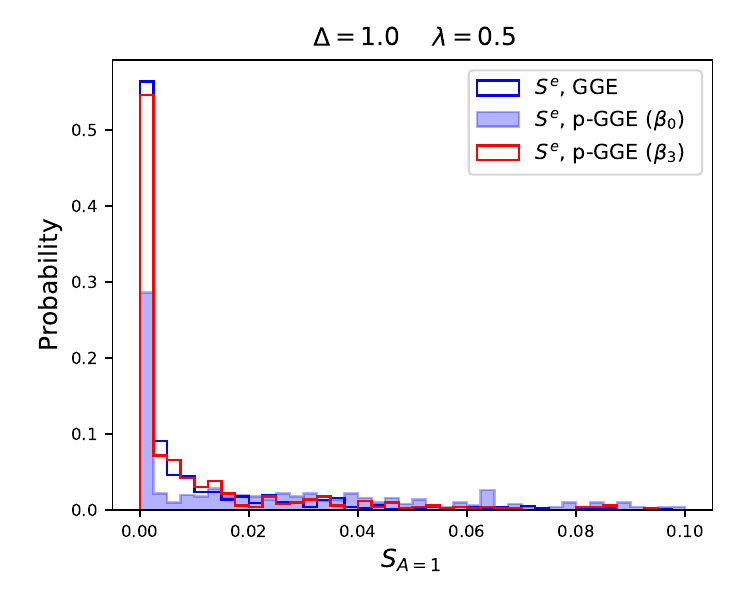}
\caption{\small Subsystem eigenstate thermalization demographics of $S^e_{A=1}$  for the XXX chain of \eq{XXX_cc} by comparing with (i) GGE states (empty blue) and (ii) p-GGE states specified by $\beta_0$ (solid blue) and (iii) p-GGE states specified by $\beta_1$ (empty red).
The subsystem thermalization hypothesis works better for GGE than p-GGE.}
\label{sETH_XXX}
\end{figure}

Finally, we consider the subsystem ETH for the XXX chains with $SU(2)$ conserved charges. Since the $SU(2)$ charges are noncommuting, we choose the eigenstates to be specified by the energy $E$ and the $z$-component of the total spin, i.e., $S^z_{\rm tot}$. For such eigenstates, the expectation values of $S^x_{\rm tot}$ and $S^y_{\rm tot}$ vanish, as expected by the usual argument for the Stern-Gerlach experiment. Thus, it is only sensible to consider subsystem ETH with the p-GGEs specified by either $\beta_0$ or $\beta_3$ or the GGE specified by both. As usual, the subsystem thermalization hypothesis holds well for $A=1$ and worsens as $A$ increases. Thus, we only show the demographics of $S_{A=1}^e$ in Fig. \ref{sETH_XXX}. The results show that the subsystem ETH performs almost equally well for both GGE and p-GGE specified by $\beta_3$, which is better than the p-GGE specified by $\beta_0$. This is similar to the previous cases for typical states and the XXZ chain.

\section{Conclusion}\label{sec:fin}

Quantum thermalization is fascinating because it yields deep implications for the second-law perspective of pure states. Compared to thermalization in open quantum systems or classical systems, quantum thermalization can manifest even without taking the thermodynamic limit. This implies that we can implement the exact diagonalization method to examine the thermalization hypothesis of a not-so-large system, as long as the subsystem size is sufficiently small compared to the system size. This provides the starting point for the numerical plots done in this work.  The usual considerations for the thermalization hypothesis are applied to systems without conserved charges, so that the corresponding thermal ensemble for comparison is either the microcanonical or canonical ensemble. This work extensively considers the subsystem thermalization hypothesis for typical states or eigenstates of systems with conserved charges of $Z_2$, $U(1)$, or $SU(2)$ symmetries.
Furthermore, we generalize the thermal ensembles beyond the generalized Gibbs ensemble (GGE).  We refer to these thermal ensembles as the partial-GGE (p-GGE), from which some conserved charges are excluded. Moreover, in the framework of p-GGEs, the Hamiltonian and other conserved charges are treated on equal footing. 
Thus, when considering the thermalization hypothesis, one can decide which information to discard by choosing the corresponding p-GGE.

Based on the framework of the subsystem thermalization hypothesis with p-GGEs, we can quantify the validity of this scheme by the smallness of the relative entropy between the subsystem's reduced states obtained from the typical states/eigenstates and the corresponding p-GGEs.  We can then examine the universality of quantum thermalization with various p-GGEs by numerically calculating the demographics of relative entropies. Our results on thermalization demographics demonstrate that the subsystem thermalization hypothesis, which applies to typical states (and their fine-grained versions) and eigenstates compared to p-GGEs, holds quite generally as long as the subsystem size is small enough. We have also performed a size-scaling study in Appendix \ref{App_B} to demonstrate that the thermalization hypothesis is more effective for larger systems.

However, there are some cases where the subsystem typical-state thermalization hypothesis fails, for example, in the p-GGEs of the $SU(2)$ case with one or two conserved charges, especially when the Hamiltonian is included. 
Moreover, in some cases. e.g., comparing Fig. \ref{ass_1_XXZ} and Fig. \ref{ass_0_XXZ}, or see Fig. \ref{SU2_typical_one_charge} and Fig.\ref{SU2_typical_two_charges},  we observe that the p-GGEs with $H$ excluded yield sharper demographic around $S^t_A\simeq 0$ than the ones with $H$. In these cases, the thermalization hypothesis works better for the $H$-excluded p-GGEs than the ones with $H$ included. This indicates that the thermalization hypothesis can be extended to p-GGEs even without including the Hamiltonian. This significantly expands the scope of quantum thermalization and warrants future studies for more general systems with conserved charges and larger sizes.

\begin{acknowledgments}
FLL and JJH are supported by Taiwan's NSTC with Grant No.~109-2112-M- 003-007-MY3 and 112-2112-M-003-006-MY3. CYH is supported by Taiwan's National Science and Technology Council (NSTC) with Grant No. 113-2112-M-029 -002 -.  
\end{acknowledgments}


\appendix

{
\section{Size-scaling study of typical-state thermalization of the quantum Ising chain}\label{App_A}

\begin{figure}
\includegraphics[width=0.35\textwidth]{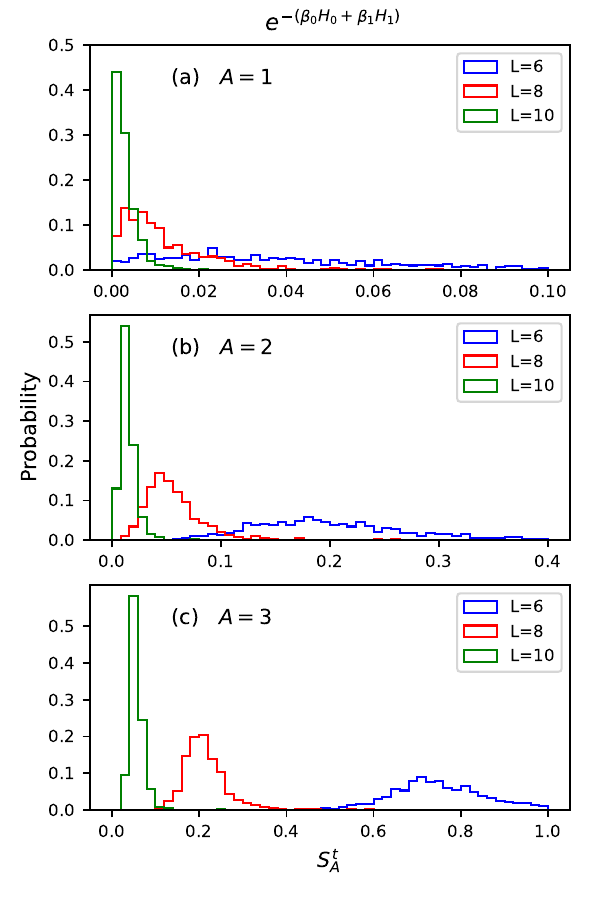}
\caption{\small 
Size-scaling of typical-state thermalization demographics of the population ratios of $S^t_{A=1,2,3}$ for different lattice sizes $L=6,8,10$ of a GGE state for the Ising chain of \eq{Ising_cc} with conserved charges $Q_{0,1}$ given by \eq{Ising_ch}.  }
\label{Ising_different_L}
\end{figure} 

\begin{figure}
\includegraphics[width=0.35\textwidth]{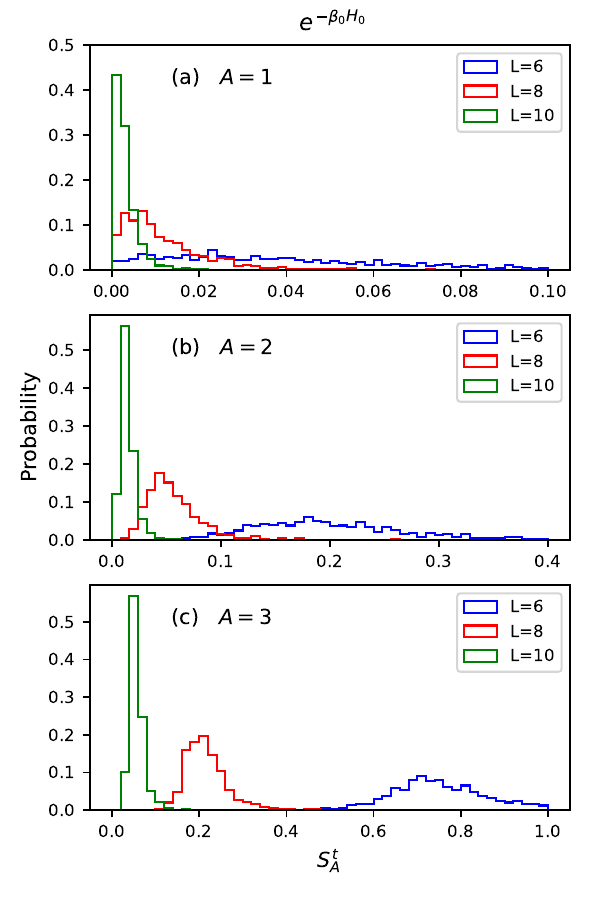}
\caption{\small 
{Size-scaling of typical-state thermalization demographics of the population ratios of $S^t_{A=1,2,3}$ for different lattice sizes $L=6,8,10$ of a p-GGE state for the Ising chain of \eq{Ising_cc} by including only the conserved charges $Q_{0}=H$.}  }
\label{Ising_different_L_B0}
\end{figure} 

\begin{figure}
\includegraphics[width=0.35\textwidth]{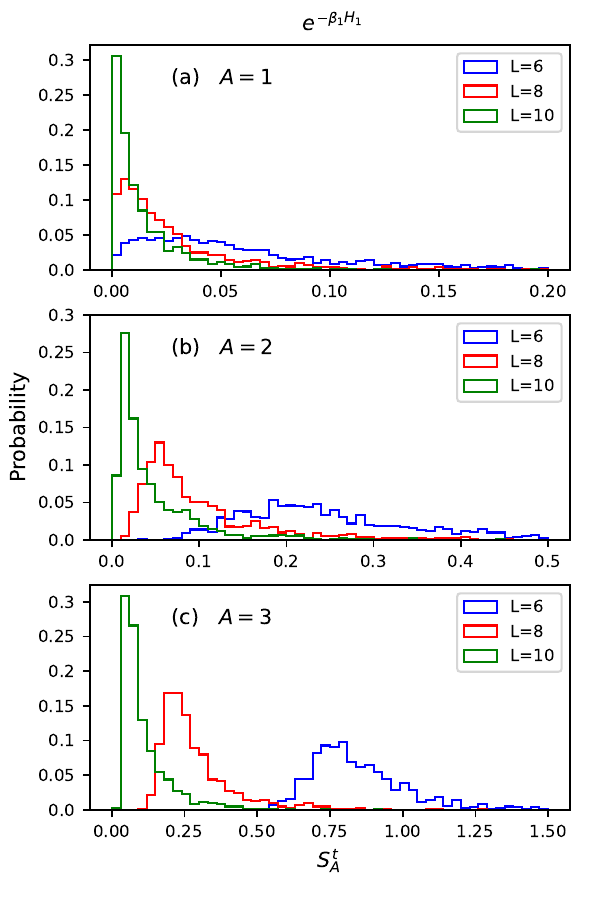}
\caption{\small 
{Size-scaling of typical-state thermalization demographics of the population ratios of $S^t_{A=1,2,3}$ for different lattice sizes $L=6,8,10$ of a p-GGE state for the Ising chain of \eq{Ising_cc} by including only conserved charges $Q_{1}=\Pi$.}  }
\label{Ising_different_L_B1}
\end{figure} 

Fig.~\ref{Ising_different_L} shows the relative entropy results for different subsystem sizes $A=1,2,3$ and lattice sizes $L=6,8,10$ in the 1D Quantum Ising model. The simulations are performed with transverse and longitudinal fields $h_x =1.05$, $h_z=0.5$, and consider two conserved symmetries: the Hamiltonian $Q_0=H$ and the parity operator $Q_1= \Pi$.
Similar to the fig.~\ref{Ising_different_L}, Fig.~\ref{Ising_different_L_B0} considers $Q_0=H$ , while Fig.~\ref{Ising_different_L_B1} considers $Q_1= \Pi$ only.
In Fig.~\ref{Ising_different_L} ~\ref{Ising_different_L_B0} ~\ref{Ising_different_L_B1}(a), it shows the relative entropy probability distributions for the size of the subsystem $A=1$, comparing the three sizes of the system.  Fig.~\ref{Ising_different_L} ~\ref{Ising_different_L_B0} ~\ref{Ising_different_L_B1}(b) and (c) extend this comparison to $A=2$ and $A=3$, respectively. 

As the total system size $L$ increases, the distributions of relative entropy become sharper and more peaked, indicating a more well-defined thermal behavior. For smaller systems $(L=6)$, the distributions of the relative entropy are significantly broader. In contrast, for $L=10$, the relative entropy distributions display a more concentrated profile, consistent with stronger thermalization.  This observation indicates that larger systems exhibit more pronounced thermalization behavior in their subsystems. 
Thermalization becomes more evident as the lattice size increases, which is consistent with the eigenstate thermalization hypothesis framework applied to integrable and near-integrable systems.
}

\section{Typical-state thermalization of the XXZ chain with $U(1)$ conserved charges}\label{App_B}

\begin{figure*} 
\centering
\subfigure[Full Spectrum]{  \includegraphics[width=0.32\textwidth]{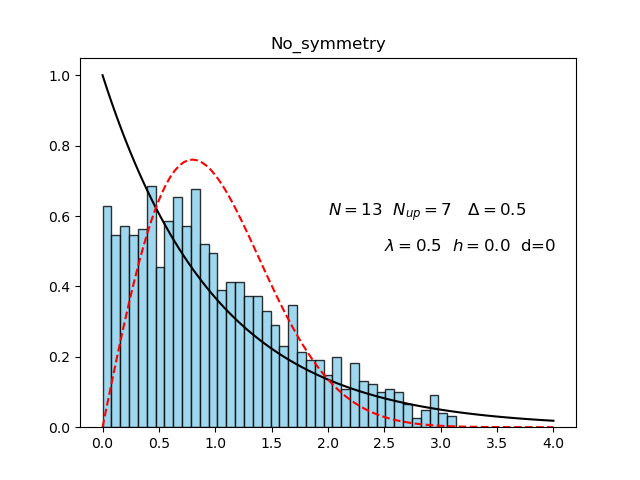}}
\subfigure[P=+1]{      \includegraphics[width=0.30\textwidth]{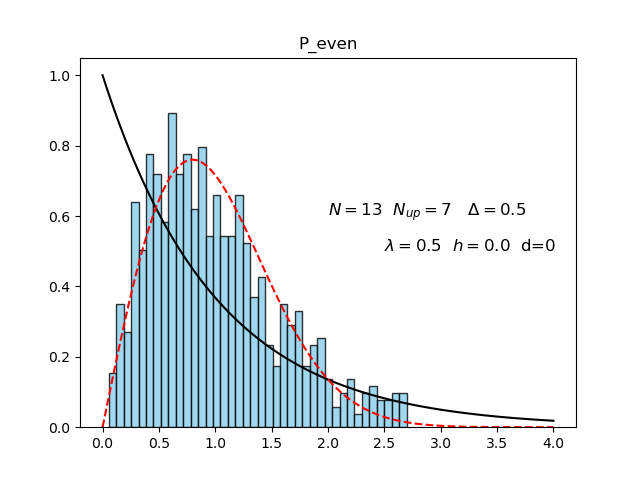}}
\subfigure[P=-1]{.     \includegraphics[width=0.32\textwidth]{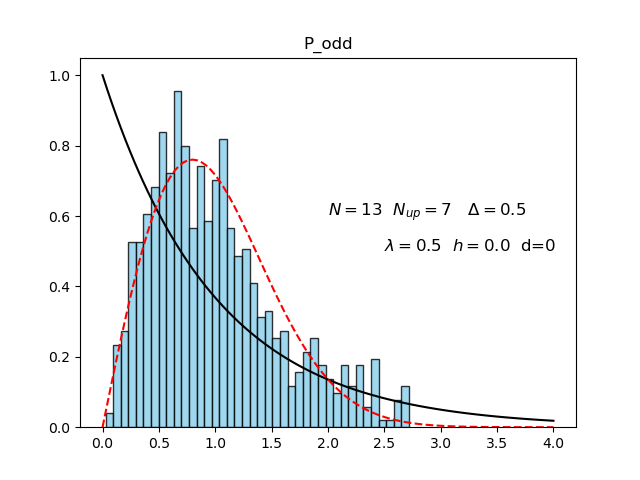}}
\caption{ \small  Level spacing demographics of energy and parity eigenstates for a $L=13$ quantum XXZ open chain for (a) full spectrum, (b) even-parity spectrum, and (c) odd-parity spectrum. The XXZ chain considered is described by \eq{H_XXZ} with $N_{up} = 7$,  $d=h = 0$, $\Delta= \lambda = 0.5$ and $J=1$. This model is non-integrable and has a $U(1)$ symmetry of $S^z_{\rm tot}$ and a parity symmetry of the $\pi$-rotation symmetry around the x-axis. We show that the level-spacing distributions for spectra of superselection sectors fit the Wigner-Dyon distribution (red dashed curve) but not the Poisson distribution (solid black curve) to which the one of the full spectrum is closer. This is consistent with the expectation of the thermalization hypothesis.}
\label{fig:level_spacing}
\end{figure*}

As mentioned in the main text, this model is chaotic if the NNN coupling $0 < \lambda <1$ and $h=0$ (no random disorder) for both XXZ ($\Delta \neq 0$) and XX ($\Delta = 0$) models. We verify this by fitting the level spacing statistics to the Wigner-Dyson distribution, as shown in Fig.~\ref{fig:level_spacing}. This is consistent with the thermalization hypothesis. For readers' interest, we also present the density plots of three quantities to characterize the GGE thermal states associated with the same numerically 
model considered in Fig.~\ref{fig:level_spacing}.

\begin{figure}
\includegraphics[width=0.35\textwidth]{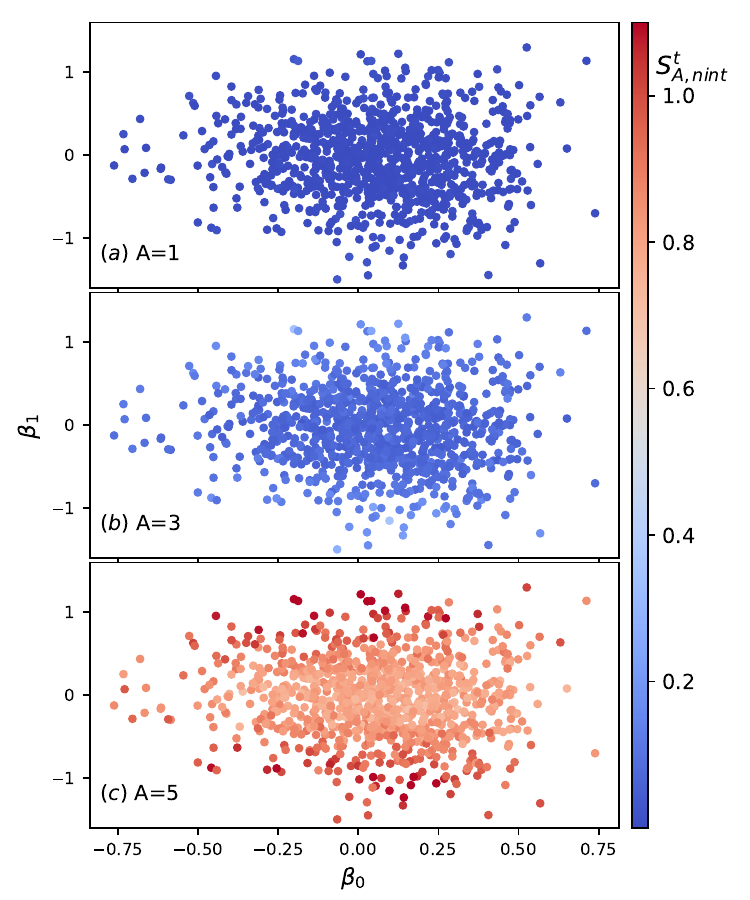}
\caption{\small 
Density plots of typical-state thermalization demographics: $S^t_{A=1,3,5}$ vs. $(\beta_0$, $\beta_1)$ for the XXZ chain of \eq{XXZ_cc} when comparing with the GGE states specified by $(\beta_0,\beta_1)$, i.e., respectively the inverse temperature and the chemical potential to the $Q_1=S^z_{\rm tot}$.
The thermalization hypothesis works well for $A=1$ (a) but worsens as $A$ increases, as seen for $A=3$ (b) and $A=5$ (c). 
}
\label{SA_gge_nint_XXZ}
\end{figure}

\begin{figure}\includegraphics[width=0.35\textwidth]{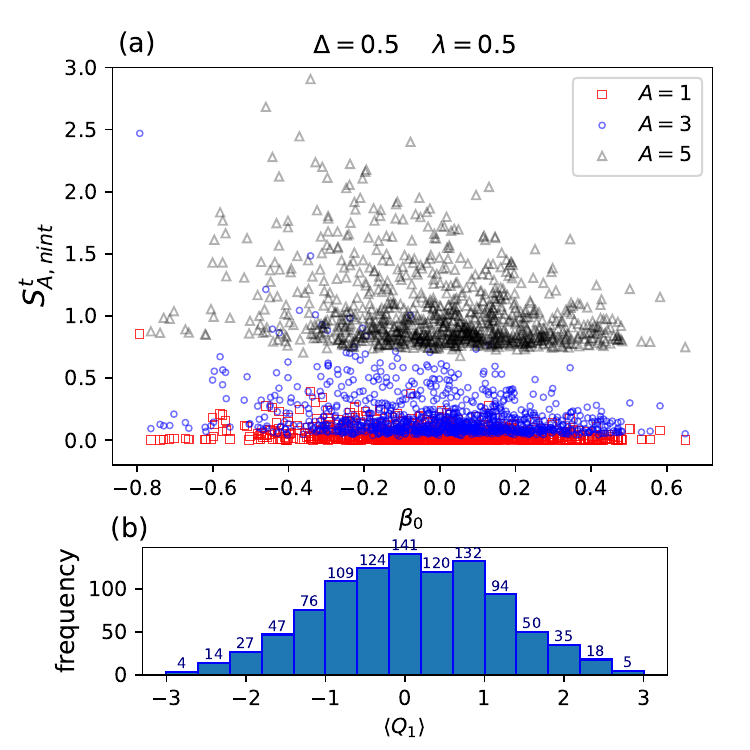}
\caption{\small 
(a) Typical-state thermalization demographics: $S^t_{A=1,3,5}$ v.s. $\beta_0$ for the XXZ chain of \eq{XXZ_cc} by comparing with a p-GGE specified only by $\beta_0$. The thermalization hypothesis also works well for $A=1$ (red squares), but it worsens as $A$ increases, as seen for $A=3$ (blue circles) and $A=5$ (black triangles).  (b) Demographics of the typical states classified by the $\langle Q_0 \rangle$.  }
\label{SA_B0_pgge_nint_XXZ}
\end{figure}

\begin{figure}
\includegraphics[width=0.40\textwidth]{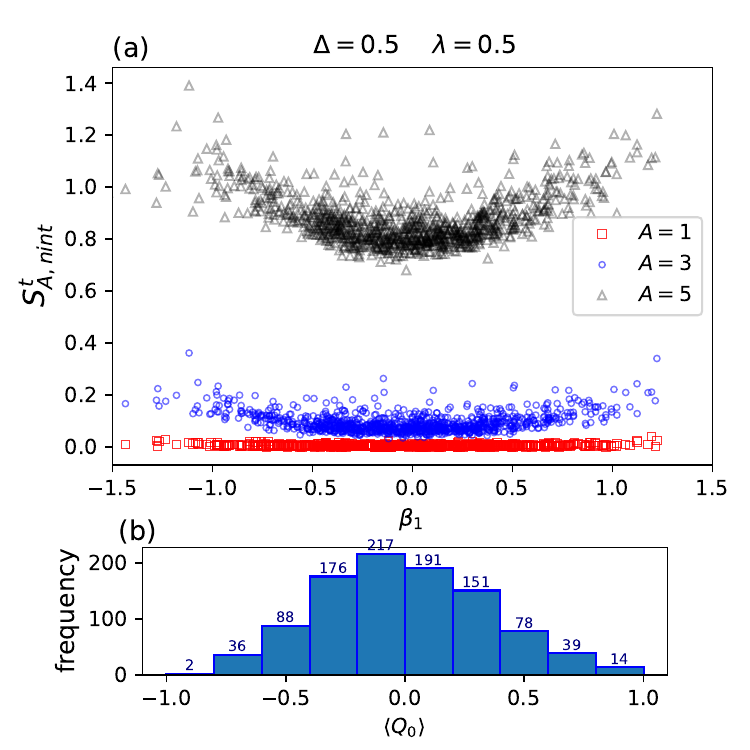}
\caption{\small 
(a) Typical-state thermalization demographics: $S^t_{A=1,3,5}$ v.s. $\beta_1$ for the XXZ spin chain of \eq{XXZ_cc} by comparing with a p-GGE specified only by $\beta_1$. The thermalization hypothesis also works well for $A=1$ (red squares), but it worsens as $A$ increases, as seen for $A=3$ (blue circles) and $A=5$ (black triangles).  (b) Demographics of the typical states classified by the $\langle Q_0 \rangle$. }
\label{SA_B1_pgge_nint_XXZ}
\end{figure}

We now consider the demographics of the thermalization hypothesis. As we have presented the plots for $S^{a, e}_A$ in the main text, we will only show the ones for the typical states. Although we consider the $U(1)$ charge, rather than the $Z_2$ of this XXZ model, for the GGE or p-GGE states, the typical-state thermalization demographics exhibit similar patterns and features to those in the Ising case, with only $Z_Z$ conserved charge. The XXZ model we consider is described by the Hamiltonian \eq{H_XXZ} with the coupling parameters given by $L=10$, $\Delta = \lambda = 0.5$,  $h=d=0$, and $J=1$. The demographics for GGE specified by $(\beta_0, \beta_1)$, and p-GGE specified by $\beta_0$ and $\beta_1$ are given in the Fig. \ref{SA_gge_nint_XXZ}, \ref{SA_B0_pgge_nint_XXZ} and \ref{SA_B1_pgge_nint_XXZ}, respectively.  The thermalization hypothesis seems to work better in Fig. \ref{SA_B1_pgge_nint_XXZ} than in Fig. \ref{SA_B0_pgge_nint_XXZ}.

The results show that the subsystem typical-state thermalization hypothesis works well for $A=1$ and worsens as $A$ increases, similar to the Ising cases. Despite that, it is interesting to compare the demographic patterns of Fig. \ref{SA_gge_nint_XXZ} with its Ising counterpart shown in Fig. \ref{SA_gge_nint_Ising}. We see that the current patterns are more uniform on the $(\beta_0,\beta_1)$-plane due to the continuous symmetry instead of the discrete one.  

The fine-grained versions of Fig. \ref{SA_B0_pgge_nint_XXZ} and Fig. \ref{SA_B1_pgge_nint_XXZ} as the corresponding almost-superselection-state thermalization demographics shown in Fig. \ref{ass_0_XXZ} and Fig. \ref{ass_1_XXZ}, respectively. We see that the almost-superselection-state thermalization hypothesis for $A=1$ works well for the p-GGE specified by $\beta_1$, not for the p-GGE specified by $\beta_0$, for which the thermalization hypothesis worsens as $\langle Q_1 \rangle$ increases away from zero. Interestingly, the almost-superselection-state thermalization demographics of quantum Ising chains, as shown in Fig. \ref{Ising_ass_0} and \ref{Ising_ass_1}, work for the p-GGE specified by $\beta_0$, but not for the p-GGE specified by $\beta_1$. It is the opposite for the ZZX chains.


{
\section{Typical-state thermalization  of the XXZ chain with parity symmetry }\label{App_C}
\begin{figure}
\includegraphics[width=0.40\textwidth]{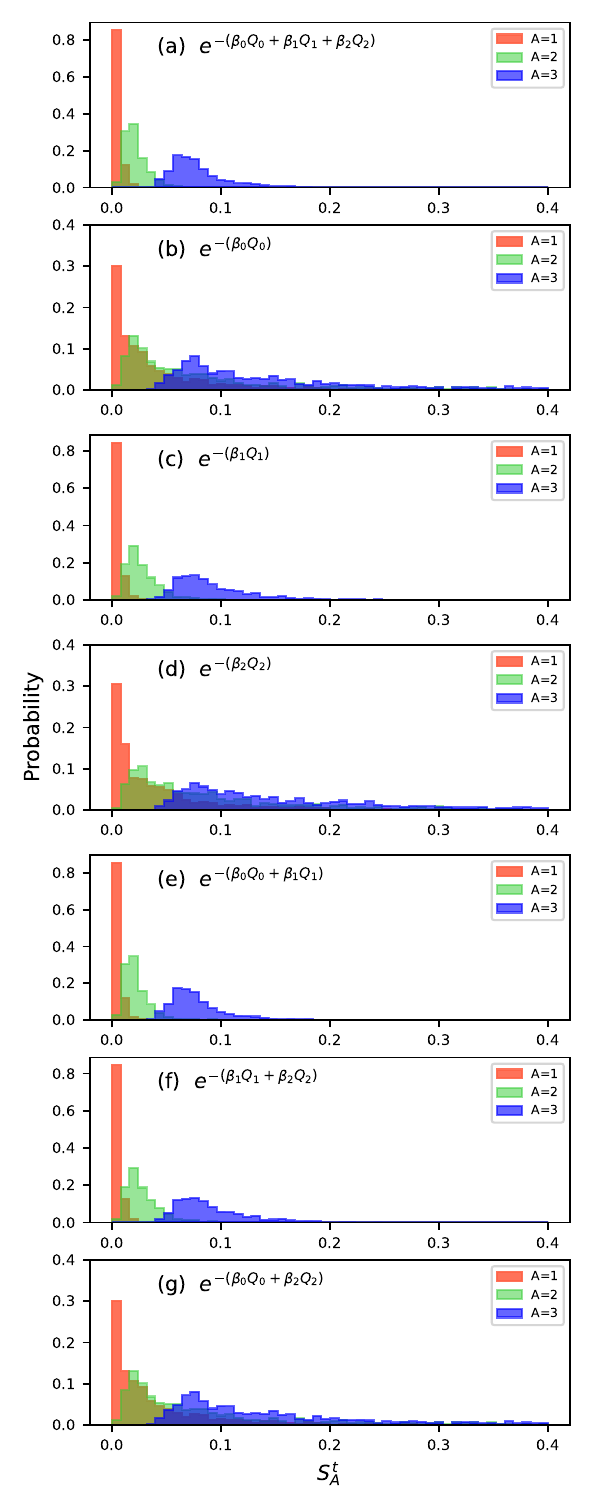}
\caption{\small 
 Typical-state thermalization demographics of the population ratios of $S^t_{A=1,2,3}$ for the XXZ chain of \eq{XXZ_cc} for the p-GGE of three, two, and one charges (Here $Q_0=H$, $Q_1=S^z_{\rm tot}$, $Q_2=\Pi$.): 
(a) $S^t_{A=1,2,3}$  v.s. $(\beta_0, \beta_1, \beta_2)$, 
(b) $S^t_{A=1,2,3}$  v.s. $(\beta_0)$,
(c) $S^t_{A=1,2,3}$  v.s. $(\beta_1)$,
(d) $S^t_{A=1,2,3}$  v.s. $(\beta_2)$, 
(e) $S^t_{A=1,2,3}$  v.s. $(\beta_0, \beta_1)$,
(f) $S^t_{A=1,2,3}$  v.s. $(\beta_1, \beta_2)$, and 
(g) $S^t_{A=1,2,3}$  v.s. $(\beta_0, \beta_2)$. }
\label{XXZ_with_parity}
\end{figure} 

Fig.~\ref{XXZ_with_parity} shows the relative entropy results for different subsystem sizes $A=1,2,3$ in the 1D Quantum XXZ model with lattice length $L=10$. 
The system is analyzed under the parameter conditions $\Delta =\lambda=0.5$ and $J=1$, taking into account three conserved charges denoted by $Q_0=H, \;\;  Q_1=S_{total}^z,$ and $Q_2=\Pi$.
We consider the subsystem thermalization hypothesis for typical states and energy eigenstates by comparing them locally with GGE states and various types of p-GGE states.

Relative entropy is used here as a diagnostic tool to examine the thermalization behavior of subsystems. It measures the distinguishability between the actual reduced density matrix of a subsystem and its corresponding thermal (or generalized Gibbs ensemble) reference state. A lower relative entropy suggests that the subsystem is effectively thermalized, whereas higher values indicate deviations from thermal equilibrium.
}

\clearpage


\begin{thebibliography}{65}%
\makeatletter
\providecommand \@ifxundefined [1]{%
 \@ifx{#1\undefined}
}%
\providecommand \@ifnum [1]{%
 \ifnum #1\expandafter \@firstoftwo
 \else \expandafter \@secondoftwo
 \fi
}%
\providecommand \@ifx [1]{%
 \ifx #1\expandafter \@firstoftwo
 \else \expandafter \@secondoftwo
 \fi
}%
\providecommand \natexlab [1]{#1}%
\providecommand \enquote  [1]{``#1''}%
\providecommand \bibnamefont  [1]{#1}%
\providecommand \bibfnamefont [1]{#1}%
\providecommand \citenamefont [1]{#1}%
\providecommand \href@noop [0]{\@secondoftwo}%
\providecommand \href [0]{\begingroup \@sanitize@url \@href}%
\providecommand \@href[1]{\@@startlink{#1}\@@href}%
\providecommand \@@href[1]{\endgroup#1\@@endlink}%
\providecommand \@sanitize@url [0]{\catcode `\\12\catcode `\$12\catcode
  `\&12\catcode `\#12\catcode `\^12\catcode `\_12\catcode `\%12\relax}%
\providecommand \@@startlink[1]{}%
\providecommand \@@endlink[0]{}%
\providecommand \url  [0]{\begingroup\@sanitize@url \@url }%
\providecommand \@url [1]{\endgroup\@href {#1}{\urlprefix }}%
\providecommand \urlprefix  [0]{URL }%
\providecommand \Eprint [0]{\href }%
\providecommand \doibase [0]{https://doi.org/}%
\providecommand \selectlanguage [0]{\@gobble}%
\providecommand \bibinfo  [0]{\@secondoftwo}%
\providecommand \bibfield  [0]{\@secondoftwo}%
\providecommand \translation [1]{[#1]}%
\providecommand \BibitemOpen [0]{}%
\providecommand \bibitemStop [0]{}%
\providecommand \bibitemNoStop [0]{.\EOS\space}%
\providecommand \EOS [0]{\spacefactor3000\relax}%
\providecommand \BibitemShut  [1]{\csname bibitem#1\endcsname}%
\let\auto@bib@innerbib\@empty
\bibitem [{\citenamefont {von Neumann}(2010)}]{von2010proof}%
  \BibitemOpen
  \bibfield  {author} {\bibinfo {author} {\bibfnamefont {J.}~\bibnamefont {von
  Neumann}},\ }\bibfield  {title} {\bibinfo {title} {Proof of the ergodic
  theorem and the h-theorem in quantum mechanics},\ }\href
  {https://doi.org/10.1140/epjh/e2010-00008-5} {\bibfield  {journal} {\bibinfo
  {journal} {The European Physical Journal H}\ }\textbf {\bibinfo {volume}
  {35}},\ \bibinfo {pages} {201} (\bibinfo {year} {2010})}\BibitemShut
  {NoStop}%
\bibitem [{\citenamefont {Deutsch}(1991)}]{Deutsch:1991msp}%
  \BibitemOpen
  \bibfield  {author} {\bibinfo {author} {\bibfnamefont {J.~M.}\ \bibnamefont
  {Deutsch}},\ }\bibfield  {title} {\bibinfo {title} {{Quantum statistical
  mechanics in a closed system}},\ }\href
  {https://doi.org/10.1103/PhysRevA.43.2046} {\bibfield  {journal} {\bibinfo
  {journal} {Phys. Rev. A}\ }\textbf {\bibinfo {volume} {43}},\ \bibinfo
  {pages} {2046} (\bibinfo {year} {1991})}\BibitemShut {NoStop}%
\bibitem [{\citenamefont {Srednicki}(1994)}]{Srednicki:1994mfb}%
  \BibitemOpen
  \bibfield  {author} {\bibinfo {author} {\bibfnamefont {M.}~\bibnamefont
  {Srednicki}},\ }\bibfield  {title} {\bibinfo {title} {{Chaos and Quantum
  Thermalization}},\ }\href {https://link.aps.org/doi/10.1103/PhysRevE.50.888}
  {\bibfield  {journal} {\bibinfo  {journal} {Phys. Rev. E}\ }\textbf {\bibinfo
  {volume} {50}} (\bibinfo {year} {1994})},\ \Eprint
  {https://arxiv.org/abs/cond-mat/9403051} {arXiv:cond-mat/9403051}
  \BibitemShut {NoStop}%
\bibitem [{\citenamefont {Srednicki}(1996)}]{Srednicki:1995pt}%
  \BibitemOpen
  \bibfield  {author} {\bibinfo {author} {\bibfnamefont {M.}~\bibnamefont
  {Srednicki}},\ }\bibfield  {title} {\bibinfo {title} {{Thermal fluctuations
  in quantized chaotic systems}},\ }\href
  {https://doi.org/10.1088/0305-4470/29/4/003} {\bibfield  {journal} {\bibinfo
  {journal} {J. Phys. A}\ }\textbf {\bibinfo {volume} {29}},\ \bibinfo {pages}
  {L75} (\bibinfo {year} {1996})},\ \Eprint
  {https://arxiv.org/abs/chao-dyn/9511001} {arXiv:chao-dyn/9511001}
  \BibitemShut {NoStop}%
\bibitem [{\citenamefont {Srednicki}(1999)}]{Srednicki_1999}%
  \BibitemOpen
  \bibfield  {author} {\bibinfo {author} {\bibfnamefont {M.}~\bibnamefont
  {Srednicki}},\ }\bibfield  {title} {\bibinfo {title} {The approach to thermal
  equilibrium in quantized chaotic systems},\ }\href
  {https://doi.org/10.1088/0305-4470/32/7/007} {\bibfield  {journal} {\bibinfo
  {journal} {Journal of Physics A: Mathematical and General}\ }\textbf
  {\bibinfo {volume} {32}},\ \bibinfo {pages} {1163–1175} (\bibinfo {year}
  {1999})}\BibitemShut {NoStop}%
\bibitem [{\citenamefont {Rigol}\ \emph {et~al.}(2007)\citenamefont {Rigol},
  \citenamefont {Dunjko},\ and\ \citenamefont
  {Olshanii}}]{rigol2008thermalization}%
  \BibitemOpen
  \bibfield  {author} {\bibinfo {author} {\bibfnamefont {M.}~\bibnamefont
  {Rigol}}, \bibinfo {author} {\bibfnamefont {V.}~\bibnamefont {Dunjko}},\ and\
  \bibinfo {author} {\bibfnamefont {M.}~\bibnamefont {Olshanii}},\ }\bibfield
  {title} {\bibinfo {title} {Thermalization and its mechanism for generic
  isolated quantum systems},\ }\href
  {https://www.nature.com/articles/nature06838} {\bibfield  {journal} {\bibinfo
   {journal} {Nature}\ }\textbf {\bibinfo {volume} {452}},\ \bibinfo {pages}
  {854} (\bibinfo {year} {2007})}\BibitemShut {NoStop}%
\bibitem [{\citenamefont {D'Alessio}\ \emph {et~al.}(2016)\citenamefont
  {D'Alessio}, \citenamefont {Kafri}, \citenamefont {Polkovnikov},\ and\
  \citenamefont {Rigol}}]{DAlessio:2015qtq}%
  \BibitemOpen
  \bibfield  {author} {\bibinfo {author} {\bibfnamefont {L.}~\bibnamefont
  {D'Alessio}}, \bibinfo {author} {\bibfnamefont {Y.}~\bibnamefont {Kafri}},
  \bibinfo {author} {\bibfnamefont {A.}~\bibnamefont {Polkovnikov}},\ and\
  \bibinfo {author} {\bibfnamefont {M.}~\bibnamefont {Rigol}},\ }\bibfield
  {title} {\bibinfo {title} {{From quantum chaos and eigenstate thermalization
  to statistical mechanics and thermodynamics}},\ }\href
  {https://doi.org/10.1080/00018732.2016.1198134} {\bibfield  {journal}
  {\bibinfo  {journal} {Adv. Phys.}\ }\textbf {\bibinfo {volume} {65}},\
  \bibinfo {pages} {239} (\bibinfo {year} {2016})},\ \Eprint
  {https://arxiv.org/abs/1509.06411} {arXiv:1509.06411 [cond-mat.stat-mech]}
  \BibitemShut {NoStop}%
\bibitem [{\citenamefont {Gogolin}\ and\ \citenamefont
  {Eisert}(2016)}]{Gogolin:2015gts}%
  \BibitemOpen
  \bibfield  {author} {\bibinfo {author} {\bibfnamefont {C.}~\bibnamefont
  {Gogolin}}\ and\ \bibinfo {author} {\bibfnamefont {J.}~\bibnamefont
  {Eisert}},\ }\bibfield  {title} {\bibinfo {title} {{Equilibration,
  thermalisation, and the emergence of statistical mechanics in closed quantum
  systems}},\ }\href {https://doi.org/10.1088/0034-4885/79/5/056001} {\bibfield
   {journal} {\bibinfo  {journal} {Rept. Prog. Phys.}\ }\textbf {\bibinfo
  {volume} {79}},\ \bibinfo {pages} {056001} (\bibinfo {year} {2016})},\
  \Eprint {https://arxiv.org/abs/1503.07538} {arXiv:1503.07538 [quant-ph]}
  \BibitemShut {NoStop}%
\bibitem [{\citenamefont {Mori}\ \emph {et~al.}(2018)\citenamefont {Mori},
  \citenamefont {Ikeda}, \citenamefont {Kaminishi},\ and\ \citenamefont
  {Ueda}}]{Mori:2017qhg}%
  \BibitemOpen
  \bibfield  {author} {\bibinfo {author} {\bibfnamefont {T.}~\bibnamefont
  {Mori}}, \bibinfo {author} {\bibfnamefont {T.~N.}\ \bibnamefont {Ikeda}},
  \bibinfo {author} {\bibfnamefont {E.}~\bibnamefont {Kaminishi}},\ and\
  \bibinfo {author} {\bibfnamefont {M.}~\bibnamefont {Ueda}},\ }\bibfield
  {title} {\bibinfo {title} {{Thermalization and prethermalization in isolated
  quantum systems: a theoretical overview}},\ }\href
  {https://doi.org/10.1088/1361-6455/aabcdf} {\bibfield  {journal} {\bibinfo
  {journal} {J. Phys. B}\ }\textbf {\bibinfo {volume} {51}},\ \bibinfo {pages}
  {112001} (\bibinfo {year} {2018})},\ \Eprint
  {https://arxiv.org/abs/1712.08790} {arXiv:1712.08790 [cond-mat.stat-mech]}
  \BibitemShut {NoStop}%
\bibitem [{\citenamefont {Deutsch}(2018)}]{deutsch2018eigenstate}%
  \BibitemOpen
  \bibfield  {author} {\bibinfo {author} {\bibfnamefont {J.~M.}\ \bibnamefont
  {Deutsch}},\ }\bibfield  {title} {\bibinfo {title} {Eigenstate thermalization
  hypothesis},\ }\href {https://doi.org/10.1088/1361-6633/aac9f1} {\bibfield
  {journal} {\bibinfo  {journal} {Reports on Progress in Physics}\ }\textbf
  {\bibinfo {volume} {81}} (\bibinfo {year} {2018})}\BibitemShut {NoStop}%
\bibitem [{\citenamefont {Dymarsky}\ \emph {et~al.}(2018)\citenamefont
  {Dymarsky}, \citenamefont {Lashkari},\ and\ \citenamefont
  {Liu}}]{Dymarsky:2016ntg}%
  \BibitemOpen
  \bibfield  {author} {\bibinfo {author} {\bibfnamefont {A.}~\bibnamefont
  {Dymarsky}}, \bibinfo {author} {\bibfnamefont {N.}~\bibnamefont {Lashkari}},\
  and\ \bibinfo {author} {\bibfnamefont {H.}~\bibnamefont {Liu}},\ }\bibfield
  {title} {\bibinfo {title} {{Subsystem ETH}},\ }\href
  {https://doi.org/10.1103/PhysRevE.97.012140} {\bibfield  {journal} {\bibinfo
  {journal} {Phys. Rev. E}\ }\textbf {\bibinfo {volume} {97}},\ \bibinfo
  {pages} {012140} (\bibinfo {year} {2018})},\ \Eprint
  {https://arxiv.org/abs/1611.08764} {arXiv:1611.08764 [cond-mat.stat-mech]}
  \BibitemShut {NoStop}%
\bibitem [{\citenamefont {Lashkari}\ \emph {et~al.}(2018)\citenamefont
  {Lashkari}, \citenamefont {Dymarsky},\ and\ \citenamefont
  {Liu}}]{Lashkari:2016vgj}%
  \BibitemOpen
  \bibfield  {author} {\bibinfo {author} {\bibfnamefont {N.}~\bibnamefont
  {Lashkari}}, \bibinfo {author} {\bibfnamefont {A.}~\bibnamefont {Dymarsky}},\
  and\ \bibinfo {author} {\bibfnamefont {H.}~\bibnamefont {Liu}},\ }\bibfield
  {title} {\bibinfo {title} {{Eigenstate Thermalization Hypothesis in Conformal
  Field Theory}},\ }\href {https://doi.org/10.1088/1742-5468/aab020} {\bibfield
   {journal} {\bibinfo  {journal} {J. Stat. Mech.}\ }\textbf {\bibinfo {volume}
  {1803}},\ \bibinfo {pages} {033101} (\bibinfo {year} {2018})},\ \Eprint
  {https://arxiv.org/abs/1610.00302} {arXiv:1610.00302 [hep-th]} \BibitemShut
  {NoStop}%
\bibitem [{\citenamefont {He}\ \emph {et~al.}(2017{\natexlab{a}})\citenamefont
  {He}, \citenamefont {Lin},\ and\ \citenamefont {Zhang}}]{He:2017vyf}%
  \BibitemOpen
  \bibfield  {author} {\bibinfo {author} {\bibfnamefont {S.}~\bibnamefont
  {He}}, \bibinfo {author} {\bibfnamefont {F.-L.}\ \bibnamefont {Lin}},\ and\
  \bibinfo {author} {\bibfnamefont {J.-j.}\ \bibnamefont {Zhang}},\ }\bibfield
  {title} {\bibinfo {title} {{Subsystem eigenstate thermalization hypothesis
  for entanglement entropy in CFT}},\ }\href
  {https://doi.org/10.1007/JHEP08(2017)126} {\bibfield  {journal} {\bibinfo
  {journal} {Journal of High Energy Physics}\ }\textbf {\bibinfo {volume}
  {08}},\ \bibinfo {pages} {126} (\bibinfo {year} {2017}{\natexlab{a}})},\
  \Eprint {https://arxiv.org/abs/1703.08724} {arXiv:1703.08724 [hep-th]}
  \BibitemShut {NoStop}%
\bibitem [{\citenamefont {Goldstein}\ \emph {et~al.}(2006)\citenamefont
  {Goldstein}, \citenamefont {Lebowitz}, \citenamefont {Tumulka},\ and\
  \citenamefont {Zanghi}}]{Goldstein:2005aib}%
  \BibitemOpen
  \bibfield  {author} {\bibinfo {author} {\bibfnamefont {S.}~\bibnamefont
  {Goldstein}}, \bibinfo {author} {\bibfnamefont {J.~L.}\ \bibnamefont
  {Lebowitz}}, \bibinfo {author} {\bibfnamefont {R.}~\bibnamefont {Tumulka}},\
  and\ \bibinfo {author} {\bibfnamefont {N.}~\bibnamefont {Zanghi}},\
  }\bibfield  {title} {\bibinfo {title} {{Canonical Typicality}},\ }\href
  {https://doi.org/10.1103/PhysRevLett.96.050403} {\bibfield  {journal}
  {\bibinfo  {journal} {Phys. Rev. Lett.}\ }\textbf {\bibinfo {volume} {96}},\
  \bibinfo {pages} {050403} (\bibinfo {year} {2006})},\ \Eprint
  {https://arxiv.org/abs/cond-mat/0511091} {arXiv:cond-mat/0511091}
  \BibitemShut {NoStop}%
\bibitem [{\citenamefont {Popescu}\ \emph {et~al.}(2006)\citenamefont
  {Popescu}, \citenamefont {Short},\ and\ \citenamefont
  {Winter}}]{popescu2006entanglement}%
  \BibitemOpen
  \bibfield  {author} {\bibinfo {author} {\bibfnamefont {S.}~\bibnamefont
  {Popescu}}, \bibinfo {author} {\bibfnamefont {A.~J.}\ \bibnamefont {Short}},\
  and\ \bibinfo {author} {\bibfnamefont {A.}~\bibnamefont {Winter}},\
  }\bibfield  {title} {\bibinfo {title} {Entanglement and the foundations of
  statistical mechanics},\ }\href {https://doi.org/10.1038/nphys444} {\bibfield
   {journal} {\bibinfo  {journal} {Nature Physics}\ }\textbf {\bibinfo {volume}
  {2}},\ \bibinfo {pages} {754} (\bibinfo {year} {2006})}\BibitemShut {NoStop}%
\bibitem [{\citenamefont {Mueller}\ \emph {et~al.}(2015)\citenamefont
  {Mueller}, \citenamefont {Adlam}, \citenamefont {Masanes},\ and\
  \citenamefont {Wiebe}}]{Mueller:2013bww}%
  \BibitemOpen
  \bibfield  {author} {\bibinfo {author} {\bibfnamefont {M.~P.}\ \bibnamefont
  {Mueller}}, \bibinfo {author} {\bibfnamefont {E.}~\bibnamefont {Adlam}},
  \bibinfo {author} {\bibfnamefont {L.}~\bibnamefont {Masanes}},\ and\ \bibinfo
  {author} {\bibfnamefont {N.}~\bibnamefont {Wiebe}},\ }\bibfield  {title}
  {\bibinfo {title} {{Thermalization and canonical typicality in
  translation-invariant quantum lattice systems}},\ }\href
  {https://doi.org/10.1007/s00220-015-2473-y} {\bibfield  {journal} {\bibinfo
  {journal} {Commun. Math. Phys.}\ }\textbf {\bibinfo {volume} {340}},\
  \bibinfo {pages} {499} (\bibinfo {year} {2015})},\ \Eprint
  {https://arxiv.org/abs/1312.7420} {arXiv:1312.7420 [quant-ph]} \BibitemShut
  {NoStop}%
\bibitem [{\citenamefont {Hayden}\ and\ \citenamefont
  {Preskill}(2007)}]{Hayden:2007cs}%
  \BibitemOpen
  \bibfield  {author} {\bibinfo {author} {\bibfnamefont {P.}~\bibnamefont
  {Hayden}}\ and\ \bibinfo {author} {\bibfnamefont {J.}~\bibnamefont
  {Preskill}},\ }\bibfield  {title} {\bibinfo {title} {{Black holes as mirrors:
  Quantum information in random subsystems}},\ }\href
  {https://doi.org/10.1088/1126-6708/2007/09/120} {\bibfield  {journal}
  {\bibinfo  {journal} {JHEP}\ }\textbf {\bibinfo {volume} {09}},\ \bibinfo
  {pages} {120}},\ \Eprint {https://arxiv.org/abs/0708.4025} {arXiv:0708.4025
  [hep-th]} \BibitemShut {NoStop}%
\bibitem [{\citenamefont {Sekino}\ and\ \citenamefont
  {Susskind}(2008)}]{Sekino:2008he}%
  \BibitemOpen
  \bibfield  {author} {\bibinfo {author} {\bibfnamefont {Y.}~\bibnamefont
  {Sekino}}\ and\ \bibinfo {author} {\bibfnamefont {L.}~\bibnamefont
  {Susskind}},\ }\bibfield  {title} {\bibinfo {title} {{Fast Scramblers}},\
  }\href {https://doi.org/10.1088/1126-6708/2008/10/065} {\bibfield  {journal}
  {\bibinfo  {journal} {JHEP}\ }\textbf {\bibinfo {volume} {10}},\ \bibinfo
  {pages} {065}},\ \Eprint {https://arxiv.org/abs/0808.2096} {arXiv:0808.2096
  [hep-th]} \BibitemShut {NoStop}%
\bibitem [{\citenamefont {Santos}\ and\ \citenamefont
  {Rigol}(2010)}]{santos2010onset}%
  \BibitemOpen
  \bibfield  {author} {\bibinfo {author} {\bibfnamefont {L.~F.}\ \bibnamefont
  {Santos}}\ and\ \bibinfo {author} {\bibfnamefont {M.}~\bibnamefont {Rigol}},\
  }\bibfield  {title} {\bibinfo {title} {Onset of quantum chaos in
  one-dimensional bosonic and fermionic systems and its relation to
  thermalization},\ }\href@noop {} {\bibfield  {journal} {\bibinfo  {journal}
  {Physical Review E—Statistical, Nonlinear, and Soft Matter Physics}\
  }\textbf {\bibinfo {volume} {81}},\ \bibinfo {pages} {036206} (\bibinfo
  {year} {2010})}\BibitemShut {NoStop}%
\bibitem [{\citenamefont {Kaufman}\ \emph
  {et~al.}(2016{\natexlab{a}})\citenamefont {Kaufman}, \citenamefont {Tai},
  \citenamefont {Lukin}, \citenamefont {Rispoli}, \citenamefont {Schittko},
  \citenamefont {Preiss},\ and\ \citenamefont {Greiner}}]{kaufman2016quantum}%
  \BibitemOpen
  \bibfield  {author} {\bibinfo {author} {\bibfnamefont {A.~M.}\ \bibnamefont
  {Kaufman}}, \bibinfo {author} {\bibfnamefont {M.~E.}\ \bibnamefont {Tai}},
  \bibinfo {author} {\bibfnamefont {A.}~\bibnamefont {Lukin}}, \bibinfo
  {author} {\bibfnamefont {M.}~\bibnamefont {Rispoli}}, \bibinfo {author}
  {\bibfnamefont {R.}~\bibnamefont {Schittko}}, \bibinfo {author}
  {\bibfnamefont {P.~M.}\ \bibnamefont {Preiss}},\ and\ \bibinfo {author}
  {\bibfnamefont {M.}~\bibnamefont {Greiner}},\ }\bibfield  {title} {\bibinfo
  {title} {Quantum thermalization through entanglement in an isolated many-body
  system},\ }\href@noop {} {\bibfield  {journal} {\bibinfo  {journal}
  {Science}\ }\textbf {\bibinfo {volume} {353}},\ \bibinfo {pages} {794}
  (\bibinfo {year} {2016}{\natexlab{a}})}\BibitemShut {NoStop}%
\bibitem [{\citenamefont {Zurek}(1981)}]{Zurek:1981xq}%
  \BibitemOpen
  \bibfield  {author} {\bibinfo {author} {\bibfnamefont {W.~H.}\ \bibnamefont
  {Zurek}},\ }\bibfield  {title} {\bibinfo {title} {{Pointer Basis of Quantum
  Apparatus: Into What Mixture Does the Wave Packet Collapse?}},\ }\href
  {https://doi.org/10.1103/PhysRevD.24.1516} {\bibfield  {journal} {\bibinfo
  {journal} {Phys. Rev. D}\ }\textbf {\bibinfo {volume} {24}},\ \bibinfo
  {pages} {1516} (\bibinfo {year} {1981})}\BibitemShut {NoStop}%
\bibitem [{\citenamefont {Biroli}\ \emph {et~al.}(2010)\citenamefont {Biroli},
  \citenamefont {Kollath},\ and\ \citenamefont {L\"auchli}}]{biroli2010effect}%
  \BibitemOpen
  \bibfield  {author} {\bibinfo {author} {\bibfnamefont {G.}~\bibnamefont
  {Biroli}}, \bibinfo {author} {\bibfnamefont {C.}~\bibnamefont {Kollath}},\
  and\ \bibinfo {author} {\bibfnamefont {A.~M.}\ \bibnamefont {L\"auchli}},\
  }\bibfield  {title} {\bibinfo {title} {Effect of rare fluctuations on the
  thermalization of isolated quantum systems},\ }\href
  {https://doi.org/10.1103/PhysRevLett.105.250401} {\bibfield  {journal}
  {\bibinfo  {journal} {Phys. Rev. Lett.}\ }\textbf {\bibinfo {volume} {105}},\
  \bibinfo {pages} {250401} (\bibinfo {year} {2010})}\BibitemShut {NoStop}%
\bibitem [{\citenamefont {Mori}()}]{mori2016weak}%
  \BibitemOpen
  \bibfield  {author} {\bibinfo {author} {\bibfnamefont {T.}~\bibnamefont
  {Mori}},\ }\bibfield  {title} {\bibinfo {title} {Weak eigenstate
  thermalization with large deviation bound},\ }\href@noop {} {\ }\Eprint
  {https://arxiv.org/abs/1609.09776} {arXiv:1609.09776 [cond-mat.stat-mech]}
  \BibitemShut {NoStop}%
\bibitem [{\citenamefont {Iyoda}\ \emph {et~al.}(2017)\citenamefont {Iyoda},
  \citenamefont {Kaneko},\ and\ \citenamefont {Sagawa}}]{iyoda2017fluctuation}%
  \BibitemOpen
  \bibfield  {author} {\bibinfo {author} {\bibfnamefont {E.}~\bibnamefont
  {Iyoda}}, \bibinfo {author} {\bibfnamefont {K.}~\bibnamefont {Kaneko}},\ and\
  \bibinfo {author} {\bibfnamefont {T.}~\bibnamefont {Sagawa}},\ }\bibfield
  {title} {\bibinfo {title} {Fluctuation theorem for many-body pure quantum
  states},\ }\href {https://doi.org/10.1103/PhysRevLett.119.100601} {\bibfield
  {journal} {\bibinfo  {journal} {Phys. Rev. Lett.}\ }\textbf {\bibinfo
  {volume} {119}},\ \bibinfo {pages} {100601} (\bibinfo {year}
  {2017})}\BibitemShut {NoStop}%
\bibitem [{\citenamefont {He}\ \emph {et~al.}(2017{\natexlab{b}})\citenamefont
  {He}, \citenamefont {Lin},\ and\ \citenamefont {Zhang}}]{He:2017txy}%
  \BibitemOpen
  \bibfield  {author} {\bibinfo {author} {\bibfnamefont {S.}~\bibnamefont
  {He}}, \bibinfo {author} {\bibfnamefont {F.-L.}\ \bibnamefont {Lin}},\ and\
  \bibinfo {author} {\bibfnamefont {J.-j.}\ \bibnamefont {Zhang}},\ }\bibfield
  {title} {\bibinfo {title} {{Dissimilarities of reduced density matrices and
  eigenstate thermalization hypothesis}},\ }\href
  {https://doi.org/10.1007/JHEP12(2017)073} {\bibfield  {journal} {\bibinfo
  {journal} {Journal of High Energy Physics}\ }\textbf {\bibinfo {volume}
  {12}},\ \bibinfo {pages} {073} (\bibinfo {year} {2017}{\natexlab{b}})},\
  \Eprint {https://arxiv.org/abs/1708.05090} {arXiv:1708.05090 [hep-th]}
  \BibitemShut {NoStop}%
\bibitem [{\citenamefont {Basu}\ \emph {et~al.}(2017)\citenamefont {Basu},
  \citenamefont {Das}, \citenamefont {Datta},\ and\ \citenamefont
  {Pal}}]{Basu:2017kzo}%
  \BibitemOpen
  \bibfield  {author} {\bibinfo {author} {\bibfnamefont {P.}~\bibnamefont
  {Basu}}, \bibinfo {author} {\bibfnamefont {D.}~\bibnamefont {Das}}, \bibinfo
  {author} {\bibfnamefont {S.}~\bibnamefont {Datta}},\ and\ \bibinfo {author}
  {\bibfnamefont {S.}~\bibnamefont {Pal}},\ }\bibfield  {title} {\bibinfo
  {title} {{Thermality of eigenstates in conformal field theories}},\ }\href
  {https://doi.org/10.1103/PhysRevE.96.022149} {\bibfield  {journal} {\bibinfo
  {journal} {Phys. Rev. E}\ }\textbf {\bibinfo {volume} {96}},\ \bibinfo
  {pages} {022149} (\bibinfo {year} {2017})},\ \Eprint
  {https://arxiv.org/abs/1705.03001} {arXiv:1705.03001 [hep-th]} \BibitemShut
  {NoStop}%
\bibitem [{\citenamefont {Guo}\ \emph {et~al.}(2019)\citenamefont {Guo},
  \citenamefont {Lin},\ and\ \citenamefont {Zhang}}]{Guo:2018pvi}%
  \BibitemOpen
  \bibfield  {author} {\bibinfo {author} {\bibfnamefont {W.-Z.}\ \bibnamefont
  {Guo}}, \bibinfo {author} {\bibfnamefont {F.-L.}\ \bibnamefont {Lin}},\ and\
  \bibinfo {author} {\bibfnamefont {J.}~\bibnamefont {Zhang}},\ }\bibfield
  {title} {\bibinfo {title} {{Note on ETH of descendant states in 2D CFT}},\
  }\href {https://doi.org/10.1007/JHEP01(2019)152} {\bibfield  {journal}
  {\bibinfo  {journal} {Journal of High Energy Physics}\ }\textbf {\bibinfo
  {volume} {01}},\ \bibinfo {pages} {152} (\bibinfo {year} {2019})},\ \Eprint
  {https://arxiv.org/abs/1810.01258} {arXiv:1810.01258 [hep-th]} \BibitemShut
  {NoStop}%
\bibitem [{\citenamefont {Muralidharan}\ \emph {et~al.}(2018)\citenamefont
  {Muralidharan}, \citenamefont {Lochan},\ and\ \citenamefont
  {Shankaranarayanan}}]{Muralidharan:2016acb}%
  \BibitemOpen
  \bibfield  {author} {\bibinfo {author} {\bibfnamefont {S.}~\bibnamefont
  {Muralidharan}}, \bibinfo {author} {\bibfnamefont {K.}~\bibnamefont
  {Lochan}},\ and\ \bibinfo {author} {\bibfnamefont {S.}~\bibnamefont
  {Shankaranarayanan}},\ }\bibfield  {title} {\bibinfo {title} {{Generalized
  thermalization for integrable system under quantum quench}},\ }\href
  {https://doi.org/10.1103/PhysRevE.97.012142} {\bibfield  {journal} {\bibinfo
  {journal} {Phys. Rev. E}\ }\textbf {\bibinfo {volume} {97}},\ \bibinfo
  {pages} {012142} (\bibinfo {year} {2018})},\ \Eprint
  {https://arxiv.org/abs/1611.03205} {arXiv:1611.03205 [quant-ph]} \BibitemShut
  {NoStop}%
\bibitem [{\citenamefont {Byju}\ \emph {et~al.}(2023)\citenamefont {Byju},
  \citenamefont {Lochan},\ and\ \citenamefont
  {Shankaranarayanan}}]{Byju:2018eyb}%
  \BibitemOpen
  \bibfield  {author} {\bibinfo {author} {\bibfnamefont {S.}~\bibnamefont
  {Byju}}, \bibinfo {author} {\bibfnamefont {K.}~\bibnamefont {Lochan}},\ and\
  \bibinfo {author} {\bibfnamefont {S.}~\bibnamefont {Shankaranarayanan}},\
  }\bibfield  {title} {\bibinfo {title} {{Quenched Kitaev chain: Analogous
  model of gravitational collapse}},\ }\href
  {https://doi.org/10.1103/PhysRevD.107.105020} {\bibfield  {journal} {\bibinfo
   {journal} {Phys. Rev. D}\ }\textbf {\bibinfo {volume} {107}},\ \bibinfo
  {pages} {105020} (\bibinfo {year} {2023})},\ \Eprint
  {https://arxiv.org/abs/1808.07742} {arXiv:1808.07742 [cond-mat.quant-gas]}
  \BibitemShut {NoStop}%
\bibitem [{\citenamefont {Abanin}\ \emph {et~al.}(2019)\citenamefont {Abanin},
  \citenamefont {Altman}, \citenamefont {Bloch},\ and\ \citenamefont
  {Serbyn}}]{Abanin_2019}%
  \BibitemOpen
  \bibfield  {author} {\bibinfo {author} {\bibfnamefont {D.~A.}\ \bibnamefont
  {Abanin}}, \bibinfo {author} {\bibfnamefont {E.}~\bibnamefont {Altman}},
  \bibinfo {author} {\bibfnamefont {I.}~\bibnamefont {Bloch}},\ and\ \bibinfo
  {author} {\bibfnamefont {M.}~\bibnamefont {Serbyn}},\ }\bibfield  {title}
  {\bibinfo {title} {Colloquium : Many-body localization, thermalization, and
  entanglement},\ }\bibfield  {journal} {\bibinfo  {journal} {Reviews of Modern
  Physics}\ }\textbf {\bibinfo {volume} {91}},\ \href
  {https://doi.org/10.1103/revmodphys.91.021001} {10.1103/revmodphys.91.021001}
  (\bibinfo {year} {2019})\BibitemShut {NoStop}%
\bibitem [{\citenamefont {Nandkishore}\ and\ \citenamefont
  {Huse}(2015)}]{Nandkishore_2015}%
  \BibitemOpen
  \bibfield  {author} {\bibinfo {author} {\bibfnamefont {R.}~\bibnamefont
  {Nandkishore}}\ and\ \bibinfo {author} {\bibfnamefont {D.~A.}\ \bibnamefont
  {Huse}},\ }\bibfield  {title} {\bibinfo {title} {Many-body localization and
  thermalization in quantum statistical mechanics},\ }\href
  {https://doi.org/10.1146/annurev-conmatphys-031214-014726} {\bibfield
  {journal} {\bibinfo  {journal} {Annual Review of Condensed Matter Physics}\
  }\textbf {\bibinfo {volume} {6}},\ \bibinfo {pages} {15–38} (\bibinfo
  {year} {2015})}\BibitemShut {NoStop}%
\bibitem [{\citenamefont {Sierant}\ \emph {et~al.}(2025)\citenamefont
  {Sierant}, \citenamefont {Lewenstein}, \citenamefont {Scardicchio},
  \citenamefont {Vidmar},\ and\ \citenamefont {Zakrzewski}}]{Sierant_2025}%
  \BibitemOpen
  \bibfield  {author} {\bibinfo {author} {\bibfnamefont {P.}~\bibnamefont
  {Sierant}}, \bibinfo {author} {\bibfnamefont {M.}~\bibnamefont {Lewenstein}},
  \bibinfo {author} {\bibfnamefont {A.}~\bibnamefont {Scardicchio}}, \bibinfo
  {author} {\bibfnamefont {L.}~\bibnamefont {Vidmar}},\ and\ \bibinfo {author}
  {\bibfnamefont {J.}~\bibnamefont {Zakrzewski}},\ }\bibfield  {title}
  {\bibinfo {title} {Many-body localization in the age of classical
  computing*},\ }\href {https://doi.org/10.1088/1361-6633/ad9756} {\bibfield
  {journal} {\bibinfo  {journal} {Reports on Progress in Physics}\ }\textbf
  {\bibinfo {volume} {88}},\ \bibinfo {pages} {026502} (\bibinfo {year}
  {2025})}\BibitemShut {NoStop}%
\bibitem [{\citenamefont {Protopopov}\ \emph {et~al.}(2020)\citenamefont
  {Protopopov}, \citenamefont {Panda}, \citenamefont {Parolini}, \citenamefont
  {Scardicchio}, \citenamefont {Demler},\ and\ \citenamefont
  {Abanin}}]{Protopopov_2020}%
  \BibitemOpen
  \bibfield  {author} {\bibinfo {author} {\bibfnamefont {I.~V.}\ \bibnamefont
  {Protopopov}}, \bibinfo {author} {\bibfnamefont {R.~K.}\ \bibnamefont
  {Panda}}, \bibinfo {author} {\bibfnamefont {T.}~\bibnamefont {Parolini}},
  \bibinfo {author} {\bibfnamefont {A.}~\bibnamefont {Scardicchio}}, \bibinfo
  {author} {\bibfnamefont {E.}~\bibnamefont {Demler}},\ and\ \bibinfo {author}
  {\bibfnamefont {D.~A.}\ \bibnamefont {Abanin}},\ }\bibfield  {title}
  {\bibinfo {title} {Non-abelian symmetries and disorder: A broad nonergodic
  regime and anomalous thermalization},\ }\bibfield  {journal} {\bibinfo
  {journal} {Physical Review X}\ }\textbf {\bibinfo {volume} {10}},\ \href
  {https://doi.org/10.1103/physrevx.10.011025} {10.1103/physrevx.10.011025}
  (\bibinfo {year} {2020})\BibitemShut {NoStop}%
\bibitem [{\citenamefont {Serbyn}\ \emph {et~al.}(2021)\citenamefont {Serbyn},
  \citenamefont {Abanin},\ and\ \citenamefont {Papić}}]{Serbyn_2021}%
  \BibitemOpen
  \bibfield  {author} {\bibinfo {author} {\bibfnamefont {M.}~\bibnamefont
  {Serbyn}}, \bibinfo {author} {\bibfnamefont {D.~A.}\ \bibnamefont {Abanin}},\
  and\ \bibinfo {author} {\bibfnamefont {Z.}~\bibnamefont {Papić}},\
  }\bibfield  {title} {\bibinfo {title} {Quantum many-body scars and weak
  breaking of ergodicity},\ }\href {https://doi.org/10.1038/s41567-021-01230-2}
  {\bibfield  {journal} {\bibinfo  {journal} {Nature Physics}\ }\textbf
  {\bibinfo {volume} {17}},\ \bibinfo {pages} {675–685} (\bibinfo {year}
  {2021})}\BibitemShut {NoStop}%
\bibitem [{\citenamefont {Moudgalya}\ \emph {et~al.}(2018)\citenamefont
  {Moudgalya}, \citenamefont {Regnault},\ and\ \citenamefont
  {Bernevig}}]{Moudgalya_2018}%
  \BibitemOpen
  \bibfield  {author} {\bibinfo {author} {\bibfnamefont {S.}~\bibnamefont
  {Moudgalya}}, \bibinfo {author} {\bibfnamefont {N.}~\bibnamefont
  {Regnault}},\ and\ \bibinfo {author} {\bibfnamefont {B.~A.}\ \bibnamefont
  {Bernevig}},\ }\bibfield  {title} {\bibinfo {title} {Entanglement of exact
  excited states of affleck-kennedy-lieb-tasaki models: Exact results,
  many-body scars, and violation of the strong eigenstate thermalization
  hypothesis},\ }\bibfield  {journal} {\bibinfo  {journal} {Physical Review B}\
  }\textbf {\bibinfo {volume} {98}},\ \href
  {https://doi.org/10.1103/physrevb.98.235156} {10.1103/physrevb.98.235156}
  (\bibinfo {year} {2018})\BibitemShut {NoStop}%
\bibitem [{\citenamefont {Moudgalya}\ \emph {et~al.}(2021)\citenamefont
  {Moudgalya}, \citenamefont {Prem}, \citenamefont {Nandkishore}, \citenamefont
  {Regnault},\ and\ \citenamefont {Bernevig}}]{Moudgalya_2021}%
  \BibitemOpen
  \bibfield  {author} {\bibinfo {author} {\bibfnamefont {S.}~\bibnamefont
  {Moudgalya}}, \bibinfo {author} {\bibfnamefont {A.}~\bibnamefont {Prem}},
  \bibinfo {author} {\bibfnamefont {R.}~\bibnamefont {Nandkishore}}, \bibinfo
  {author} {\bibfnamefont {N.}~\bibnamefont {Regnault}},\ and\ \bibinfo
  {author} {\bibfnamefont {B.~A.}\ \bibnamefont {Bernevig}},\ }\bibinfo {title}
  {Thermalization and its absence within krylov subspaces of a constrained
  hamiltonian},\ in\ \href {https://doi.org/10.1142/9789811231711_0009} {\emph
  {\bibinfo {booktitle} {Memorial Volume for Shoucheng Zhang}}}\ (\bibinfo
  {publisher} {WORLD SCIENTIFIC},\ \bibinfo {year} {2021})\ p.\ \bibinfo
  {pages} {147–209}\BibitemShut {NoStop}%
\bibitem [{\citenamefont {Shiraishi}\ and\ \citenamefont
  {Mori}(2017)}]{Shiraishi_2017}%
  \BibitemOpen
  \bibfield  {author} {\bibinfo {author} {\bibfnamefont {N.}~\bibnamefont
  {Shiraishi}}\ and\ \bibinfo {author} {\bibfnamefont {T.}~\bibnamefont
  {Mori}},\ }\bibfield  {title} {\bibinfo {title} {Systematic construction of
  counterexamples to the eigenstate thermalization hypothesis},\ }\bibfield
  {journal} {\bibinfo  {journal} {Physical Review Letters}\ }\textbf {\bibinfo
  {volume} {119}},\ \href {https://doi.org/10.1103/physrevlett.119.030601}
  {10.1103/physrevlett.119.030601} (\bibinfo {year} {2017})\BibitemShut
  {NoStop}%
\bibitem [{\citenamefont {Moudgalya}\ \emph {et~al.}(2022)\citenamefont
  {Moudgalya}, \citenamefont {Bernevig},\ and\ \citenamefont
  {Regnault}}]{Moudgalya_2022}%
  \BibitemOpen
  \bibfield  {author} {\bibinfo {author} {\bibfnamefont {S.}~\bibnamefont
  {Moudgalya}}, \bibinfo {author} {\bibfnamefont {B.~A.}\ \bibnamefont
  {Bernevig}},\ and\ \bibinfo {author} {\bibfnamefont {N.}~\bibnamefont
  {Regnault}},\ }\bibfield  {title} {\bibinfo {title} {Quantum many-body scars
  and hilbert space fragmentation: a review of exact results},\ }\href
  {https://doi.org/10.1088/1361-6633/ac73a0} {\bibfield  {journal} {\bibinfo
  {journal} {Reports on Progress in Physics}\ }\textbf {\bibinfo {volume}
  {85}},\ \bibinfo {pages} {086501} (\bibinfo {year} {2022})}\BibitemShut
  {NoStop}%
\bibitem [{\citenamefont {Chandran}\ \emph {et~al.}(2023)\citenamefont
  {Chandran}, \citenamefont {Iadecola}, \citenamefont {Khemani},\ and\
  \citenamefont {Moessner}}]{Chandran_2023}%
  \BibitemOpen
  \bibfield  {author} {\bibinfo {author} {\bibfnamefont {A.}~\bibnamefont
  {Chandran}}, \bibinfo {author} {\bibfnamefont {T.}~\bibnamefont {Iadecola}},
  \bibinfo {author} {\bibfnamefont {V.}~\bibnamefont {Khemani}},\ and\ \bibinfo
  {author} {\bibfnamefont {R.}~\bibnamefont {Moessner}},\ }\bibfield  {title}
  {\bibinfo {title} {Quantum many-body scars: A quasiparticle perspective},\
  }\href {https://doi.org/10.1146/annurev-conmatphys-031620-101617} {\bibfield
  {journal} {\bibinfo  {journal} {Annual Review of Condensed Matter Physics}\
  }\textbf {\bibinfo {volume} {14}},\ \bibinfo {pages} {443–469} (\bibinfo
  {year} {2023})}\BibitemShut {NoStop}%
\bibitem [{\citenamefont {Dymarsky}\ and\ \citenamefont
  {Pavlenko}(2019)}]{Dymarsky:2019etq}%
  \BibitemOpen
  \bibfield  {author} {\bibinfo {author} {\bibfnamefont {A.}~\bibnamefont
  {Dymarsky}}\ and\ \bibinfo {author} {\bibfnamefont {K.}~\bibnamefont
  {Pavlenko}},\ }\bibfield  {title} {\bibinfo {title} {{Generalized Eigenstate
  Thermalization Hypothesis in 2D Conformal Field Theories}},\ }\href
  {https://doi.org/10.1103/PhysRevLett.123.111602} {\bibfield  {journal}
  {\bibinfo  {journal} {Phys. Rev. Lett.}\ }\textbf {\bibinfo {volume} {123}},\
  \bibinfo {pages} {111602} (\bibinfo {year} {2019})},\ \Eprint
  {https://arxiv.org/abs/1903.03559} {arXiv:1903.03559 [hep-th]} \BibitemShut
  {NoStop}%
\bibitem [{\citenamefont {Chen}\ \emph {et~al.}(2024)\citenamefont {Chen},
  \citenamefont {Dymarsky}, \citenamefont {Tian},\ and\ \citenamefont
  {Wang}}]{chen2024subsystem}%
  \BibitemOpen
  \bibfield  {author} {\bibinfo {author} {\bibfnamefont {L.}~\bibnamefont
  {Chen}}, \bibinfo {author} {\bibfnamefont {A.}~\bibnamefont {Dymarsky}},
  \bibinfo {author} {\bibfnamefont {J.}~\bibnamefont {Tian}},\ and\ \bibinfo
  {author} {\bibfnamefont {H.}~\bibnamefont {Wang}},\ }\bibfield  {title}
  {\bibinfo {title} {{Subsystem entropy in 2d CFT and KdV ETH}},\ }\href@noop
  {} {\  (\bibinfo {year} {2024})},\ \Eprint {https://arxiv.org/abs/2409.19046}
  {arXiv:2409.19046 [hep-th]} \BibitemShut {NoStop}%
\bibitem [{\citenamefont {Fagotti}\ and\ \citenamefont
  {Essler}(2013)}]{fagotti2013reduced}%
  \BibitemOpen
  \bibfield  {author} {\bibinfo {author} {\bibfnamefont {M.}~\bibnamefont
  {Fagotti}}\ and\ \bibinfo {author} {\bibfnamefont {F.~H.}\ \bibnamefont
  {Essler}},\ }\bibfield  {title} {\bibinfo {title} {Reduced density matrix
  after a quantum quench},\ }\href@noop {} {\bibfield  {journal} {\bibinfo
  {journal} {Physical Review B—Condensed Matter and Materials Physics}\
  }\textbf {\bibinfo {volume} {87}},\ \bibinfo {pages} {245107} (\bibinfo
  {year} {2013})}\BibitemShut {NoStop}%
\bibitem [{\citenamefont {Essler}\ \emph {et~al.}(2017)\citenamefont {Essler},
  \citenamefont {Mussardo},\ and\ \citenamefont
  {Panfil}}]{essler2017truncated}%
  \BibitemOpen
  \bibfield  {author} {\bibinfo {author} {\bibfnamefont {F.}~\bibnamefont
  {Essler}}, \bibinfo {author} {\bibfnamefont {G.}~\bibnamefont {Mussardo}},\
  and\ \bibinfo {author} {\bibfnamefont {M.}~\bibnamefont {Panfil}},\
  }\bibfield  {title} {\bibinfo {title} {On truncated generalized gibbs
  ensembles in the ising field theory},\ }\href@noop {} {\bibfield  {journal}
  {\bibinfo  {journal} {Journal of Statistical Mechanics: Theory and
  Experiment}\ }\textbf {\bibinfo {volume} {2017}},\ \bibinfo {pages} {013103}
  (\bibinfo {year} {2017})}\BibitemShut {NoStop}%
\bibitem [{\citenamefont {Piroli}\ \emph {et~al.}(2017)\citenamefont {Piroli},
  \citenamefont {Pozsgay},\ and\ \citenamefont
  {Vernier}}]{piroli2017integrable}%
  \BibitemOpen
  \bibfield  {author} {\bibinfo {author} {\bibfnamefont {L.}~\bibnamefont
  {Piroli}}, \bibinfo {author} {\bibfnamefont {B.}~\bibnamefont {Pozsgay}},\
  and\ \bibinfo {author} {\bibfnamefont {E.}~\bibnamefont {Vernier}},\
  }\bibfield  {title} {\bibinfo {title} {What is an integrable quench?},\
  }\href@noop {} {\bibfield  {journal} {\bibinfo  {journal} {Nuclear Physics
  B}\ }\textbf {\bibinfo {volume} {925}},\ \bibinfo {pages} {362} (\bibinfo
  {year} {2017})}\BibitemShut {NoStop}%
\bibitem [{\citenamefont {Yunger~Halpern}\ \emph {et~al.}(2016)\citenamefont
  {Yunger~Halpern}, \citenamefont {Faist}, \citenamefont {Oppenheim},\ and\
  \citenamefont {Winter}}]{yunger2016microcanonical}%
  \BibitemOpen
  \bibfield  {author} {\bibinfo {author} {\bibfnamefont {N.}~\bibnamefont
  {Yunger~Halpern}}, \bibinfo {author} {\bibfnamefont {P.}~\bibnamefont
  {Faist}}, \bibinfo {author} {\bibfnamefont {J.}~\bibnamefont {Oppenheim}},\
  and\ \bibinfo {author} {\bibfnamefont {A.}~\bibnamefont {Winter}},\
  }\bibfield  {title} {\bibinfo {title} {{Microcanonical and resource-theoretic
  derivations of the thermal state of a quantum system with noncommuting
  charges}},\ }\href {https://doi.org/10.1038/ncomms12051} {\bibfield
  {journal} {\bibinfo  {journal} {Nature Commun.}\ }\textbf {\bibinfo {volume}
  {7}},\ \bibinfo {pages} {12051} (\bibinfo {year} {2016})}\BibitemShut
  {NoStop}%
\bibitem [{\citenamefont {{Fukai}}\ \emph {et~al.}(2020)\citenamefont
  {{Fukai}}, \citenamefont {{Nozawa}}, \citenamefont {{Kawahara}},\ and\
  \citenamefont {{Ikeda}}}]{2020PhRvR...2c3403F}%
  \BibitemOpen
  \bibfield  {author} {\bibinfo {author} {\bibfnamefont {K.}~\bibnamefont
  {{Fukai}}}, \bibinfo {author} {\bibfnamefont {Y.}~\bibnamefont {{Nozawa}}},
  \bibinfo {author} {\bibfnamefont {K.}~\bibnamefont {{Kawahara}}},\ and\
  \bibinfo {author} {\bibfnamefont {T.~N.}\ \bibnamefont {{Ikeda}}},\
  }\bibfield  {title} {\bibinfo {title} {{Noncommutative generalized Gibbs
  ensemble in isolated integrable quantum systems}},\ }\href
  {https://doi.org/10.1103/PhysRevResearch.2.033403} {\bibfield  {journal}
  {\bibinfo  {journal} {Physical Review Research}\ }\textbf {\bibinfo {volume}
  {2}},\ \bibinfo {eid} {033403} (\bibinfo {year} {2020})},\ \Eprint
  {https://arxiv.org/abs/2003.00022} {arXiv:2003.00022 [cond-mat.stat-mech]}
  \BibitemShut {NoStop}%
\bibitem [{\citenamefont {Murthy}\ \emph {et~al.}(2023)\citenamefont {Murthy},
  \citenamefont {Babakhani}, \citenamefont {Iniguez}, \citenamefont
  {Srednicki},\ and\ \citenamefont {Halpern}}]{Murthy:2022dao}%
  \BibitemOpen
  \bibfield  {author} {\bibinfo {author} {\bibfnamefont {C.}~\bibnamefont
  {Murthy}}, \bibinfo {author} {\bibfnamefont {A.}~\bibnamefont {Babakhani}},
  \bibinfo {author} {\bibfnamefont {F.}~\bibnamefont {Iniguez}}, \bibinfo
  {author} {\bibfnamefont {M.}~\bibnamefont {Srednicki}},\ and\ \bibinfo
  {author} {\bibfnamefont {N.~Y.}\ \bibnamefont {Halpern}},\ }\bibfield
  {title} {\bibinfo {title} {{Non-Abelian Eigenstate Thermalization
  Hypothesis}},\ }\href {https://doi.org/10.1103/PhysRevLett.130.140402}
  {\bibfield  {journal} {\bibinfo  {journal} {Phys. Rev. Lett.}\ }\textbf
  {\bibinfo {volume} {130}},\ \bibinfo {pages} {140402} (\bibinfo {year}
  {2023})},\ \Eprint {https://arxiv.org/abs/2206.05310} {arXiv:2206.05310
  [quant-ph]} \BibitemShut {NoStop}%
\bibitem [{\citenamefont {Yunger~Halpern}\ \emph {et~al.}(2020)\citenamefont
  {Yunger~Halpern}, \citenamefont {Beverland},\ and\ \citenamefont
  {Kalev}}]{PhysRevE.101.042117}%
  \BibitemOpen
  \bibfield  {author} {\bibinfo {author} {\bibfnamefont {N.}~\bibnamefont
  {Yunger~Halpern}}, \bibinfo {author} {\bibfnamefont {M.~E.}\ \bibnamefont
  {Beverland}},\ and\ \bibinfo {author} {\bibfnamefont {A.}~\bibnamefont
  {Kalev}},\ }\bibfield  {title} {\bibinfo {title} {Noncommuting conserved
  charges in quantum many-body thermalization},\ }\href
  {https://doi.org/10.1103/PhysRevE.101.042117} {\bibfield  {journal} {\bibinfo
   {journal} {Phys. Rev. E}\ }\textbf {\bibinfo {volume} {101}},\ \bibinfo
  {pages} {042117} (\bibinfo {year} {2020})}\BibitemShut {NoStop}%
\bibitem [{\citenamefont {Majidy}\ \emph {et~al.}(2023)\citenamefont {Majidy},
  \citenamefont {Braasch}, \citenamefont {Lasek}, \citenamefont {Upadhyaya},
  \citenamefont {Kalev},\ and\ \citenamefont {Halpern}}]{Majidy:2023xhm}%
  \BibitemOpen
  \bibfield  {author} {\bibinfo {author} {\bibfnamefont {S.}~\bibnamefont
  {Majidy}}, \bibinfo {author} {\bibfnamefont {W.~F.}\ \bibnamefont {Braasch}},
  \bibinfo {author} {\bibfnamefont {A.}~\bibnamefont {Lasek}}, \bibinfo
  {author} {\bibfnamefont {T.}~\bibnamefont {Upadhyaya}}, \bibinfo {author}
  {\bibfnamefont {A.}~\bibnamefont {Kalev}},\ and\ \bibinfo {author}
  {\bibfnamefont {N.~Y.}\ \bibnamefont {Halpern}},\ }\bibfield  {title}
  {\bibinfo {title} {{Noncommuting conserved charges in quantum thermodynamics
  and beyond}},\ }\href {https://doi.org/10.1038/s42254-023-00641-9} {\bibfield
   {journal} {\bibinfo  {journal} {Nature Rev. Phys.}\ }\textbf {\bibinfo
  {volume} {5}},\ \bibinfo {pages} {689} (\bibinfo {year} {2023})},\ \Eprint
  {https://arxiv.org/abs/2306.00054} {arXiv:2306.00054 [quant-ph]} \BibitemShut
  {NoStop}%
\bibitem [{\citenamefont {Lasek}\ \emph {et~al.}(2024)\citenamefont {Lasek},
  \citenamefont {Noh}, \citenamefont {LeSchack},\ and\ \citenamefont
  {Halpern}}]{Lasek:2024ess}%
  \BibitemOpen
  \bibfield  {author} {\bibinfo {author} {\bibfnamefont {A.}~\bibnamefont
  {Lasek}}, \bibinfo {author} {\bibfnamefont {J.~D.}\ \bibnamefont {Noh}},
  \bibinfo {author} {\bibfnamefont {J.}~\bibnamefont {LeSchack}},\ and\
  \bibinfo {author} {\bibfnamefont {N.~Y.}\ \bibnamefont {Halpern}},\
  }\bibfield  {title} {\bibinfo {title} {{Numerical evidence for the
  non-Abelian eigenstate thermalization hypothesis}},\ }\href@noop {} {\
  (\bibinfo {year} {2024})},\ \Eprint {https://arxiv.org/abs/2412.07838}
  {arXiv:2412.07838 [quant-ph]} \BibitemShut {NoStop}%
\bibitem [{\citenamefont {Upadhyaya}\ \emph {et~al.}(2024)\citenamefont
  {Upadhyaya}, \citenamefont {Braasch}, \citenamefont {Landi},\ and\
  \citenamefont {Halpern}}]{Upadhyaya:2023kjv}%
  \BibitemOpen
  \bibfield  {author} {\bibinfo {author} {\bibfnamefont {T.}~\bibnamefont
  {Upadhyaya}}, \bibinfo {author} {\bibfnamefont {W.~F.}\ \bibnamefont
  {Braasch}, \bibfnamefont {Jr.}}, \bibinfo {author} {\bibfnamefont {G.~T.}\
  \bibnamefont {Landi}},\ and\ \bibinfo {author} {\bibfnamefont {N.~Y.}\
  \bibnamefont {Halpern}},\ }\bibfield  {title} {\bibinfo {title} {{Non-Abelian
  Transport Distinguishes Three Usually Equivalent Notions of Entropy
  Production}},\ }\href {https://doi.org/10.1103/PRXQuantum.5.030355}
  {\bibfield  {journal} {\bibinfo  {journal} {PRX Quantum}\ }\textbf {\bibinfo
  {volume} {5}},\ \bibinfo {pages} {030355} (\bibinfo {year} {2024})},\ \Eprint
  {https://arxiv.org/abs/2305.15480} {arXiv:2305.15480 [quant-ph]} \BibitemShut
  {NoStop}%
\bibitem [{\citenamefont {Kaufman}\ \emph
  {et~al.}(2016{\natexlab{b}})\citenamefont {Kaufman}, \citenamefont {Tai},
  \citenamefont {Lukin}, \citenamefont {Rispoli}, \citenamefont {Schittko},
  \citenamefont {Preiss},\ and\ \citenamefont {Greiner}}]{Kaufman2016}%
  \BibitemOpen
  \bibfield  {author} {\bibinfo {author} {\bibfnamefont {A.~M.}\ \bibnamefont
  {Kaufman}}, \bibinfo {author} {\bibfnamefont {M.~E.}\ \bibnamefont {Tai}},
  \bibinfo {author} {\bibfnamefont {A.}~\bibnamefont {Lukin}}, \bibinfo
  {author} {\bibfnamefont {M.}~\bibnamefont {Rispoli}}, \bibinfo {author}
  {\bibfnamefont {R.}~\bibnamefont {Schittko}}, \bibinfo {author}
  {\bibfnamefont {P.~M.}\ \bibnamefont {Preiss}},\ and\ \bibinfo {author}
  {\bibfnamefont {M.}~\bibnamefont {Greiner}},\ }\bibfield  {title} {\bibinfo
  {title} {Quantum thermalization through entanglement in an isolated many-body
  system},\ }\href {https://doi.org/10.1126/science.aaf6725} {\bibfield
  {journal} {\bibinfo  {journal} {Science}\ }\textbf {\bibinfo {volume}
  {353}},\ \bibinfo {pages} {794} (\bibinfo {year} {2016}{\natexlab{b}})},\
  \Eprint
  {https://arxiv.org/abs/https://www.science.org/doi/pdf/10.1126/science.aaf6725}
  {https://www.science.org/doi/pdf/10.1126/science.aaf6725} \BibitemShut
  {NoStop}%
\bibitem [{\citenamefont {Clos}\ \emph {et~al.}(2016)\citenamefont {Clos},
  \citenamefont {Porras}, \citenamefont {Warring},\ and\ \citenamefont
  {Schaetz}}]{Govinda2016}%
  \BibitemOpen
  \bibfield  {author} {\bibinfo {author} {\bibfnamefont {G.}~\bibnamefont
  {Clos}}, \bibinfo {author} {\bibfnamefont {D.}~\bibnamefont {Porras}},
  \bibinfo {author} {\bibfnamefont {U.}~\bibnamefont {Warring}},\ and\ \bibinfo
  {author} {\bibfnamefont {T.}~\bibnamefont {Schaetz}},\ }\bibfield  {title}
  {\bibinfo {title} {Time-resolved observation of thermalization in an isolated
  quantum system},\ }\href {https://doi.org/10.1103/PhysRevLett.117.170401}
  {\bibfield  {journal} {\bibinfo  {journal} {Phys. Rev. Lett.}\ }\textbf
  {\bibinfo {volume} {117}},\ \bibinfo {pages} {170401} (\bibinfo {year}
  {2016})}\BibitemShut {NoStop}%
\bibitem [{\citenamefont {Neill}\ and\ \citenamefont
  {Martinis}(2016)}]{Neill2016}%
  \BibitemOpen
  \bibfield  {author} {\bibinfo {author} {\bibfnamefont {R.~P. F. M. C. Y. K.
  M. C. Z. M. A. B. R. C. B. C. B. D. A. J. E. K. J. M. J. O. P. J. J. Q. C. S.
  D. V. A. W. J. W. T. C. P.~A.}\ \bibnamefont {Neill}, \bibfnamefont {C.}}\
  and\ \bibinfo {author} {\bibfnamefont {J.~M.}\ \bibnamefont {Martinis}},\
  }\bibfield  {title} {\bibinfo {title} {Ergodic dynamics and thermalization in
  an isolated quantum system},\ }\href {https://doi.org/10.1038/nphys3830}
  {\bibfield  {journal} {\bibinfo  {journal} {Nature Physics}\ }\textbf
  {\bibinfo {volume} {12}},\ \bibinfo {pages} {1037} (\bibinfo {year}
  {2016})}\BibitemShut {NoStop}%
\bibitem [{\citenamefont {Kinoshita}\ \emph {et~al.}(2004)\citenamefont
  {Kinoshita}, \citenamefont {Wenger},\ and\ \citenamefont
  {Weiss}}]{Kinoshita:2004fnt}%
  \BibitemOpen
  \bibfield  {author} {\bibinfo {author} {\bibfnamefont {T.}~\bibnamefont
  {Kinoshita}}, \bibinfo {author} {\bibfnamefont {T.}~\bibnamefont {Wenger}},\
  and\ \bibinfo {author} {\bibfnamefont {D.~S.}\ \bibnamefont {Weiss}},\
  }\bibfield  {title} {\bibinfo {title} {{Observation of a One-Dimensional
  Tonks-Girardeau Gas}},\ }\href {https://doi.org/10.1126/science.1100700}
  {\bibfield  {journal} {\bibinfo  {journal} {Science}\ }\textbf {\bibinfo
  {volume} {305}},\ \bibinfo {pages} {1100700} (\bibinfo {year}
  {2004})}\BibitemShut {NoStop}%
\bibitem [{\citenamefont {Gring}\ \emph {et~al.}(2012)\citenamefont {Gring},
  \citenamefont {Kuhnert}, \citenamefont {Langen}, \citenamefont {Kitagawa},
  \citenamefont {Rauer}, \citenamefont {Schreitl}, \citenamefont {Mazets},
  \citenamefont {Smith}, \citenamefont {Demler},\ and\ \citenamefont
  {Schmiedmayer}}]{Gring2012}%
  \BibitemOpen
  \bibfield  {author} {\bibinfo {author} {\bibfnamefont {M.}~\bibnamefont
  {Gring}}, \bibinfo {author} {\bibfnamefont {M.}~\bibnamefont {Kuhnert}},
  \bibinfo {author} {\bibfnamefont {T.}~\bibnamefont {Langen}}, \bibinfo
  {author} {\bibfnamefont {T.}~\bibnamefont {Kitagawa}}, \bibinfo {author}
  {\bibfnamefont {B.}~\bibnamefont {Rauer}}, \bibinfo {author} {\bibfnamefont
  {M.}~\bibnamefont {Schreitl}}, \bibinfo {author} {\bibfnamefont
  {I.}~\bibnamefont {Mazets}}, \bibinfo {author} {\bibfnamefont {D.~A.}\
  \bibnamefont {Smith}}, \bibinfo {author} {\bibfnamefont {E.}~\bibnamefont
  {Demler}},\ and\ \bibinfo {author} {\bibfnamefont {J.}~\bibnamefont
  {Schmiedmayer}},\ }\bibfield  {title} {\bibinfo {title} {Relaxation and
  prethermalization in an isolated quantum system},\ }\href
  {https://doi.org/10.1126/science.1224953} {\bibfield  {journal} {\bibinfo
  {journal} {Science}\ }\textbf {\bibinfo {volume} {337}},\ \bibinfo {pages}
  {1318} (\bibinfo {year} {2012})},\ \Eprint
  {https://arxiv.org/abs/https://www.science.org/doi/pdf/10.1126/science.1224953}
  {https://www.science.org/doi/pdf/10.1126/science.1224953} \BibitemShut
  {NoStop}%
\bibitem [{\citenamefont {Langen}\ \emph {et~al.}(2015)\citenamefont {Langen},
  \citenamefont {Erne}, \citenamefont {Geiger}, \citenamefont {Rauer},
  \citenamefont {Schweigler}, \citenamefont {Kuhnert}, \citenamefont
  {Rohringer}, \citenamefont {Mazets}, \citenamefont {Gasenzer},\ and\
  \citenamefont {Schmiedmayer}}]{Langen2015}%
  \BibitemOpen
  \bibfield  {author} {\bibinfo {author} {\bibfnamefont {T.}~\bibnamefont
  {Langen}}, \bibinfo {author} {\bibfnamefont {S.}~\bibnamefont {Erne}},
  \bibinfo {author} {\bibfnamefont {R.}~\bibnamefont {Geiger}}, \bibinfo
  {author} {\bibfnamefont {B.}~\bibnamefont {Rauer}}, \bibinfo {author}
  {\bibfnamefont {T.}~\bibnamefont {Schweigler}}, \bibinfo {author}
  {\bibfnamefont {M.}~\bibnamefont {Kuhnert}}, \bibinfo {author} {\bibfnamefont
  {W.}~\bibnamefont {Rohringer}}, \bibinfo {author} {\bibfnamefont {I.~E.}\
  \bibnamefont {Mazets}}, \bibinfo {author} {\bibfnamefont {T.}~\bibnamefont
  {Gasenzer}},\ and\ \bibinfo {author} {\bibfnamefont {J.}~\bibnamefont
  {Schmiedmayer}},\ }\bibfield  {title} {\bibinfo {title} {Experimental
  observation of a generalized gibbs ensemble},\ }\href
  {https://doi.org/10.1126/science.1257026} {\bibfield  {journal} {\bibinfo
  {journal} {Science}\ }\textbf {\bibinfo {volume} {348}},\ \bibinfo {pages}
  {207} (\bibinfo {year} {2015})},\ \Eprint
  {https://arxiv.org/abs/https://www.science.org/doi/pdf/10.1126/science.1257026}
  {https://www.science.org/doi/pdf/10.1126/science.1257026} \BibitemShut
  {NoStop}%
\bibitem [{\citenamefont {Scholl}\ \emph {et~al.}(2022)\citenamefont {Scholl},
  \citenamefont {Williams}, \citenamefont {Bornet}, \citenamefont {Wallner},
  \citenamefont {Barredo}, \citenamefont {Henriet}, \citenamefont {Signoles},
  \citenamefont {Hainaut}, \citenamefont {Franz}, \citenamefont {Geier},
  \citenamefont {Tebben}, \citenamefont {Salzinger}, \citenamefont {Z\"urn},
  \citenamefont {Lahaye}, \citenamefont {Weidem\"uller},\ and\ \citenamefont
  {Browaeys}}]{Scholl2022}%
  \BibitemOpen
  \bibfield  {author} {\bibinfo {author} {\bibfnamefont {P.}~\bibnamefont
  {Scholl}}, \bibinfo {author} {\bibfnamefont {H.~J.}\ \bibnamefont
  {Williams}}, \bibinfo {author} {\bibfnamefont {G.}~\bibnamefont {Bornet}},
  \bibinfo {author} {\bibfnamefont {F.}~\bibnamefont {Wallner}}, \bibinfo
  {author} {\bibfnamefont {D.}~\bibnamefont {Barredo}}, \bibinfo {author}
  {\bibfnamefont {L.}~\bibnamefont {Henriet}}, \bibinfo {author} {\bibfnamefont
  {A.}~\bibnamefont {Signoles}}, \bibinfo {author} {\bibfnamefont
  {C.}~\bibnamefont {Hainaut}}, \bibinfo {author} {\bibfnamefont
  {T.}~\bibnamefont {Franz}}, \bibinfo {author} {\bibfnamefont
  {S.}~\bibnamefont {Geier}}, \bibinfo {author} {\bibfnamefont
  {A.}~\bibnamefont {Tebben}}, \bibinfo {author} {\bibfnamefont
  {A.}~\bibnamefont {Salzinger}}, \bibinfo {author} {\bibfnamefont
  {G.}~\bibnamefont {Z\"urn}}, \bibinfo {author} {\bibfnamefont
  {T.}~\bibnamefont {Lahaye}}, \bibinfo {author} {\bibfnamefont
  {M.}~\bibnamefont {Weidem\"uller}},\ and\ \bibinfo {author} {\bibfnamefont
  {A.}~\bibnamefont {Browaeys}},\ }\bibfield  {title} {\bibinfo {title}
  {Microwave engineering of programmable $xxz$ hamiltonians in arrays of
  rydberg atoms},\ }\href {https://doi.org/10.1103/PRXQuantum.3.020303}
  {\bibfield  {journal} {\bibinfo  {journal} {PRX Quantum}\ }\textbf {\bibinfo
  {volume} {3}},\ \bibinfo {pages} {020303} (\bibinfo {year}
  {2022})}\BibitemShut {NoStop}%
\bibitem [{\citenamefont {Cover}\ and\ \citenamefont
  {Thomas}(2006)}]{Cover2006}%
  \BibitemOpen
  \bibfield  {author} {\bibinfo {author} {\bibfnamefont {T.~M.}\ \bibnamefont
  {Cover}}\ and\ \bibinfo {author} {\bibfnamefont {J.~A.}\ \bibnamefont
  {Thomas}},\ }\href@noop {} {\emph {\bibinfo {title} {Elements of Information
  Theory 2nd Edition (Wiley Series in Telecommunications and Signal
  Processing)}}}\ (\bibinfo  {publisher} {Wiley-Interscience},\ \bibinfo {year}
  {2006})\BibitemShut {NoStop}%
\bibitem [{\citenamefont {C\'aceres}\ \emph {et~al.}()\citenamefont
  {C\'aceres}, \citenamefont {Eccles}, \citenamefont {Pollack},\ and\
  \citenamefont {Racz}}]{caceres2024generic}%
  \BibitemOpen
  \bibfield  {author} {\bibinfo {author} {\bibfnamefont {E.}~\bibnamefont
  {C\'aceres}}, \bibinfo {author} {\bibfnamefont {S.}~\bibnamefont {Eccles}},
  \bibinfo {author} {\bibfnamefont {J.}~\bibnamefont {Pollack}},\ and\ \bibinfo
  {author} {\bibfnamefont {S.}~\bibnamefont {Racz}},\ }\bibfield  {title}
  {\bibinfo {title} {{Generic ETH: Eigenstate Thermalization beyond the
  Microcanonical}},\ }\href@noop {} {\ }\Eprint
  {https://arxiv.org/abs/2403.05197} {arXiv:2403.05197 [quant-ph]} \BibitemShut
  {NoStop}%
\bibitem [{\citenamefont {Guryanova}\ \emph {et~al.}(2016)\citenamefont
  {Guryanova}, \citenamefont {Popescu}, \citenamefont {Short}, \citenamefont
  {Silva},\ and\ \citenamefont {Skrzypczyk}}]{guryanova2016thermodynamics}%
  \BibitemOpen
  \bibfield  {author} {\bibinfo {author} {\bibfnamefont {Y.}~\bibnamefont
  {Guryanova}}, \bibinfo {author} {\bibfnamefont {S.}~\bibnamefont {Popescu}},
  \bibinfo {author} {\bibfnamefont {A.~J.}\ \bibnamefont {Short}}, \bibinfo
  {author} {\bibfnamefont {R.}~\bibnamefont {Silva}},\ and\ \bibinfo {author}
  {\bibfnamefont {P.}~\bibnamefont {Skrzypczyk}},\ }\bibfield  {title}
  {\bibinfo {title} {Thermodynamics of quantum systems with multiple conserved
  quantities},\ }\href {https://doi.org/10.1038/ncomms12049} {\bibfield
  {journal} {\bibinfo  {journal} {Nature communications}\ }\textbf {\bibinfo
  {volume} {7}},\ \bibinfo {pages} {12049} (\bibinfo {year}
  {2016})}\BibitemShut {NoStop}%
\bibitem [{\citenamefont {Ng}\ and\ \citenamefont
  {Woods}(2019)}]{ng2019resource}%
  \BibitemOpen
  \bibfield  {author} {\bibinfo {author} {\bibfnamefont {N.~H.~Y.}\
  \bibnamefont {Ng}}\ and\ \bibinfo {author} {\bibfnamefont {M.~P.}\
  \bibnamefont {Woods}},\ }\bibfield  {title} {\bibinfo {title} {Resource
  theory of quantum thermodynamics: Thermal operations and second laws},\ }in\
  \href {https://link.springer.com/chapter/10.1007/978-3-319-99046-0_26} {\emph
  {\bibinfo {booktitle} {Thermodynamics in the quantum regime: Fundamental
  aspects and new directions}}}\ (\bibinfo  {publisher} {Springer},\ \bibinfo
  {year} {2019})\ pp.\ \bibinfo {pages} {625--650}\BibitemShut {NoStop}%
\bibitem [{\citenamefont {Lin}\ and\ \citenamefont
  {Huang}(2024)}]{lin2024work}%
  \BibitemOpen
  \bibfield  {author} {\bibinfo {author} {\bibfnamefont {F.-L.}\ \bibnamefont
  {Lin}}\ and\ \bibinfo {author} {\bibfnamefont {C.-Y.}\ \bibnamefont
  {Huang}},\ }\bibfield  {title} {\bibinfo {title} {{Work statistics for
  quantum spin chains: Characterizing quantum phase transitions, benchmarking
  time evolution, and examining passivity of quantum states}},\ }\href
  {https://doi.org/10.1103/PhysRevResearch.6.023169} {\bibfield  {journal}
  {\bibinfo  {journal} {Phys. Rev. Res.}\ }\textbf {\bibinfo {volume} {6}},\
  \bibinfo {pages} {023169} (\bibinfo {year} {2024})},\ \Eprint
  {https://arxiv.org/abs/2308.13366} {arXiv:2308.13366 [cond-mat.stat-mech]}
  \BibitemShut {NoStop}%
\bibitem [{\citenamefont {Yuste}\ \emph {et~al.}(2018)\citenamefont {Yuste},
  \citenamefont {Cartwright}, \citenamefont {Chiara},\ and\ \citenamefont
  {Sanpera}}]{Yuste_2018}%
  \BibitemOpen
  \bibfield  {author} {\bibinfo {author} {\bibfnamefont {A.}~\bibnamefont
  {Yuste}}, \bibinfo {author} {\bibfnamefont {C.}~\bibnamefont {Cartwright}},
  \bibinfo {author} {\bibfnamefont {G.~D.}\ \bibnamefont {Chiara}},\ and\
  \bibinfo {author} {\bibfnamefont {A.}~\bibnamefont {Sanpera}},\ }\bibfield
  {title} {\bibinfo {title} {Entanglement scaling at first order quantum phase
  transitions},\ }\href {https://doi.org/10.1088/1367-2630/aab2db} {\bibfield
  {journal} {\bibinfo  {journal} {New Journal of Physics}\ }\textbf {\bibinfo
  {volume} {20}},\ \bibinfo {pages} {043006} (\bibinfo {year}
  {2018})}\BibitemShut {NoStop}%
\bibitem [{\citenamefont {Santos}\ and\ \citenamefont
  {Torres-Herrera}(2018)}]{santos2018nonequilibrium}%
  \BibitemOpen
  \bibfield  {author} {\bibinfo {author} {\bibfnamefont {L.~F.}\ \bibnamefont
  {Santos}}\ and\ \bibinfo {author} {\bibfnamefont {E.~J.}\ \bibnamefont
  {Torres-Herrera}},\ }\bibfield  {title} {\bibinfo {title} {Nonequilibrium
  quantum dynamics of many-body systems},\ }\href@noop {} {\bibfield  {journal}
  {\bibinfo  {journal} {Chaotic, Fractional, and Complex Dynamics: New Insights
  and Perspectives}\ ,\ \bibinfo {pages} {231}} (\bibinfo {year} {2018})},\
  \Eprint {https://arxiv.org/abs/1706.02031} {arXiv:1706.02031} \BibitemShut
  {NoStop}%
\end{thebibliography}
%

\end{document}